\documentclass[twocolumn,pre,showpacs,preprintnumbers,amsmath,amssymb,nofootinbib,superscriptaddress,floatfix]{revtex4-1}
\usepackage{amsfonts}
\usepackage{amsmath}
\usepackage{amssymb}
\usepackage{graphicx}
\usepackage{dcolumn}
\usepackage{bm}
\usepackage{longtable}
\usepackage[usenames, dvipsnames, table]{xcolor}
\begin{document}
%\preprint{}
\title{Community detection for correlation matrices}

%\author{}
%\thanks{}
%\affiliation{}
\author{Mel MacMahon}
%\email[]{}
\affiliation{Lorentz Institute for Theoretical Physics, University of Leiden (The Netherlands)}
\author{Diego Garlaschelli}
\affiliation{Lorentz Institute for Theoretical Physics, University of Leiden (The Netherlands)}
\affiliation{CABDyN Complexity Centre, Sa\"{i}d Business School, University of Oxford (UK)}
\date{\today}
\begin{abstract}
A challenging problem in the study of complex systems is that of resolving, without prior information, the emergent, mesoscopic organization determined by groups of units whose dynamical activity is more strongly correlated internally than with the rest of the system. 
The existing techniques to filter correlations are not explicitly oriented towards identifying such modules and can suffer from an unavoidable information loss. 
A promising alternative is that of employing community detection techniques developed in network theory. Unfortunately, this approach has focused predominantly on replacing network data with correlation matrices, a procedure that tends to be intrinsically biased due to its inconsistency with the null hypotheses underlying the existing algorithms.
Here we introduce, via a consistent redefinition of null models based on random matrix theory, the appropriate correlation-based counterparts of the most popular community detection techniques.
Our methods can filter out both unit-specific noise and system-wide dependencies, and the resulting communities are internally correlated and mutually anti-correlated.
We also implement multiresolution and multifrequency approaches revealing hierarchically nested sub-communities with `hard' cores and `soft' peripheries.
We apply our techniques to several financial time series and identify mesoscopic groups of stocks which are irreducible to a standard, sectorial taxonomy, detect `soft stocks' that alternate between communities, and discuss implications for portfolio optimization and risk management. 
\end{abstract}
\pacs{89.65.Gh; 89.75.Hc; 05.45.Tp; 02.50.Sk}
%\keywords{}
\maketitle
\section{Introduction}
Over the past couple of decades the amount of raw data available has started to grow at an exponential rate, doubling approximately every 12 months \cite{HowMuchInfo}, while the amount of data being consumed by users remains linear \cite{DataGrowth}. 
The so-called `Big Data' phenomenon imposes an urgent need to develop, possibly with the aid of high-speed computing and cheap data storage, efficient pattern detection methods and data mining techniques aimed at identifying a few but highly relevant pieces of information in an ever-increasing noisy or irrelevant background.
One of the most important and widespread examples of the Big Data phenomenon is time series data, as witnessed by the impressive growth of databases of electronic and mobile-device communication patterns in large social systems, financial returns in stock markets, physiological signals such as heartbeat and brain dynamics, gene expression profiles, and finally climate, weather and earthquake activity.
In all these examples, high-dimensional (multiple) time series originate from the dynamical activity of the constituent units (such as stocks, people, neurons, genes, etc.) of large systems with complicated internal interactions. 
For this reason, `Big time series Data' offer an unprecedented empirical resource for the science of complex systems.

Multiple time series are in fact the key ingredient required in order to face one of the main challenges for our modern understanding of real-world complex systems: the identification of an emergent, mesoscopic level of dynamical organization which is intermediate between the microscopic dynamics of indivual units (e.g. neurons) and the macroscopic dynamics of the system as a whole (e.g. the brain). 
Many complex systems are indeed organized in a modular way, with functionally related units being correlated with each other, while at the same time being relatively less (or even negatively) correlated with functionally dissimilar ones.
While the existence of such a modular organization is intuitively plausible, its empirical identification is still an open problem, complicated by the fact that modules are typically emergent, in the sense that they are not evident \emph{a priori} from a local inspection of static, or even dynamic, similarities or connections among individual units. 
In neuroscience, for instance, `functional brain networks' are precisely defined by the correlated dynamical activity of neurons, as opposed to `structural brain networks' which are instead defined by static neuronal connections \cite{neuro}.
Remarkably, it has been proposed that the observed divergence between functional and structural  brain networks represents a signature of the brain's many-to-one (degenerate) function-structure relationships which allow diverse functions to arise from a static neuronal anatomy \cite{neuro}.
Similarly, in the analysis of financial markets it has been observed that groups of correlated stocks evolve in time and only partially overlap with industrial sectors, implying that the (static) industrial classification fails to capture the dynamical modularity of real markets~\cite {Econophysics,FinancialRisk,PhysRevE.70.026110,2005AcPPB..36.2767P,PhysRevE.65.066126}. 

The approaches proposed so far to infer some form of modular or hierarchical organization from multiple time series are based on (necessarily arbitrary) criteria used to filter information \cite{mantegna,Econophysics,FinancialRisk}. 
As we discuss in more detail below, these filtering criteria are either the introduction of thresholds or a geometric embedding in some metric space with pre-defined properties. 
Our aim in the present paper is that of going beyond the limitations imposed by these arbitrary criteria. 
We propose that, both conceptually and algorithmically, the identification of mesocopic modules whose dynamical activity is more correlated internally than with that of other modules, requires iterated recursions into many attempted partitions of the system, an inherently non-local operation.
By their nature, threshold-based or geometric methods are unfortunately not suited to deal with this sort of iterative partitioning problem\footnote{One might think of community detection itself as a geometric method, but this is not what we mean here. The geometric methods we are referring to consist in embedding techniques that reduce the complexity of the original system by projecting the latter into some metric space of low dimensionality.}.

Our strategy towards a solution is the adaptation of a different class of rapidly developing techniques, specifically those aimed at identifying the static mesoscopic organization in complex networks, a problem known as \emph{community detection} \cite{Fortunato_2010,NewmanNetworks}. 
Communities within networks are groups of nodes that are more densely connected to each other than would be expected under a suitable null hypothesis. 
Additionally, the nodes within a community are less connected to the nodes within other communities of the same network.
Several methods have been proposed over the last decade in order to empirically detect communities within networks.
Different techniques have explored different ways to optimize the search over all possible partitions of the system.
Conceptually, these methods contain precisely the ingredients that we need in order to solve our problem of identifying the hidden mesoscopic organization encoded within multiple time series.
Adapting the existing community detection techniques to deal with time series data is the main goal of this paper.

While the idea of using community detection algorithms in order to analyse time series data has been already exploited a few times in the past \cite{saramaki_correlations,mason_correlations,isogai}, the attempts made so far have basically replaced network data with cross-correlation matrices.
Here we show that this procedure suffers from the limitation that the underlying null hypotheses used in network-based community detection algorithms are inconsistent with the properties of correlation matrices.
We illustrate that one of the undesired consequences is a systematic bias in the search over partitions, that becomes stronger as the heterogeneity of the size of the `true' communities increases.

Here we propose a solution to this problem by introducing appropriate redefinitions of the so-called \emph{modularity} \cite{Fortunato_2010}, the core quantity that most methods aim at maximizing when searching the space of possible partitions. 
While in ordinary community detection methods the modularity is defined in terms of a \emph{null model} that is (approximately) correct for networks, in the methods we propose the modularity is defined in terms of different null models that are appropriate for time series data and therefore dictated by random matrix theory (RMT) \cite{1955,2005AcPPB..36.2767P,RMT}.
We also adapt three popular algorithms that have been proposed to find the optimal partition (in networks), i.e. the one that maximizes the modularity. 
As a result, we end up with three community detection algorithms that are consistent with time series data and represent the counterparts of the most popular techniques used in network analysis.
We also provide extensions to resolve hierarchically nested subcommunities (multiresolution community detection) and `hard' cores versus `soft' peripheries inside communities (multifrequency and time dependent community detection).

After introducing our theoretical framework, we put special emphasis on financial applications, where the units of the system are assets and the corresponding time series are sequences of logarithmic price increments \cite{mantegna,Econophysics,FinancialRisk}. 
Even though advanced techniques to analyse correlations have been developed in other fields as well, financial time series analysis is one of the most active domains in this respect (another important example is that of functional brain networks, as we have already mentioned).
We show that our methods allow us to efficiently probe the mesoscopic structure of different financial markets and ascertain communities of corporations, based on the time series of their daily stock returns. 
We uncover a variety of correlations between stocks of different industry sectors, not intuitively obvious from the sectorial taxonomy alone, thus confirming in a more rigorous manner the aforementioned result that market correlations only partially overlap with industry classifications.
More importantly, the communities we detect after removing noisy and market-wide dependencies turn out to be internally correlated and mutually anti-correlated, a feature of particular relevance for portfolio optimization and risk management.
We also analyse the stability of communities over different frequency resolutions and time horizons, thereby identifying groups of `hard stocks' that reside stably in the core of communities and groups of `soft stocks' that alternate between communities.

The rest of the paper is organized as follows: in section \ref{sec:existing} we briefly describe the most important approaches that have been proposed in order to filter correlation matrices and highlight their issues with characterizing the modular properties of systems described by multiple time series.
In section \ref{sec:inconsistencies} we show that the existing community detection algorithms are based on a null hypothesis that is inconsistent for time series data, making these methods inadequate as well.
In sec. \ref{sec:methods} we then introduce alternative and appropriate null models based on RMT, and exploit them in order to redefine three of the most popular community detection algorithms, in a way that makes them consistent with time series data.
In sec. \ref{sec:results} we apply our methods to several time series of daily stock returns, from various financial markets around the globe. 
In sec. \ref{sec:resol} we analyse the dependence of community structure on the temporal resolution (i.e. the frequency) of the original time series.
In sec. \ref{sec:dyn} we investigate the evolution of community structure over time.
Finally, in sec. \ref{sec:conclusions} we summarize our results and provide some conclusions.

\section{Existing approaches\label{sec:existing}}
We start by introducing some useful notation. Let us consider a system with $N$  units. 
The single time series 
\begin{equation}
X_i\equiv\{x_i(1),x_i(2),\dots, x_i(T)\}
\end{equation}
represents the temporally ordered activity of the $i$-th unit of the system over $T$ timesteps. 
In the case of financial markets, $i$ is typically one particular stock and $x_i(t)$ is the `log-return' of stock $i$, i.e. the difference between the logarithms of the price of $i$ at times $t$ and $t-1$ (more details will be given later). 

The whole set of $N$ time series, denoted by $\{X_1,X_2,\dots, X_N\}$, describes the synchronous activity of all the units of the system.
The vast majority of the available techniques aimed at quantifying the level of mutual dependency within such a set of multiple time series exploit the information encoded in the $N\times N$ \emph{cross-correlation matrix}.
The cross-correlation matrix $\mathbf{C}$ measures the mutual dependencies among $N$ time series on a scale between $-1$ and $1$. 
The $ij^{th}$ entry of $\mathbf{C}$ is defined as the Pearson correlation coefficient
\begin{equation}
{C}_{ij} \equiv \textrm{Corr}[X_i,X_j] \equiv\frac{\textrm{Cov}[X_i,X_j]}{\sqrt{\textrm{Var}[X_i]\cdot \textrm{Var}[X_j]}},
\label{eq:corrg}
\end{equation}
where 
\begin{equation}
\textrm{Cov}[X_i,X_j]\equiv \overline{X_i X_j}-\overline{X_i}\cdot\overline{X_j}
\label{eq:cov}
\end{equation}
is the covariance of $X_i$ and $X_j$ and 
\begin{equation}
\textrm{Var}[X_i]\equiv\sigma^2_i\equiv\overline{X_i^2}-\overline{X_i}^2=\textrm{Cov}[X_i,X_i]
\end{equation}
is the variance of $X_i$.
In the above equations, the bar denotes a temporal average, i.e.
\begin{eqnarray}
\overline{X_i}&\equiv& T^{-1}\sum_{t=1}^T x_i(t),\\
\overline{X^2_i}&\equiv& T^{-1}\sum_{t=1}^T x^2_i(t),\\
\overline{X_i X_j} &\equiv&  T^{-1}{\sum_{t=1}^T x_i(t)x_j(t)}.
\end{eqnarray}
Clearly, the diagonal entries of the correlation matrix are ${C}_{ii} =1$.

We will assume, as routinely done in order to filter out the intrinsic heterogeneity of time series, that each series $X_i$ has been \emph{standardized} by subtracting out the temporal average $\overline{X_i}$ and dividing the result by the  standard deviation $\sigma_i$, i.e. that $X_i$ has been redefined to $(X_i-\overline{X_i})/\sigma_i$.
Then the following expressions hold:
\begin{eqnarray}
&\overline{X_i}= 0,&\\
&\textrm{Var}[X_i]= \overline{X^2_i}=1,&\\
&{C}_{ij}=\textrm{Cov}[X_i,X_j] = \overline{X_i X_j}.&
\label{eq:corrs}
\end{eqnarray}

Note that, despite in statistics the notation $\textrm{Corr}[X_i,X_j]$, $\textrm{Cov}[X_i,X_j]$ or $\textrm{Var}[X_i]$ usually denotes a \emph{population} value, i.e. a theoretical value calculated using the knowledge of the (joint) probability distributions for $X_i$ and $X_j$, all quantities we have defined so far are instead \emph{sample} quantities, i.e. measured on the specific realized values of a set of time series.
Our choice of a somewhat unconventional notation is merely due to the fact that it allow us to describe various operations more compactly.
We will need to denote the population value of a quantity only in a few cases, and when this happens such population value will coincide with the expected value $\langle f(X,Y,\dots)\rangle$ (over the joint probability distribution of the random variables $X,Y,\dots$ involved) of the corresponding sample quantity $f(X,Y,\dots)$. We will therefore directly express population quantities in terms of expected values when necessary.

We stress that empirical cross-correlation matrices are intrinsically limited by the fact that they assume \emph{temporally stationary} and \emph{linearly interdependent} time series. 
Clearly, both assumptions are in general violated in real financial markets and many other complex systems.
Nonetheless, cross-correlations are still the most widely used quantity. Improving the definition of correlations is a very important open problem, but is beyond the scope of this paper. 
Here, we want to overcome the limitations encountered when the methods introduced so far to process or filter correlation matrices are used in order to identify a mesoscopic modular structure. 
These current limitations are in place even when correlations are an appropriate measure, i.e. for stationary and linearly interdependent time series.
Therefore, our goal is that of introducing a consistent methodology that makes optimal use of correlation matrices in order to resolve the mesoscopic organization of complex systems.
If improved measures of interdependency are introduced, our approach will still represent a valuable guideline in order to implement a consistent community detection framework in that case as well.

In what follows, we review the most important correlation-based approaches and their limitations. 
We will put special emphasis on financial time series, even if our discussion is more general.

\subsection{Asset Graphs\label{sec:AG}}
Among the proposed approaches to filter cross-correlation matrices, the simplest one is perhaps that of focusing on the strongest (off-diagonal) correlations by introducing a threshold value and discarding all the correlations below the threshold. 
The result can be represented as a network, also known as an \emph{Asset Graph} (AG) in the Econophysics literature \cite{ag1,maxspanasset,Econophysics}, connecting the nodes whose time series are more strongly correlated.
Since the method entirely depends on the choice of the threshold, one usually investigates how the properties of the AG change as the threshold is varied.
The method is quite robust to noise, precisely  because it discards the weakest correlations that are more subject to random fluctuations.
However, for the same reason it fails in detecting a mesoscopic organization (if present) of the system. 
In fact, the use of a global threshold prevents the identification of modules whose internal correlations, even if below the threshold because they are weak with respect to the strongest ones, are still significantly stronger than the external correlations with different modules.
Therefore, while valuable as a filtering technique, the AG discards a significant amount of information and is not best suited to detect emergent groups of  correlated time series.
We provide additional information about AGs, along with an explicit example, when we analyze real financial data in sec.\ref{sec:standard}.

\subsection{Minimal Spanning Trees}
Another filtering approach looks for the \emph{Minimal Spanning Tree} (MST) obtained again from the strongest correlations, but now retaining only the $N-1$ correlations that are required for each node to be reachable from any other node via a connected path, while discarding those that produce loops \cite{mantegna}.
This procedure automatically produces an agglomerative hierarchical clustering (a dendrogram) of the original time series and requires that the correlation matrix is `renormalized' at each iteration of the clustering according to some protocol (the one having some distinct theoretical advantage is the so-called Single-Linkage clustering algorithm \cite{mantegna}), until a final filtered matrix is obtained.

The MST method does not require the introduction of an arbitrary threshold, but it assumes that the original correlations are well approximated by the filtered ones.
At a geometrical level, this corresponds to the assumption that the metric space in which the original time series are embedded (via the definition of a proper correlation-based distance) effectively reduces to a so-called \emph{ultrametric} space where well-separated clusters of points are hierarchically nested within larger well-separated clusters \cite{ultrametricity}. 

Even if the method exploits the correlations required for the MST to span the entire set of time series, it discards all the weaker correlations.
Moreover, the approximating (renormalized) correlations are progressively more distant from the original ones as higher and higher levels of the taxonomic tree are resolved.
This means that the method is more reliable when using the strongest correlations to determine the low-level structure of the taxonomic tree (small clusters of time series), while it is progressively less reliable when using the weaker correlations to determine the high-level taxonomy (medium-sized and large clusters). 

With the above warning in mind, the method allows one to identify correlated groups of stocks lying on separate `branches' of the MST or that become disconnected when the associated dendrogram is cut at some level.
However, this comes at the price of introducing an arbitrary threshold on the value of the correlation again. 
Moreover, just like the AG technique, the MST one does not compare internal and cross-group correlations (possibly with the aid of a null model) in order to identify emergent mesoscopic modules. 

\subsection{Planar Maximally Filtered Graphs}

An alternative approach, which is similar in spirit to the MST but discards less information, is the so-called \emph{Planar Maximally Filtered Graph} (PMFG) \cite{planar0,planar}.
This method allows one to retain not just the correlations required to form the MST, but also a number of additional ones, provided that the resulting structure is a \emph{planar graph} (a network that can be drawn on a plane without creating intersecting links).

A nice feature of the PMFG is that it always contains the entire MST, so that the former provides additional, and not just different, information with respect to the latter. 
However, also this method is affected by some degree of arbitrariness, which lies again in the properties of the postulated, approximating structure. There is no obvious reason why stocks (or other time series) should find a natural embedding in a bidimensional plane.
In fact, the PMFG has also been described as the simplest case of a more general procedure based on the embedding of high-dimensional data in lower-dimensional manifolds with a controllable \emph{genus} (number of `handles' or `holes') \cite{planar}.
The PMFG corresponds to the case when the genus is zero.
So the arbitrariness of the method can be rephrased as its dependence on some value of the genus that must be fixed \emph{a priori}.

The method has been extended in a variety of ways in order to produce a nested hierarchy of time series by exploiting the properties of the embedding space \cite{tiziana1,tiziana2,tiziana3}.
However, as with the MST, the target of these methods is that of finding the postulated approximating structure, rather than optimizing the search of groups of time series that are more correlated internally than with each other.

\subsection{Random matrix theory\label{sec:RMT}}
We finally mention an important technique, based on random matrix theory (RMT)~\cite{1955,2005AcPPB..36.2767P,RMT}, which is widely used in order to identify the non-random properties of empirical correlation matrices.
We will use this technique extensively in the paper.
A correlation matrix constructed from $N$ completely random time  series of duration $T$ has (in the limits $N\to+\infty$ and $T\to+\infty$ with $1<T/N<+\infty$) a very specific distribution of its eigenvalues, known as the Marcenko-Pastur or Sengupta-Mitra distribution~\cite{PhysRevLett.83.1471,PhysRevLett.83.1467}. This distribution reads
\begin{equation}\label{RMTDensity}
\rho(\lambda) = \frac{T}{N}\frac{\sqrt{(\lambda_{+} - \lambda)(\lambda - \lambda_{-})}}{2\pi\lambda}\quad \textrm{if}\quad \lambda_{+} \le \lambda\le \lambda_{-}
\end{equation}
and $\rho(\lambda) =0$ otherwise, where the maximum ($\lambda_{+}$) and minimum ($\lambda_{-}$) eigenvalues are given by 
\begin {equation}
\lambda_{\pm} = \left[1 \pm \sqrt{\frac{N}{T}}\right]^2.
\label{eq:lambda+-}
\end{equation}
The bulk of the eigenvalues of an empirical correlation matrix that fall within the range $[\lambda_-,\lambda_+]$ can be considered to be mostly due to random noise. Thus, any eigenvalues larger than the maximum eigenvalue $\lambda_{+}$ predicted by the Marcenko-Pastur distribution are deemed to represent meaningful structure in the data ~\cite {PhysRevE.70.026110,2005AcPPB..36.2767P,PhysRevE.65.066126}. That being the case, any empirical correlation matrix $\mathbf{C}$ can be decomposed as the sum of two matrices: 
\begin{equation}
\mathbf{C}=\mathbf{C}^{(r)}
+\mathbf{C}^{(s)},
\label{eq:rmt+}
\end{equation}
where (using $\langle bra |$ and $| ket\rangle$ notation)
\begin{equation}
\mathbf{C}^{(r)}\equiv\sum_{i:\lambda_i\leq\lambda_+}\lambda_i |v_i\rangle\langle v_i|
\label{eq:cr}
\end{equation}
is the `random' component constituted from the eigenvalues $\{\lambda_i\}$ less than or equal to $\lambda_{+}$ (usually, the eigenvalues smaller than $\lambda_-$ are included as well) and their corresponding eigenvectors $\{|v_i\rangle\}$, and $\mathbf{C}^{(s)}=\mathbf{C}-\mathbf{C}^{(r)}$ is the `structured' component constituted from the remaining eigenvalues corresponding to eigenvalues larger than $\lambda_+$. 

\begin{figure}[t]
\centerline{\includegraphics[width = .48\textwidth]{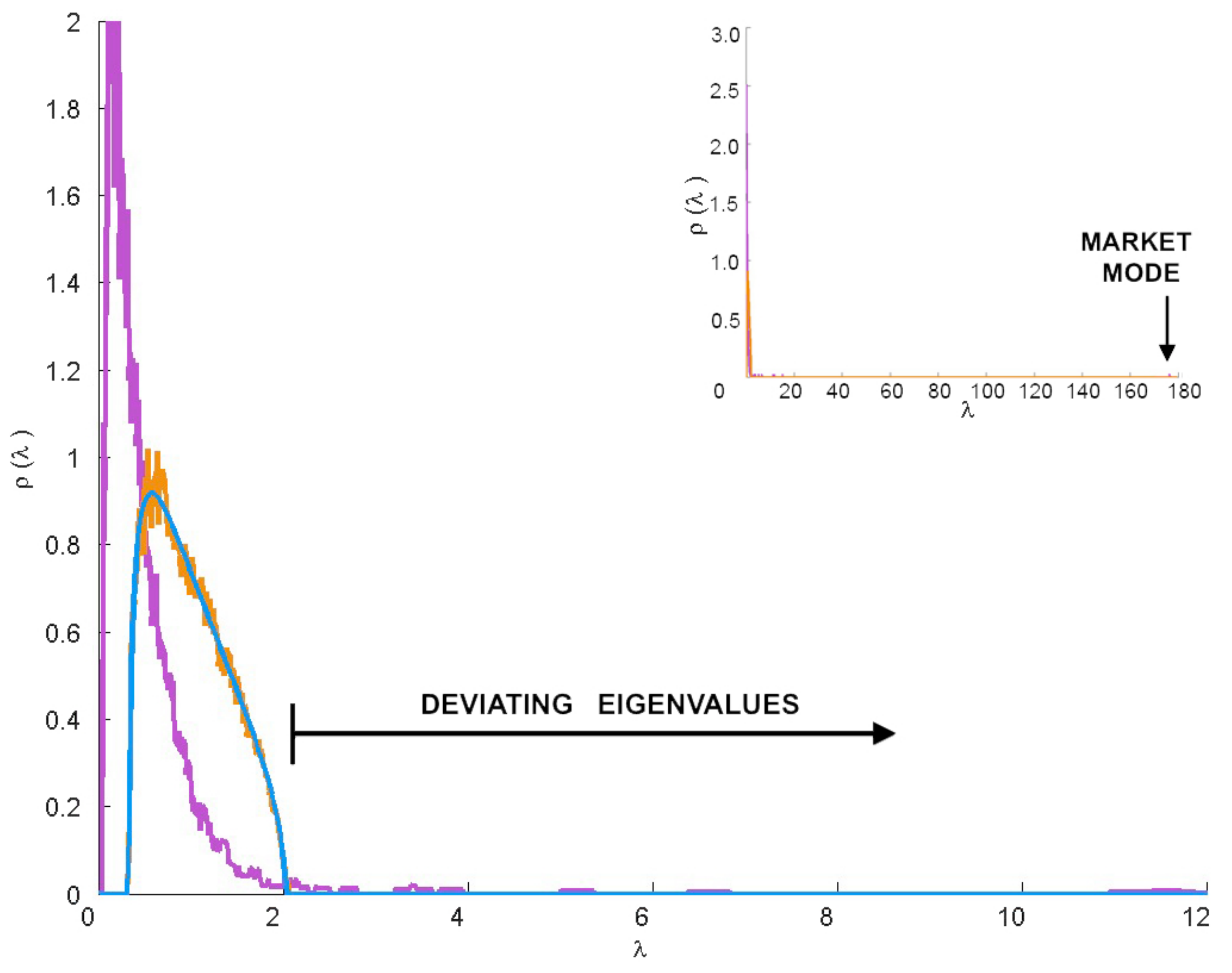}}
\caption{The eigenvalue density of the empirical correlation matrix of $T=2500$ log-returns (of daily closing prices from 2001Q4 to 2011Q3) for $N=445$ stocks of the S\&P500 index (purple) and the Marcenko-Pastur prediction for a random correlation matrix with the same values of $N$ and $T$ (blue), denoting a maximum expected eigenvalue of approximately 2. The orange plot is the eigenvalue density obtained by randomly shuffling (i.e. permuting with uniform probability) the empirical increments within each of the observed time series, confirming the agreement with random matrix theory for uncorrelated data. The inset is the fully zoomed-out version of the plot, showing that the empirical correlation matrix has a maximum eigenvalue of about 175 (`market mode'), as well as a handful of other leading eigenvalues above the predicted maximum value.}
\label{fig:eigendensity}
\end{figure}

The deviation of the spectra of real correlation matrices from the RMT prediction provides an effective way to filter out noise from empirical data, and also illustrates some robust property of financial markets. 
For instance, in Fig. \ref{fig:eigendensity} we superimpose the eigenvalue density of the empirical correlation matrix obtained from $T=2500$ log-returns of daily closing prices of $N=445$ stocks of the S\&P 500 index (from 2001 to 2011) and the corresponding expectation given by the Marcenko-Pastur distribution with the same values of $N$ and $T$.
As also observed in a multitude of previous studies \cite{Econophysics,FinancialRisk}, a typical feature of the spectrum of empirical correlation matrices is that the largest observed eigenvalue $\lambda_m$ is much larger than all other eigenvalues (see inset of fig. \ref{fig:eigendensity}). 
The corresponding eigenvector $|v_m\rangle$ has all positive signs and one can therefore identify this eigencomponent of the correlations as the so-called \emph{market mode} ~\cite{Econophysics,FinancialRisk}, i.e. a common factor influencing all stocks within a given market. 
Interpreting this, the bulk of the correlation between pairs of stocks is attributed to a single common factor, much as all boats in a harbor will rise and fall with the tide. 
%%%%%%
%%%%%%%%%

In order to clearly see which `boats' are rising and falling relative to one another, one must subtract out the common `tide', which in terms of the correlation matrix leads to the further decomposition
\begin {equation}
\mathbf{C} =  \mathbf{C}^{(r)} + \mathbf{C}^{(g)} + \mathbf{C}^{(m)}, 
\label{eq:rmt++}
\end {equation}
where we have rewritten the structured component as $\mathbf{C}^{(s)}=\mathbf{C}^{(g)} + \mathbf{C}^{(m)}$, with
\begin{equation}
\mathbf{C}^{(m)}\equiv\lambda_m |v_m\rangle\langle v_m|
\label{eq:cm}
\end{equation}
(representing the `market' mode) and
\begin{equation}
\mathbf{C}^{(g)}\equiv\sum_{i:\lambda_+<\lambda_i<\lambda_m}\lambda_i |v_i\rangle\langle v_i|
\label{eq:cg}
\end{equation}
(representing the remaining correlations).

The correlations embodied by $\mathbf{C}^{(g)}$ act neither at the level of individual stocks (uncorrelated noise), nor at that of the entire market.
Such correlations act at the level of sub-groups of stocks within a market, and they are often referred to as the `group' mode \cite{PhysRevLett.83.1467,Econophysics}.
The eigenvectors contributing to $\mathbf{C}^{(g)}$ have alternating signs, and this allows the identification of groups of stocks that are influenced in a similar manner by one or more common factors ~\cite {PhysRevE.70.026110,2005AcPPB..36.2767P,PhysRevE.65.066126}. 
Broadly speaking, these groups are expected to reflect some sectorial or sub-sectorial classification of stocks according to their industrial category, however the overlap between nominal asset classes and groups of empirically correlated stocks is only partial~\cite {PhysRevE.70.026110,2005AcPPB..36.2767P,PhysRevE.65.066126}.

We should at this point stress that the above discussion makes some strong assumptions, which have been recently criticized. 
In particular, the interpretation of the largest eigenvalue in terms of a `market' mode and the assumption that the elimination of the market and `noise' modes does not alter the information present in the remaining subspace are not correct in general, and sometimes only approximate  \cite{tumminello_spectral,tumminello_nested}. 
Moreover, the eigencomponents of the correlation matrix, and consequently the filtered correlation matrix itself, can end up being not proper correlation matrices, and alternative constructions enforcing the required properties have been proposed \cite{higham,rebonato,simonian,shrinkage}.
Finally, the way to filter out the `market' and `noise' modes is not unique \cite{tumminello_KL}.

Bearing these limitations is mind, RMT is still to be considered a valuable tool to filter empirical correlation matrices and clean them from both stock-level (random) and market-wide fluctuations. 
However, after this pre-processing, filtered correlation data still needs to be analyzed according to the particular research question.
For instance, the matrix $\mathbf{C}^{(g)}$ is often processed further and used as an alternative, filtered input in all the algorithms (AG, MST and PMGF) described above.
So RMT alone is not enough in order to resolve the mesoscopic organization of markets, in the sense defined above. 

\section{Community detection in graphs and its inconsistency with correlation matrices \label{sec:inconsistencies}}
In the previous section, we clarified that many of the available techniques used to identify the most relevant correlations are not designed to isolate groups of time series whose dynamical activity is more correlated internally than with that of other groups. 
At an abstract level, achieving this task would require iterated recursions into many attempted partitions of the system, an inherently non-local and computationally demanding operation.
Notably, an entire branch of Network Science is devoted to an analogous problem, known as \emph{community detection} \cite{Fortunato_2010}.
In this section, we briefly illustrate the principles of community detection in networks and show how that knowledge can be in principle transferred to our initial problem, namely the identification of a mesoscopic organization across multiple time series. 
We also show that despite the many progressive inroads made in this direction so far, they often rely on an inherently biased approach.

\subsection{Community detection in networks}
In network analysis, community detection is the process of identifying relatively dense clusters of nodes.
There has been a flurry of research in the area of community detection over the last decade \cite{Fortunato_2010}. 
In this paper we focus on the method of \emph{modularity optimization} ~\cite{PhysRevE.69.026113}, which is one of the most popular methods identifying \emph{non-overlapping} communities. 
It should be noted that various alternative methods other than modularity optimization exist, including techniques that resolve overlapping communities \cite{Fortunato_2010,communities_mason}.
However, this method has the advantage of being based on a null model, acting as a community-free benchmark to which the real network is compared.
It is the appropriate modification of such benchmarks that will lead us, in sec.\ref{sec:methods}, to a redefinition of modularity optimization methods valid for correlation matrices.

We restrict ourselves to undirected networks, since they exhibit the same symmetry property as correlation matrices. 
Given a network with $N$ nodes, one can introduce a number of partitions of the $N$ nodes into non-overlapping sets. Each such partition can be mathematically represented by an $N$-dimensional vector $\vec{\sigma}$ where the $i$-th component $\sigma_i$ denotes the set in which node $i$ is placed by that particular partition.
Then, one can introduce the so-called \emph{modularity} $Q(\vec{\sigma})$ as a measure of the effectiveness of a particular partition $\vec{\sigma}$ in identifying densely connected groups of nodes. 
The process of modularity optimization seeks to find the optimal partition that maximizes the value of $Q(\vec{\sigma})$, by varying the communities to which the different nodes of the network belong. 
The modularity $Q(\vec{\sigma})$ is expressed in the form
\begin {equation} \label{QNewman}
Q (\vec{\sigma})= \frac {1}{A_{tot}}\sum_{i,j} \Big[A_{ij} -\langle A_{ij}\rangle \Big] \delta(\sigma_i,\sigma_j),
\end {equation}
where, here and throughout the paper, the sum is intended to run over \emph{all} pairs of nodes even if we are considering undirected networks, and we are also including the diagonal elements corresponding to $i=j$ since many expressions become simpler with this choice. The meaning of the different terms of the above expression is as follows.
The delta function is $\delta(\sigma_i,\sigma_j)=1$ if $\sigma_i=\sigma_j$ and $\delta(\sigma_i,\sigma_j)=0$ if $\sigma_i\ne \sigma_j$, ensuring that only nodes within the same community contribute to the sum.
For binary networks, $A_{ij}$ is the entry of the adjacency matrix representing the presence ($A_{ij}=1$) or absence ($A_{ij}=0$) of a link between nodes $i$ and $j$ in the observed network.
The initial pre-factor works to normalize the value of $Q(\vec{\sigma})$ between $-1$ and $1$, where $A_{tot}\equiv\sum_{i,j} A_{ij}=2m$ is twice the total number $m$ of links.

The term $\langle A_{ij}\rangle$ is a key element determining the outcome of the entire community detection process.
It mathematically represents a null model for the network, i.e. an expectation for $A_{ij}$ under some suitable null hypothesis.
The most popular null model for a binary network, known as the \emph{Configuration Model}, is one where the expected value $\langle k_i\rangle$ of the degree $k_i$  (number of links) of each node $i$ is equal to the value $\sum_{j=1}^N A_{ij}$ observed in the real network and where the topology is otherwise completely random.
This null hypothesis ensures that the local heterogeneity of nodes, e.g. the fact that more popular people naturally have more friends in social networks, is appropriately controlled for.
Mathematically, this model is approximately (i.e. only when the heterogeneity of the degrees is weak \cite{mylikelihood,mymethod}) represented by the expression
\begin{equation}
\langle A_{ij}\rangle =\frac{k_i k_j}{2m},
\label{eq:null}
\end{equation}
which gives a rough estimate of the probability that nodes $i$ and $j$ are connected, \emph{under the null hypothesis that the observed network's structure is completely explained on the basis of the different degrees of vertices}. 

For weighted networks, $A_{ij}$ denotes the weight of the link between nodes $i$ and $j$, $k_i$ is called the  \emph{strength} of node $i$ and $2m$ is twice the total weight (of all links in the network). Still, eq.(\ref{eq:null}) is used without modifications \cite{Fortunato_2010} to determine the (again approximate \cite{mymethod,mybosefermi}) expected edge weight \emph{under the null hypothesis that the network's structure is completely explained on the basis of the observed strengths of all vertices}. 

The accuracy and usefulness of the results obtained from the process of modularity optimization depend heavily on the choice and suitability of the null model.
When the null hypothesis is true, no higher-order patterns (including communities) are present. 
Consistently, one expects the modularity in eq.(\ref{QNewman}) to be close to zero for every partition. 
In maximizing the modularity for a network which does have community structure, the nodes that are more tightly connected than one would expect on the basis of their individual characteristics will be clustered together in the same community, while the nodes for which the opposite occurs will be placed in different communities.

It should be noted that, in the context of network analysis, the modularity function defined in eq.(\ref{QNewman}) suffers from a main drawback: it cannot resolve communities below a typical scale  \cite{resolution}.
This \emph{resolution limit} was proven to be rooted in the specific mathematical form of eq.(\ref{eq:null}) used to represent the null model. 
However, it was not proven to be due to the concept of the null model itself, i.e. to the choice of comparing the real network with an ensemble of graphs with given degrees (or strenghts). 
In particular, we stress again that eq.(\ref{eq:null}) only approximately represents such an ensemble, the exact formula being a more complicated nonlinear equation \cite{mylikelihood,mymethod,mybosefermi}.  
Whether the resolution limit disappears if the exact expression is used in place of eq.(\ref{eq:null}) has never been investigated.
Rather, it has been proposed \cite{potts} that a way to change the resolution of the community detection is the introduction of an extra resolution parameter $\phi>0$ in the null model, i.e. replacing eq.(\ref{eq:null}) with 
\begin{equation}
\langle A_{ij}\rangle =\phi\frac{k_i k_j}{2m}.
\label{eq:multinull}
\end{equation}
Many studies have indeed shown that, as $\phi$ is varied, different hierarchical levels of the community structure can be revealed, so that a so-called \emph{multiresolution method} can be obtained \cite{potts,multiarenas,multilambiotte}.
In general, multiresolution methods can resolve smaller subcommunities, which are nested inside larger communities.
One should however bear in mind that the resolution parameter was originally introduced in an \emph{ad hoc} fashion and without a theoretical foundation, its main justification being an agreement \emph{a posteriori} with the hierarchical community structure expected in some real-world networks. Only later, it was shown to have some physical interpretation in terms of an inverse time required to explore the network under certain assumptions \cite{multilambiotte}.
When extending modularity-based algorithms to the analysis of multiple time series, we will address the problem of multiresolution community detection in a fundamentally different way, which avoids \emph{ad hoc} parameters and is theoretically consistent with the properties of correlation matrices (see sec. \ref{sec:multi}).

\subsection{The inconsistency of modularity for cross-correlation matrices\label{sec:inconsistency}}
The appealing properties of community detection in networks clearly have the potential to solve our initial problem of finding groups of time series that are more correlated than we would expect. 
However, one should be very careful in identifying the correlation-based problem with the network-based one. 
A na\"ive approach would be that of treating the empirical correlation matrix $\mathbf{C}$ as a weighted network, and looking for communities using the modularity as defined in eq.(\ref{QNewman}), i.e. setting $A_{ij}=C_{ij}$.
This would result in a modularity of the form
\begin {equation} \label{eq:Qcorr}
Q(\vec{\sigma}) = \frac {1}{C_{norm}}\sum_{i,j} \Big[C_{ij} -\langle C_{ij}\rangle \Big] \delta(\sigma_i, \sigma_j),
\end{equation}
where $C_{norm}=\sum_{i,j}C_{ij}$, $\langle C_{ij}\rangle =k_i k_j/C_{norm}$ and $k_i=\sum_{j=1}^N C_{ij}$.
This idea has been recently exploited, sometimes with modifications, to study communities of interest rates \cite{mason_correlations} and stocks \cite{saramaki_correlations,isogai} in financial markets.

Unfortunately, although the above approach has made a lot of headway, it suffers from some fundamental flaws and can lead to biased results, as we now show.
The problem arises because the null model defined in eq.(\ref{eq:null}), while (approximately \cite{mylikelihood,mymethod,mybosefermi}) valid when the matrix $\mathbf{A}$ describes a network, is inconsistent if $\mathbf{A}$ is replaced by a correlation matrix $\mathbf{C}$.
To see this, note that if $A_{ij}=C_{ij}$ and if $X_i$ denotes a standardized time series $i$ (see sec. \ref{sec:existing}), then eq.(\ref{eq:corrs}) implies  
\begin{equation}
k_i\equiv\sum_{j=1}^N C_{ij}=\sum_{j=1}^N \textrm{Cov}[X_i, X_j]=\textrm{Cov}[X_i,X_{tot}],
\label{eq:ki}
\end{equation}
where $X_{tot}=\{x_{tot}(1),x_{tot}(2),\dots,x_{tot}(T)\}$ is the time series of the total increment $x_{tot}(t)\equiv \sum_{j=1}^N x_j(t)$.
Note that, even if all $X_i$'s are standardized, $X_{tot}$ has zero mean but non-unit variance, and is therefore \emph{not} standardized.
Similarly, 
\begin{equation}
2m=\sum_{i=1}^N k_i =\textrm{Cov}[X_{tot},X_{tot}]=\textrm{Var}[X_{tot}].
\label{eq:2m}
\end{equation}
It then follows that
\begin{eqnarray}
\frac{k_i k_j}{2m}&=&\frac{\textrm{Cov}[X_i,X_{tot}]\cdot \textrm{Cov}[X_i,X_{tot}]}{\textrm{Var}[X_{tot}]}\nonumber\\
&=&\textrm{Corr}[X_i,X_{tot}]\cdot \textrm{Corr}[X_j,X_{tot}].
\label{eq:badnull1}
\end{eqnarray}

We therefore arrive at an important conclusion: for correlation matrices, the `na\"ive' modularity as ordinarily defined in eq.(\ref{QNewman}) with the ordinary specification given in eq.(\ref{eq:null}) corresponds to the following null hypothesis:
\begin{equation}
\langle C_{ij}\rangle_{naive}=\textrm{Corr}[X_i,X_{tot}]\cdot \textrm{Corr}[X_j,X_{tot}].
\label{eq:badnull2}
\end{equation}
When used within the modularity function, the above null model will not necessarily give more importance to pairs of strongly correlated time series, but rather to pairs of time series whose `direct' correlation $C_{ij}$ is larger than the product of the correlations of each time series with the `common signal' $X_{tot}$.
On the other hand, if we want to detect communities of time series that are empirically more correlated than expected under the hypothesis that all time series are independent of each other, we know that the correct null model (at least for infinitely long time series, a hypothesis that we will relax later) is
\begin{equation}
\langle C_{ij}\rangle=\delta_{ij},
\label{eq:goodnull}
\end{equation}
i.e. the expected correlation matrix $\langle\mathbf{C}\rangle$ should be the $N\times N$ identity matrix $\mathbf{I}$.
Other acceptable forms of $\langle C_{ij}\rangle$ based on realistic properties of correlation matrices will be discussed later.

The origin of the problematic discrepancy between eq.(\ref{eq:badnull2}) and eq.(\ref{eq:goodnull}) is the fact that the null model defined in eq.(\ref{eq:null}) is meant to represent networks with given degrees, i.e. matrices with given column and row sums. 
Any matrix that matches this constraint is admissible, in the sense that it represents a possible\footnote{This is true as long as the degrees are \emph{graphic}, i.e. realized by at least one network. This is always the case here, since the values of the degrees used in the null model are the observed ones and are therefore realized by at least the empirical network.}  network consistent with the hypothesis that degrees are an important structural constraint.
By contrast, sums over rows or columns of correlation matrices do not represent any meaningful constraint, as evident from eq.(\ref{eq:ki}).
Moreover, not every symmetric real matrix with given row and column sums is a possible correlation matrix. Correlation matrices must also be \emph{positive-semidefinite}, i.e. have non-negative eigenvalues.
A little algebra shows that eq.\eqref{eq:badnull2} fulfills this property, but in a very extreme way: the eigenvalues of the matrix having elements $\langle C_{ij}\rangle_{naive}$ are $\lambda=0$ (with multiplicity $N-1$) and 
\begin{equation}
\lambda=\sum_{i=1}^N\big(\textrm{Corr}[X_i,X_{tot}]\big)^2
\label{eq:eigenvalue}
\end{equation}
(with multiplicity $1$). 
This result holds irrespective of the original data, e.g. also for correlated and finite-length time series. 
Our discussion of the spectrum of realistic correlation matrices in sec. \ref{sec:RMT} strongly indicates that a sensible null model for correlation matrices should feature an eigenvalue distribution that is not easily reducible to the extremely simple one found above.

Similar conceptual limitations are encountered also in more sophisticated null models, which while allowing for both positive and negative link weights \cite{arenas}, still consider all possible matrices (many of which are inconsistent with correlation matrices) with given sums over rows and columns.
More importantly, the above problems cannot be solved by the introduction of resolution parameters. If, in analogy with eq.(\ref{eq:multinull}), we consider the generalized null model
\begin{equation}
\langle C_{ij}\rangle_{naive}=\phi\cdot \textrm{Corr}[X_i,X_{tot}]\cdot \textrm{Corr}[X_j,X_{tot}]
\label{eq:badmultinull}
\end{equation}
(with $\phi>0$), we are still left with an expression that cannot be reduced to eq.(\ref{eq:goodnull}) or some other meaningful alternatives, which we will introduce later in sec. \ref{sec:ourmod}.
For instance, the eigenvalues become $\phi\lambda$, where $\lambda$ still takes only the two values shown above.
Further aspects of this limitation are explicitly illustrated in a benchmark case below, and imply that appropriate multiresolution community detection methods for correlation matrices should be implemented in a completely different way (see sec. \ref{sec:multi}).

\subsection{The bias produced by the na\"ive approach\label{sec:bias}}
To have an idea of the consequence of using the na\"ive approach, i.e. the application of a network-based modularity directly to a cross-correlation matrix, we consider an ideal benchmark case where $N$ \emph{infinitely long} time series are divided into $c$ `true' communities, specified by a `true' partition $\vec{\sigma}^*$. 
We assume that each community $A$ is made of $n_A$ standardized time series (with $\sum_{A=1}^c n_A=N$) that are perfectly correlated with each other and completely uncorrelated to the time series in other communities, i.e. 
\begin{equation}
C_{ij}=\textrm{Corr}[X_i,X_j]=\textrm{Cov}[X_i,X_j]=\delta(\sigma_i^*,\sigma_j^*).
\end{equation}
In such a case, 
\begin{equation}
\textrm{Cov}[X_i,X_{tot}]=\sum_{j=1}^N \textrm{Cov}[X_i,X_j]=n_{\sigma_i^*}
\end{equation}
(where $n_{\sigma_i^*}$ is the number of time series in the community of the time series $i$) and
\begin{equation}
\textrm{Var}[X_{tot}]=\sum_{i,j}\textrm{Cov}[X_i,X_j]=\sum_{A=1}^c n^2_{A}.
\end{equation}

\begin{figure*}
\centerline{\includegraphics[width=.82\textwidth]{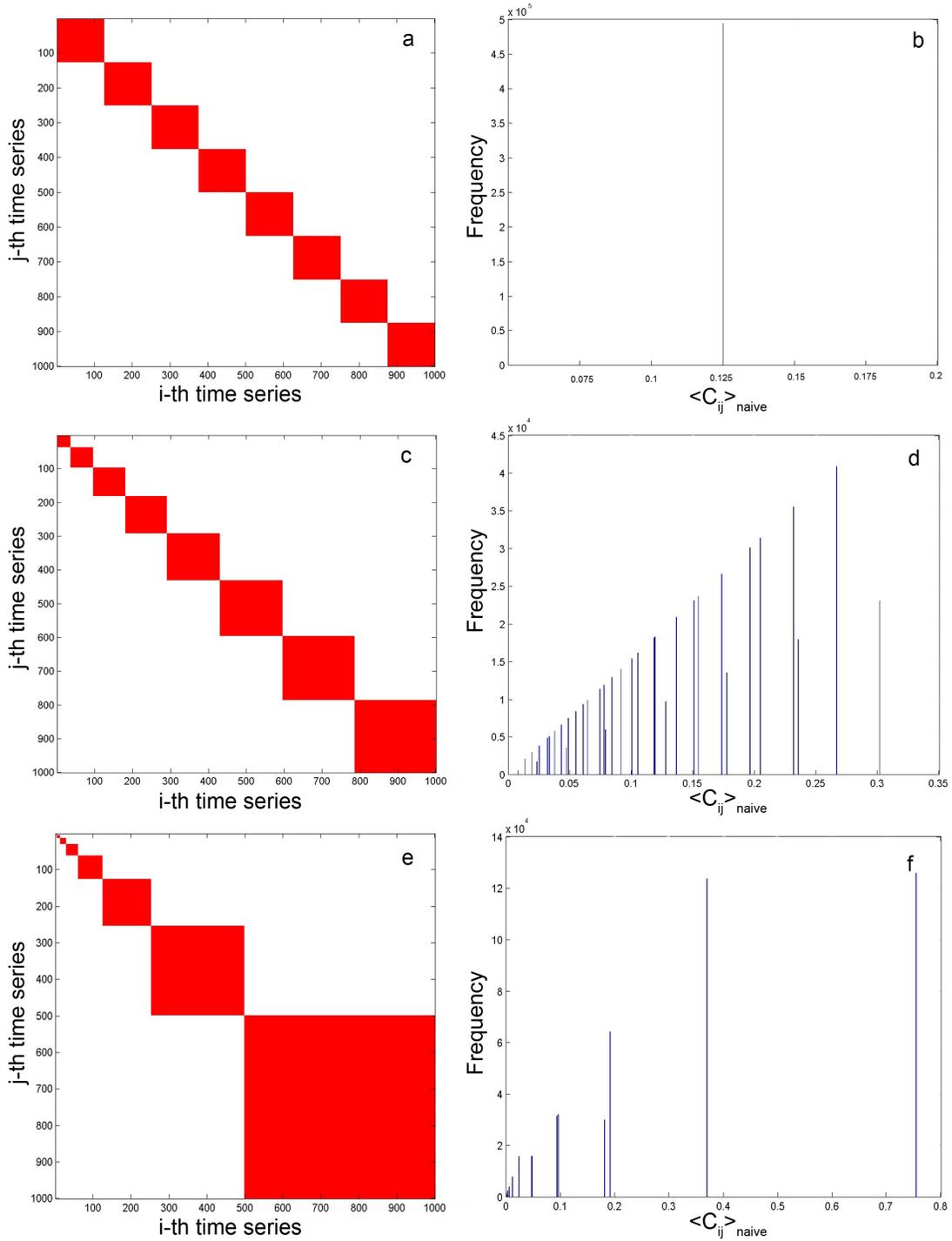}}
\caption{The biasing effect of the heterogeneity of community size on the na\"ive (network-based) modularity. The left panels show the entries $\delta(\sigma^*_i,\sigma^*_j)$ of three benchmark community matrices (white=0, red=1), each with $N=1000$ time series and $c=8$ communities of increasingly heterogeneous sizes. The right panels show the corresponding distribution of the off-diagonal entries $\langle C_{ij}\rangle_{naive}$ of the null model defined in eq.\eqref{eq:badnull3}, with $\phi=1$ (the bin size of the histograms is of the order of $10^{-5}$, which makes the distributions correctly normalized). 
For perfectly homogeneous community sizes, i.e. if each community contains exactly 125 time series (a), the distribution has a single peak at $1/c=0.125$ (b). For moderately heterogeneous sizes, i.e. if the 8 communities contain 35, 60, 85, 110, 140, 165, 190, and 215 time series respectively  (c), the distribution has several peaks (d) coming from the 64 different combinations of $n_{\sigma_i^*}n_{\sigma_j^*}$ in eq.\eqref{eq:badnull3}.
For strongly heterogeneous sizes, i.e. if the 8 communities contain 4, 8, 16, 32, 64, 128, 246, and 502 time series respectively (e), the distribution has still 64 different peaks but is much broader (f). The two dominant peaks are located at $\langle C_{ij}\rangle_{naive}=0.7536$ (coming from pairs of time series inside the largest community) and  $\langle C_{ij}\rangle_{naive}=0.3694$ (coming from pairs of time series across the largest and the second-largest communities).}\label{fig:offdiag}
\end{figure*}

From the last two equations it follows that eq.(\ref{eq:badnull2}), or more generally eq.\eqref{eq:badmultinull}, can be rewritten as
\begin{equation}
\langle C_{ij}\rangle_{naive}=\phi\frac{n_{\sigma_i^*}n_{\sigma_j^*}}{\sum_{A=1}^c n^2_{A}}
\label{eq:badnull3}
\end{equation}
(with $\phi>0$), which is the fundamental result showing the inconsistency of the na\"ive approach, and the nature of the resulting bias. Equation (\ref{eq:badnull3}) can never lead to the correct expectation (\ref{eq:goodnull})
because it cannot produce off-diagonal zeros. 
If there are $c$ equally sized communities of $n=N/c$ time series each, then $\langle C_{ij}\rangle_{naive}={\phi/c}$ for all $i,j$, i.e.  the distribution of $\langle C_{ij}\rangle_{naive}$ has a single peak and zero standard deviation. 
In this case, apart from the minor \footnote{For homogeneously sized communities, the problem of non-unit diagonal entries can be avoided if we require $i\ne j$ in the summation defining $Q$. However, for heterogeneously sized community the bias of the na\"ive approach cannot be eliminated.} problem of non-unit diagonal entries, the use of $\langle C_{ij}\rangle_{naive}$ in eq.(\ref{QNewman}) can still be justified on the basis of the fact that $\phi/c$ is a constant term having no effect on the modularity maximization.
However, for heterogeneously sized communities, eq.(\ref{eq:badnull3}) does not lead to a mere overall shift in the modularity.
As the size heterogeneity increases, the distribution of the off-diagonal entries of $\langle C_{ij}\rangle_{naive}$ will become broader.
In general, $\langle C_{ij}\rangle_{naive}$ is larger for pairs of time series belonging to larger communities. 
This effect is shown in Fig. \ref{fig:offdiag} for three choices of benchmark communities.

%%%%%%%%%%%
%%%%%%%%%%%%%%

The above consideration implies that the standard deviation (irrespective of the average) of the off-diagonal ($i\ne j$) entries of $\langle C_{ij}\rangle_{naive}$ can be taken as a quantitative measure of the \emph{bias} induced by eq.\eqref{eq:badnull3}.
This definition depends linearly on the multiresolution parameter $\phi$.
Alternatively, the \emph{coefficient of variation} (standard deviation divided by average value) of the off-diagonal entries of $\langle C_{ij}\rangle_{naive}$ is a measure of the \emph{relative bias} of the na\"ive approach, and is independent of $\phi$.
One should bear in mind that when the value of the coefficient of variation is much lower than one, the heterogeneity is moderate while when it approaches or exceeds one then the heterogeneity is such that the average value is no longer representative of the distribution.

\begin{figure}
\centerline{\includegraphics[width=.49\textwidth]{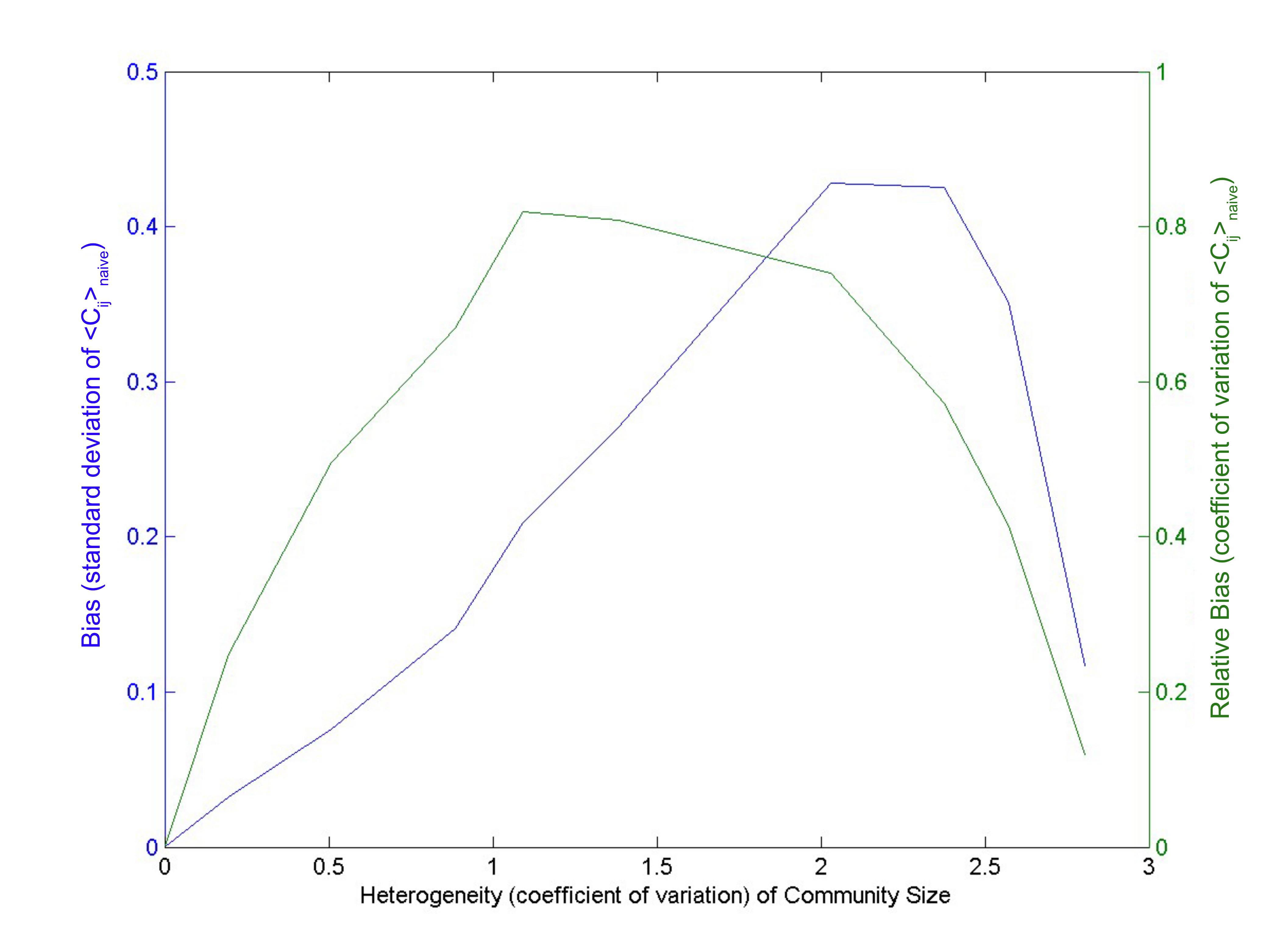}}
\caption{Dependence of the (relative) bias of the na\"ive approach on the heterogeneity (coefficient of variation) of community size, for various benchmarks with $N=1000$ time series and $c=8$ communities. 
The bias is defined as the standard deviation (coefficient of variation) of the off-diagonal entries $\langle C_{ij}\rangle_{naive}$ of the null model defined in eq.\eqref{eq:badnull3} with $\phi=1$, while the relative bias is defined as the coefficient of variation of the same entries (and is independent of $\phi$).}\label{fig:bias}
\end{figure}

In Fig. \ref{fig:bias} we show both the bias (for $\phi=1$) and the relative bias as a function of size heterogeneity, the latter being in turn defined as the coefficient of variation of community size.
We see that the (relative) bias first steadily increases as the size heterogeneity increases from zero to approximately two, and then decreases when the heterogeneity further increases. This decrease corresponds to  entering an extremely heterogeneous regime where there is a giant community of $O(N)$ nodes, and other very small communities of $O(1)$ nodes. 
In this regime, the effective number of communities is practically one and the distribution of $\langle C_{ij}\rangle_{naive}$ becomes sharp again, as most entries have the same value. 
So, for a very broad range of heterogeneity (say, when the coefficient of variation of community sizes is between 0.5 and 2.5), the (relative) bias is very strong. 
In this regime, eq.(\ref{eq:badnull3}) gives a prediction $\langle C_{ij}\rangle_{naive}\approx 0$ (close to the correct expectation) only for pairs of time series belonging to the smallest community. 
For such time series the difference $C_{ij}-\langle C_{ij}\rangle_{naive}$ is still close to one, and one therefore expects that the smallest community will be detected correctly.
However, for time series belonging to larger communities $\langle C_{ij}\rangle_{naive}$ increases, progressively biasing the community detection. 
For the largest community, the expected internal correlation is always larger than the correlation among any pair of communities, so $C_{ij}-\langle C_{ij}\rangle_{naive}$ is very low and this community is paradoxically difficult to detect. 

It should be noted that the use of the multiresolution parameter $\phi$ does not help reduce the relative bias, as the latter is independent of $\phi$.
In order to reduce the absolute bias (which for $\phi=1$ has values around $0.3$ in the relevant regime, see fig. \ref{fig:bias}) to small values (say of the order of $0.01$), $\phi$ should be set to very small values (around $0.03$), which is another way of saying that the null model in eq.\eqref{eq:badmultinull} should effectively be replaced by that in eq.\eqref{eq:goodnull} (we recall that we are referring only to the off-diagonal entries here), thus confirming our previous discussion. 

The above results lead us to conclude that the ordinary definition of modularity, even with the introduction of a multiresolution parameter, cannot properly detect communities. 
This limitation would systematically bias any modularity-based community detection algorithm.
It is therefore clear that ordinary network-based clustering methods, when used with correlation matrices, lead to incorrect results.
In the rest of the paper, we try to overcome this limitation.

%%%%%%%%%%%
%%%%%%%%%%%%%%

\section{Redefining community detection methods for multiple time series\label{sec:methods}}  

We now come to the most pertinent of our results, i.e. the  introduction of improved and consistent methods to cluster multiple time series using appropriate null models.
In sec.\ref{sec:ourmod} we give three redefinitions of the modularity $Q(\vec{\sigma})$ that make use of the results of RMT, which we summarized in sec.\ref{sec:RMT}. In sec.\ref{sec:redef} we introduce the correlation-based counterparts of three of the most popular community detection algorithms used in network analysis.
In sec.\ref{sec:multi} we discuss how these algorithms can be further extended in order to obtain appropriate, multiresolution community detection methods.
Finally, in sec.\ref{sec:bench} we benchmark our methods on various test cases.

\subsection{Correlation-based redefinitions of modularity\label{sec:ourmod}}
From our previous discussion it should be clear that simply replacing network data with correlation matrices in eq.(\ref{QNewman}) leads to eq.(\ref{eq:Qcorr}) where $C_{norm}$ is $\sum_{i,j}C_{ij}$ and the null model $\langle C_{ij}\rangle$ is incorrectly given by eq.(\ref{eq:badnull2}). 
We now introduce three redefinitions of modularity based on appropriate null models.
The end result of this redefinition will be a set of modularity functions that correctly identify communities of correlated time series. 
For compactness, we postpone the possible (re)definition of $C_{norm}$ to the end of this discussion, in sec. \ref{sec:unif}.

\subsubsection{Infinite time series without global mode}
We have already noted that, for infinitely long time series, the correct expression corresponding to the null hypothesis of independency is given by eq.(\ref{eq:goodnull}).
This leads us to a first redefinition of modularity with expectation $\langle C_{ij}\rangle_1\equiv\delta_{ij}$, i.e.
\begin {eqnarray} 
Q_1(\vec{\sigma}) &= &\frac {1}{C_{norm}}\sum_{i,j} \Big[C_{ij} - \delta_{ij}\Big] \delta(\sigma_i,\sigma_j)\nonumber\\
&= &\frac {1}{C_{norm}}\sum_{i,j} C_{ij}^{(\delta)}\delta(\sigma_i,\sigma_j),
\label{QCor}
\end {eqnarray}
where $\mathbf{C}^{(\delta)}\equiv \mathbf{C}-\mathbf{I}$ ($\mathbf{I}$ being the $N\times N$ identity matrix), so that $C_{ii}^{(\delta)}=0$.

\subsubsection{Finite time series without global mode}
For finite-length independent time series, we should further modify our null model to one which anticipates a certain amount of noise, as determined by RMT (see sec.\ref{sec:RMT}). In such a case, we know that the correct null hypothesis is $\langle C_{ij}\rangle_2\equiv C^{(r)}_{ij}$ where $\mathbf{C}^{(r)}$ is given by eq.(\ref{eq:cr}).
This gives us a second redefinition of modularity for dealing with noisy correlation matrices:
\begin {eqnarray} 
Q_2(\vec{\sigma})&=& \frac {1}{C_{norm}}\sum_{i,j} \Big[C_{ij} - C_{ij}^{(r)}\Big] \delta(\sigma_i,\sigma_j)\nonumber\\
&=& \frac {1}{C_{norm}}\sum_{i,j} C_{ij}^{(s)} \delta(\sigma_i,\sigma_j),
\label{QNoise}
\end {eqnarray}
Note that now in general $C_{ii}^{(s)}\ne 0$  as a result of the eigendecomposition defined in eq.(\ref{eq:rmt+}). 
However, the diagonal terms with $i=j$ give an irrelevant constant contribution to the modularity, due to the fact that $\delta(\sigma_i,\sigma_i)=1$ for all $i$, independently of the particular partition $\vec{\sigma}$.
This makes the above definition well defined even in the presence of non-zero diagonal entries.

\subsubsection{Finite time series with global mode}
Lastly, we consider the case where we expect an overall level of positive correlation among all time series, or `global mode'. 
For instance, we have already mentioned that in financial markets the presence of the `market mode' (see sec. \ref{sec:RMT}) generally results in a positive correlation affecting all pairs of stocks altogether. 
The corresponding dominant positive component $\mathbf{C}^{(m)}$ of $\mathbf{C}$ would make $Q(\vec{\sigma})$ be maximized by the (trivial) partition where all time series are in the same community.
In order to detect non-trivial communities, we can choose a null model that includes both the random component of the correlation matrix and the global or market mode, i.e. $\langle C_{ij}\rangle_3\equiv C^{(r)}_{ij}+C^{(m)}_{ij}$ where $\mathbf{C}^{(r)}$ and $\mathbf{C}^{(m)}$ are given by eqs.(\ref{eq:cr}) and (\ref{eq:cm}) respectively.
This yields our third and final formulation for the modularity:
\begin {eqnarray}
Q_3(\vec{\sigma})&=& \frac {1}{C_{norm}}\sum_{i,j} \Big[C_{ij} - C_{ij}^{(r)} - C_{ij}^{(m)}\Big] \delta(\sigma_i, \sigma_j)\nonumber\\
&=& \frac {1}{C_{norm}}\sum_{i,j} C_{ij}^{(g)} \delta(\sigma_i,\sigma_j), \label{QFinance}
\end {eqnarray}
In this case as well, $C_{ii}^{(g)}\ne 0$  as a result of the eigendecomposition defined in eq.(\ref{eq:rmt++}), but this does not affect the outcome of the community detection.

The above definition is now explicitly aimed at detecting mesoscopic communities, which are in between the `microscopic' level of unit-specific noise and the `macroscopic' level of system-wide fluctuations.
While the existence of the market mode is well established in finance, for other types of time series it might be inappropriate to postulate the existence of a global mode. 
However, we also expect that, whenever the use of $Q_1(\vec{\sigma})$ or $Q_2(\vec{\sigma})$ yields only a single community, the most plausible reason is the existence of a global mode.
Accordingly, we expect that the use of $Q_3(\vec{\sigma})$ might be the most appropriate way to filter out global dependencies for a variety of systems, not only for financial markets.
Moreover, as we discuss at length in sec.\ref{sec:multi}, iteratively filtering out the global mode from the correlation matrices restricted to individual communities can result in the definition of a useful multiresolution method to resolve multiple hierarchical levels of community structure, if present.

\subsubsection{A unified redefinition\label{sec:unif}}
For simplicity in what follows, it is useful to  express the three definitions of modularity we gave in eqs. (\ref{QCor}), (\ref{QNoise}) and (\ref{QFinance}) in unified form:
\begin {equation}
Q_l(\vec{\sigma})\equiv \frac {1}{C_{norm}}\sum_{i,j} C_{ij}^{(l)} \delta(\sigma_i,\sigma_j), \label{eq:Qunified}
\end {equation}
where 
\begin{equation}
\mathbf{C}^{(l)}\equiv \mathbf{C}-\langle \mathbf{C}\rangle_l =\left\{
\begin{array}{ll}\mathbf{C}^{(\delta)}&l=1\\
\mathbf{C}^{(s)}&l=2\\
\mathbf{C}^{(g)}&l=3\end{array}
\right. .
\label{eq:l}
\end{equation}
In what follows, given a choice of $l$ we will refer to $\mathbf{C}^{(l)}$ as the `filtered' correlation matrix.

The overall constant $C_{norm}$ has no role in determining the final partition, but it does have a role when different systems, or different snapshots of the same system (including dynamical analyses of community structure), are compared.
For simplicity we keep the same definition as in eq.\eqref{eq:Qcorr}, i.e.
\begin{equation}
C_{norm}\equiv \sum_{i,j}C_{ij}=\textrm{Var}[X_{tot}].
\label{eq:norm}
\end{equation}
This definition implies that the modularity is the sum of intra-community (filtered) correlations, divided by the variance of the total increment $X_{tot}$. 
This variance is a natural measure of the \emph{volatility} of the system over the considered time window, which in the case of financial time series is an important property of the market. 
In other words, eq. \eqref{eq:norm} automatically controls for the volatility of the system, a feature that is typically desirable when analysing the evolution of (the community structure of) wildly fluctuating systems.
However, in some cases it might be interesting to compare the above modularity with one calculated using a different definition of $C_{norm}$, e.g. one that does not control for the volatility.

It should be noted that the above definition is such that the typical (for real-world systems like financial markets) values of the modularity defined in eq.\eqref{eq:Qunified} will tend to be much lower than the typical (for real-world networks) values of the modularity defined in eq.\eqref{QNewman}, even for systems with well-defined communities.
One should bear this consideration in mind when interpreting the (maximized) modularity value as a measure of the strength of community structure in the system.
Unlike its network counterpart, our definition of the modularity does not quantify the strength of community structure in an absolute scale between $-1$ and $+1$. It only has a meaning in relative terms, and the more information is contained in the null model, the lower the value of the resulting modularity.

We remind the reader of the fact that, since the results of RMT used in the above definition hold only in the regime where $N$ and $T$ are both large (with $T>N$), we require the original time series to respect these conditions.
The requirement $T>N$ is sometimes referred to as the `curse of dimensionality' in the literature, since it implies that, in order to study the cross-correlations of a large set of time series, one needs to extend the time interval so much that the assumption of stationarity (implicit, as we mentioned, in the definition of cross-correlations themselves) is violated. 
On the other hand, choosing sufficiently short time intervals to make the time series approximately stationary implies that the number $N$ of time series be severely reduced. 
One should therefore choose the data in such a way that a reasonable compromise is achieved.
This is an ordinary trade-off to be made in the analysis of any empirical (financial) cross-correlation matrix.

We finally stress that the three RMT-based null models we have adopted do not represent the only possible choices. 
One might for instance exploit more sophisticated results \cite{tumminello_spectral,tumminello_nested,tumminello_KL,higham,rebonato,simonian,shrinkage} and introduce refined null models that  overcome some of the limitations of RMT that we mentioned in sec. \ref{sec:RMT}. 
These alternative choices can then be incorporated into our approach by redefining $\langle \mathbf{C}\rangle_l $ and consequently $\mathbf{C}^{(l)}$. 
Exploring the entire space of possibilities is beyond the scope of this paper.
The key point we are stressing here is that, whatever the choice of the null model, it must respect some realistic properties of correlation matrices.
The network-based definition of modularity, which has been used so far, does not do so and as such is not the best choice.
Our approach can therefore be considered as a guideline, in order to introduce improved techniques in the future.

\subsection{Maximizing the new modularity\label{sec:redef}}
The discussion so far completes our first task of introducing modularity functions which are consistent with the properties of correlation matrices. 
Our second task is that of incorporating the above definition(s) into community detection algorithms that seek to maximize the modularity.
Below, we start by briefly mentioning the algorithms we adapted in order to search for the optimal partition (more extended descriptions are in the Appendix) and then prove an important property of the optimal partition itself, namely the fact that its communities are internally positively correlated and mutually negatively correlated. 

\subsubsection{Redefining three community detection algorithms}
Given our new definition of modularity in eq.(\ref{eq:Qunified}), we cannot directly apply the traditional optimization algorithms devised for graphs, since the majority of these algorithms rely in some way or another on the properties of the original network-based definition of modularity, where the degrees of nodes are used to construct the null model. 
For this reason, we selected three of the most popular network-based community detection algorithms and reformulated them to be compatible with time series data and our new definition of modularity.
The three algorithms we selected are known as the Potts (or spin glass) method \cite{potts,Reichardt_Bornholdt_2006}, the Louvain method ~\cite{1742-5468-2008-10-P10008} and the spectral method~\cite{Newman_2006}.
Note that even if these techniques are customarily referred to as `methods', they can actually be considered as three different algorithms implementing the same method of modularity maximization.
Since the appropriate redefinition of these algorithms can require quite technical discussions, it is described in the Appendix.

We note that there exist many modularity maximization algorithms, some of which may already be much better suited to our definition of modularity. However, we wanted to choose popular algorithms whose original specifications required varying levels of rework, ranging from verification of its suitability to accommodate time series based modularity through to modifications of the underlying tenets of the algorithm itself.

Doing so, allows us the possibility to illustrate further differences between network-based and correlation-based community detection problems.
The reader is again referred to the Appendix for a detailed discussion of these differences.

\subsubsection{Identifying anti-correlated communities\label{sec:anti}}
We now prove the result that the partition maximizing the modularity (whichever method is used to search for it) is characterized by positive intra-community (filtered) correlations and negative inter-community (filtered) correlations. 

Let us first define the `renormalized' inter-community correlations (also see the Appendix)
\begin{equation}
\tilde{C}_{AB}^{(l)}\equiv\sum_{i\in A}\sum_{j\in B}C^{(l)}_{ij},
\label{eq:meatball}
\end{equation}
where the notation $i\in A$ indicates that the node $i$ belongs to the community $A$, and the sum is over all such nodes.
Now, assume that we have identified the optimal partition maximizing the modularity, and consider the modularity change $\Delta Q_l$ that would be obtained by further merging two different communities of the optimal partition, say $A$ and $B$. From eq.\eqref{eq:Qunified}, we can write this change as 
\begin{eqnarray}
\Delta Q_l &=&\big[\tilde{C}^{(l)}_{AA}+
\tilde{C}^{(l)}_{BB}+\tilde{C}^{(l)}_{AB}
+\tilde{C}^{(l)}_{BA}\big]
-\big[\tilde{C}^{(l)}_{AA}+\tilde{C}^{(l)}_{BB}\big]\nonumber\\
&=&2\tilde{C}^{(l)}_{AB}.
\end{eqnarray}
The above change cannot be positive, otherwise merging $A$ and $B$ would further increase the modularity, which is impossible since $A$ and $B$ are communities of the optimal partition.
Therefore $\Delta Q_l \le 0$ which also implies $\tilde{C}^{(l)}_{AB}\le 0$.
On the other hand, for every community $A$ of the optimal partition we must have $\tilde{C}^{(l)}_{AA}\ge 0$, otherwise $A$ would give a negative contribution to the modularity, which is impossible as the partition where all nodes of $A$ are isolated communities would have higher modularity than the optimal partition.
Taken together, these considerations imply that
\begin{equation}
\tilde{C}_{AB}^{(l)}
\left\{\begin{array}{ll}
\ge 0&\textrm{if }A=B\\
\le 0&\textrm{if }A\ne B
\end{array}\right. .
\label{eq:split}
\end{equation}
The above result follows simply from the maximization of eq.(\ref{eq:Qunified}) and will be confirmed empirically in sec.\ref{sec:anticorrelation}.

So our algorithms effectively partition the network into mutually anti-correlated communities of positively correlated time series, where it is intended that the term `(anti-)correlated' refers to the residual correlations remaining after applying the filtering procedure defined by eq.(\ref{eq:l}).
For this reason, we will sometimes use the term `residually (anti-)correlated' when referring to the sign of filtered correlations.  
As we will discuss in more detail in sec.\ref{sec:anticorrelation}, this property has important consequences for portfolio optimization and risk management.

\subsection{Multiresolution community detection \label{sec:multi}}
We now come to the problem of introducing an appropriate multiresolution method.
As we mentioned, one way to resolve a hierarchical community structure in ordinary networks using a modularity-based community detection algorithm is that of introducing a resolution parameter $\phi$ as in eq.(\ref{eq:multinull}). 
We have already noted, in our discussion of eq.(\ref{eq:badmultinull}), that the same operation would not cluster correlation matrices appropriately if applied to the na\"ive null model appearing in eq.(\ref{eq:badnull2}).
The same kind of limitation persists if we introduce a resolution parameter multiplying any of the three improved null models $\langle \mathbf{C}\rangle_l$ defining eq.(\ref{eq:Qunified}) through eq.(\ref{eq:l}).
While the range of any observed correlation coefficient $C_{ij}$ is ${[-1,+1]}$ by construction, a resolution paramater would unreasonably map the range of the expected correlation  $\langle C_{ij}\rangle$ to ${[-\phi,+\phi]}$. 
Similarly, since the null correlation matrices $\langle \mathbf{C}\rangle_l$ we introduced are obtained from the eigencomponents of the observed correlation matrix $\mathbf{C}$, rescaling them by $\phi$ is equivalent to an overall rescaling of the corresponding eigenvalues of $\mathbf{C}$, which is again an unjustified operation.

Given the above limitations, which indicate a lack of theoretical foundation for resolution parameters in the case of correlation matrices, we introduce a completely different multiresolution approach that is specifically designed for multiple time series, and has no counterpart in network analysis.
After running one of our newly introduced  community detection algorithms on the original empirical correlation matrix $\mathbf{C}$, for each community of size $s$ in the optimal partition we consider the corresponding $s\times s$ sub-matrix $\mathbf{C}_*$ of $\mathbf{C}$.
For this sub-matrix, we define the three null models $\langle \mathbf{C}_*\rangle_l$ as discussed in sec.\ref{sec:ourmod} for the original matrix $\mathbf{C}$.
By running our community detection algorithms recursively inside each of the communities, we can thus resolve subcommunities within communities.
Iterating this procedure identifies a hierarchical community structure, if present.
Within each community, the procedure stops automatically when it resolves no further subcommunities.

At each iteration, the `noise' component $\mathbf{C}^{(r)}_*$  will have the same interpretation as when it is identified on the entire correlation matrix, since  $\mathbf{C}_*$ is the sub-matrix of the original matrix $\mathbf{C}$ and \emph{not} of the filtered matrix $\mathbf{C}^{(l)}$ defined in eq.(\ref{eq:l}), so it still contains the node-specific noise component (the reason why we do not consider the sub-matrix of $\mathbf{C}^{(l)}$ is because, as we mentioned, the latter may not be a proper correlation matrix \cite{higham,rebonato,simonian,shrinkage} and cannot thus be filtered further using RMT).
The `global' mode $\mathbf{C}^{(m)}_*$ is now interpreted as the `community' mode, i.e. a common factor influencing all the time series within that particular community.
This will now include both the system-wide mode $\mathbf{C}^{(m)}$, restricted to the subspace relative to $\mathbf{C}_*$, that would be identified on the entire matrix $\mathbf{C}$ (e.g., in the case of financial time series, the market mode) \emph{and} a genuinely community-specific mode not shared with the time series in other communities.
Different communities are therefore possibly characterized by different community modes, and the fact that both this mode and the restriction of the global mode are filtered out is precisely what allows the algorithm to resolve deeper hierarchical modules.
Finally, the `group' component $\mathbf{C}^{(g)}_*$
represents the effect of subgroups nested within the specific community, if present.

It should be noted that the original correlation matrix $\mathbf{C}$ typically has large dimensionality (large $N$), a property ensuring that the results of RMT, in particular the expected eigenvalue distribution appearing in eq.\eqref{RMTDensity}, hold to a satisfactory level. 
However, when considering smaller subcommunities, RMT becomes less reliable because eq.\eqref{RMTDensity} no longer holds for small sets of nodes. 
For this reason, for small submatrices (low-dimensional $\mathbf{C}_*$) it is preferable to determine the eigenvalues $\lambda_{\pm}$ not via eq.\eqref{eq:lambda+-}, but by randomly shuffling the temporal increments of the original time series and constructing the corresponding spectrum as shown in Fig. \ref{fig:eigendensity}.

We conclude by noting that, for the particular case of multiple time series, there is another `multiresolution' character, which can be attached to the problem of community detection, namely the fact that different communities can in principle be obtained for different choices of the initial temporal resolution, i.e. for different choices of the frequency of the original time series (e.g. second, minute, or daily returns).
Note that this notion of temporal resolution is specific to correlation matrices and has no analogue in the ordinary problem of  community detection in networks.
It is also not necessarily attached to an idea of hierarchy, in the sense that we do not expect e.g. communities obtained at higher frequency to be necessarily nested within communities obtained at lower frequency (even if this can reasonably happen in some cases).
To distinguish this specific notion from the usual one of multiresolution community detection, we will refer to it as the `multifrequency' problem and address it separately in sec.\ref{sec:resol}.

%%%%%%%%%%%
\begin{figure*}
\centerline{\includegraphics[width=.99\textwidth]{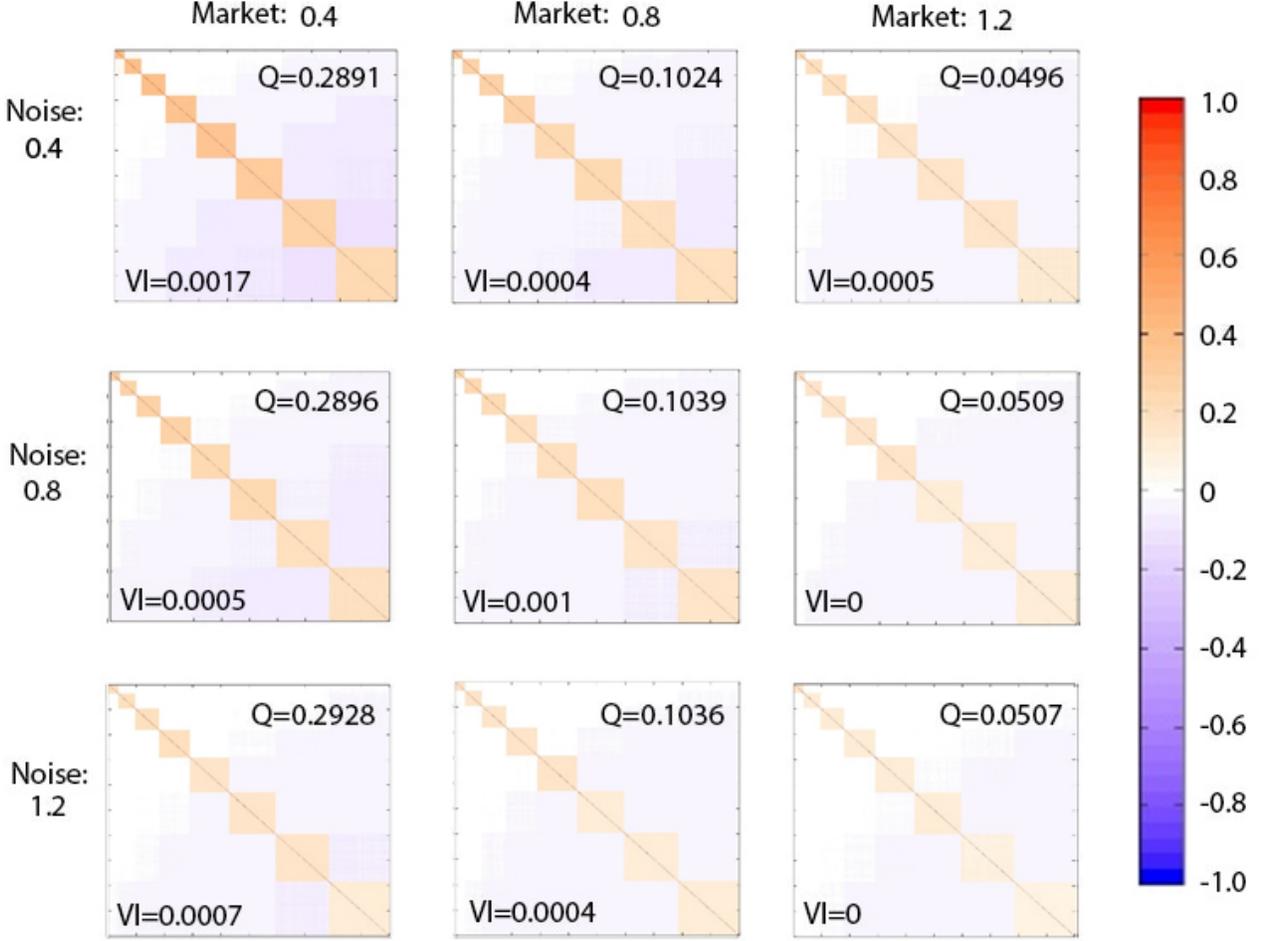}}
\caption{Performance of our method on 9 benchmark sets of correlated time series with varying levels of noise ($\nu$) and market mode ($\mu$) components.
For each combination of $\mu$ and $\nu$, $N=1000$ time series of length $T=50000$, partitioned into $c=8$ communities (always containing 35, 60, 85, 110, 140, 165, 190, and 215 time series respectively), were initially generated according to eq.\eqref{eq:bench}.
Then, the $1000\times 1000$ correlation matrix $\mathbf{C}$ was calculated.
The heat maps in this figure show the values of the entries of the filtered matrix $\mathbf{C}^{(g)}$ defined in eq.\eqref{eq:cg} and obtained by removing the noise and `market-mode' components from the original correlation matrix.
Blocks along the diagonal represent the residual correlations within each community, while off-diagonal blocks show the residual negative cross-correlations among communities. 
Our method, here using the Potts algorithm (see Appendix) to maximize the modularity $Q_3(\vec{\sigma})$ defined in eq.\eqref{QFinance}, was always able to correctly identify the target communities, even for values of $\mu$ and $\nu$ exceeding one.
This is indicated by the value of $VI$ (averaged over 10 runs of the community detection algorithm) calculated between the `true' and the detected partition in each benchmark.
The average (over multiple runs) maximum modularity value $Q_3(\vec{\sigma}^*)$ is also shown in each case.}
\label{CgHM}
\end{figure*}
%%%%%%%%%%%%%%

\subsection{Benchmarking our methods\label{sec:bench}}
Before applying our methods to the analysis of real correlation matrices, we ran a series of tests confirming that we can correctly detect correlated sets of time series in controlled benchmark cases.
Our benchmarks consist of heterogeneously sized communities of time series that are internally correlated and additionally display varying levels of noise and global signal (market mode).
The reason why we consider heterogeneous community sizes is because this is the more challenging case where we showed the na\"ive method to display a higher bias (see sec. \ref{sec:bias}). 

We constructed these benchmarks by first choosing the number $N$ of time series, the number $c$ of communities and the desired number $n_A$ of time series in each community $A$, such that $\sum_{A=1}^c n_A=N$ (as in sec. \ref{sec:bias}).
Then we generated $c$ random and uncorrelated time series (with $T>N$) with values $\gamma_A(t)$ (where $1\le A\le c$) drawn independently from a normal distribution with zero mean and unit variance.
Then, we created $n_A$ identical copies of the $A$-th time series, for all $A$. 
To each of the resulting $N$ time series, each labeled by an index $i$, we added a local noise $\beta_i(t)$ (a new normally distributed random variable with zero mean and unit variance, independent of all the other ones) multiplied by a `noise parameter' $\nu\ge 0$ and a global signal $\alpha(t)$ (again, an independent normally distributed random variable with zero mean and unit variance) multiplied by a `market-mode parameter' $\mu\ge 0$.
This resulted in a set $\{Y_1,\dots,Y_N\}$ of $N$ time series with values
\begin{equation}
y_i(t)=\mu\cdot \alpha(t) +\nu\cdot \beta_i(t)+\gamma_A(t)\quad i\in A, \quad A=1,c.
\label{eq:bench}
\end{equation}
Note that this procedure is similar to the so-called `factor models' used in financial analysis \cite{FinancialRisk,tumminello_spectral,tumminello_nested, FamaFrench, FamaFrenchII, Sharpe}.
The time series $\{Y_1,\dots,Y_N\}$ were further standardized to obtain a final set $\{X_1,\dots,X_N\}$ of $N$ time series, each with zero mean and unit variance, in compliance with the general prescription mentioned in sec. \ref{sec:existing}.

We generated several benchmarks according to the recipe described above, for various choices of $N$, $c$, $\{n_A\}$, $\mu$ and $\nu$. 
In general, when $\mu=\nu=0$ the benchmark is similar to the ideal one described in sec.\ref{sec:bias}: the communities are completely correlated internally (all the time series in the same communities are identical) and uncorrelated with the time series in other communities. 
This results in a benchmark partition $\vec{\sigma}^*$ such that, for infinite time series, $C_{ij}=\delta(\sigma_i^*,\sigma_j^*)$. However, for finite (but still such that $T>N$ as prescribed by random matrix theory, see sec.\ref{sec:RMT}) time series, $C_{ij}$ will be affected by noise.
As $\mu$ and $\nu$ increase, additional noise will be generated and the community structure will be more difficult to detect. 
If $\mu=1$ ($\nu=1$) then the amplitude of the global mode (local noise) is the same as that of the community signal.
Therefore when $\mu$ and/or $\nu$ approach or exceed one, the community detection problem becomes more challenging. 
Still, the ambition of our method is that of correctly identifying the benchmark partition $\vec{\sigma}^*$ even in this `hard' regime. 

In Fig. \ref{CgHM} we show nine benchmarks, organized in a $3\times 3$ table with different combinations of values for $\mu$ and $\nu$. In all these cases, the communities to detect are the same set of $c=8$ heterogeneously sized communities shown previously in fig.\ref{fig:offdiag}c.
The color maps show the values of the entries of the filtered correlation matrix $\mathbf{C}^{(g)}$ defined in eq.\eqref{eq:cg}, i.e. the residual correlations obtained after removing the noise and market-mode components. 
It can be seen that, even for values of $\mu$ and $\nu$ exceeding one, the filtered matrices always display a clear block-diagonal structure with a visible contrast across diagonal and off-diagonal blocks. 

In all these benchmarks we confirmed that, using the corresponding modularity $Q_3(\vec{\sigma})$ defined in eq.\eqref{QFinance}, our method succeeded in detecting the correct partition $\vec{\sigma}^*$.
We quantitatively measured the performance of our method in terms of a metric known as Variation of Information ($VI$) ~\cite{Meila2007873,978-3-540-45167-9_14}, which measures the entropy difference between two partitions of the same network, providing a rigorous way for us to quantify the similarity between the `true' partition and the one identified by our method. 
More precisely, $VI$ involves the use of Shannon's entropy to measure the amount of uncertainty that exists across the set of communities of two different partitions of the same network. 
It provides a quantitative measure of the difference between two partitions, a normalized value where zero implies the two partitions are completely identical and one implies that they are completely unrelated. 
As can be seen from Fig. \ref{CgHM}, the values of $VI$ (averaged over multiple runs of the community detection algorithm) are zero or extremely small, indicating a perfect or almost perfect performance of the method.

The average (over multiple runs) maximum modularity value $Q_3(\vec{\sigma}^*)$ obtained in the above benchmarks is also illustrated in Fig. \ref{CgHM}.
Lower values of the modularity imply that the network as a whole is more homogeneous in its construction, to the extent that the detected communities exhibit only a relatively weak increase in their collective correlation, above the ambient level.
As expected, we see that the modularity decreases for increasing levels of market mode. Increasing levels of noise however do not have such a strong effect, since noisy time series tend to diminish the strength of the intra-community correlations, which enter in both the numerator and denominator of the modularity. In contrast, the market mode has significant impact on the inter-community correlations, which primarily end up only in the denominator of the modularity. Hence the observed decrease in modularity with an increase in market mode.
The corresponding low values of the modularity confirm what we had anticipated about the effects of eq.\eqref{eq:norm}.
We should bear these effects in mind when interpreting the (low) values of the modularity arising from the partition of real financial time series, where the market mode is very strong.
The fact that our method correctly identifies the benchmark partitions even for strong market mode (and low resulting modularity) makes us confident that it will also properly detect the community structure of real markets.

\section{The mesoscopic organization of real financial markets\label{sec:results}}
Having redefined the modularity consistently with the properties of correlation matrices and appropriately reconfigured three different techniques for optimizing it, we are now in a position to apply our methodology to a variety of real-world data sets and evaluate the quality of the results. 
In particular we will apply our three algorithms and the null model expressed in eq.\eqref{QFinance} to time series representing  stock prices from a variety of stock indexes that span multiple industries and multiple countries. 
We first obtained static results, including the multiresolution community structure as introduced in sec. \ref{sec:multi}, using time series of log-returns of daily closing prices for all the three indexes. These results are shown in this section.
Then, we considered different temporal (frequency) resolutions and studied the time dynamics of community structure. These additional results are described in secs.\ref{sec:resol} and \ref{sec:dyn} respectively.

\subsection{Data and pre-processing}
The indexes we used are the S\&P 500 (US Large Cap. Stocks), the FTSE 100 (British Large Cap.) and the Nikkei 225 (Japanese Large Cap.).
For each of these indexes, we considered a period of 2500 trading days, corresponding to approximately 10 years of market activity, from 2001Q4 to 2011Q3.
We selected all stocks for which complete data are available during this period.
This resulted in the selection of 445 S\&P stocks, 78 FTSE stocks and 193 Nikkei stocks. 
All these stocks are classified within the Global Industry Classification Standard (GICS)\footnote{Standard \& Poor's developed what is known as the Global Industry Classification Standard (GCIS) as an industry taxonomy, for use by the financial sector. All stocks in the three indexes we work with are classified using this single taxonomy.}.
The complete taxonomy can be found online \footnote{http://www.standardandpoors.com/products-services/GICS/en/us}, however we briefly mention that there are ten top-level `sectors' (see table \ref{tbl:GICScolors}) split into 24 sub-categories called `industry groups', which are in turn divided into 68 `industries'.

%%%%%
\begin{table}[t]
\center
\begin{tabular}{|lc|lc|}
\hline
Consumer Discretionary: &\color{Purple} $\blacksquare$ \color{black} & Consumer Staples: &\color{Aquamarine}$\blacksquare$ \color{black}\\
Energy: &\color{CadetBlue} $\blacksquare$ \color{black} & Financials: &\color{Green} $\blacksquare$ \color{black}\\
Health Care: &\color{Red} $\blacksquare$ \color{black} & Industrials: &\color{Orange} $\blacksquare$ \color{black}\\
Information Technology: &\color{Blue} $\blacksquare$ \color{black} & Materials: &\color{Yellow} $\blacksquare$ \color{black}\\
Telecom. Services: &\color{Magenta} $\blacksquare$ \color{black} &  Utilities: &\color{Brown} $\blacksquare$ \color{black}\\
\hline
\end{tabular}
\caption{The 10 industry sectors in the Global Industry Classification Standard (GICS), with the color representation used to highlight the sectors in the following figures.}
\label{tbl:GICScolors}
\end{table}
%%%%%%%%%%%

It is important to note that, although we would expect stocks within certain industry sectors to be correlated with each other, we do not expect to observe this effect within and throughout all industry sectors. 
Previous research in the area of stock clustering  \cite{Econophysics,FinancialRisk,maxspanasset,PhysRevE.72.046133,PhysRevE.70.026110,PhysRevE.76.046116} (see also our discussion in sec.\ref{sec:existing}) has shown some relationships between the industry sectors and clusters of stocks identified by the various methods.
We therefore expect to find a certain degree of overlap with this research. However, our choice of null models in conjunction with our tailored community detection algorithms is designed to uncover nontrivial correlations, beyond a direct mapping to industry sectors, such as finding stocks from different industry sectors that tend to move together, and even in opposition to other stocks in their own sector. 
It is therefore useful to use industry sectors not as a target, but as a baseline to highlight important and non-trivial deviations identified by the community detection algorithms. 

As with the benchmarks described in sec.\ref{sec:bench}, each set of financial time series was used to initially create a correlation matrix $\mathbf{C}$ that was then filtered to produce the matrix $\mathbf{C}^{(g)}$ following the procedure described in secs. \ref{sec:RMT} and \ref{sec:ourmod}.
Each such matrix was then operated on individually by the three community detection algorithms described in sec. \ref{sec:methods}. 
We found that all algorithms always generate very similar partitions. 
This important result, which for the sake of exposition, is postponed to sec.  \ref{sec:comparative}, implies that we can refrain from showing the results of every algorithm.
For brevity, we will instead select representative exemplars, with the understanding that any one of the algorithms would generate very similar results.

\subsection{Standard approaches\label{sec:standard}}
Before showing the main results of our own methodology, as a preliminary study we illustrate what would be obtained using some of the standard approaches available, in particular the correlation thresholding described in sec.\ref{sec:AG} and the community detection built on the network-based modularity, described in secs.\ref{sec:inconsistency} and \ref{sec:bias}.

%%%%%%%%%%%%%%%%%
\begin{figure*}
\includegraphics[width = \textwidth]{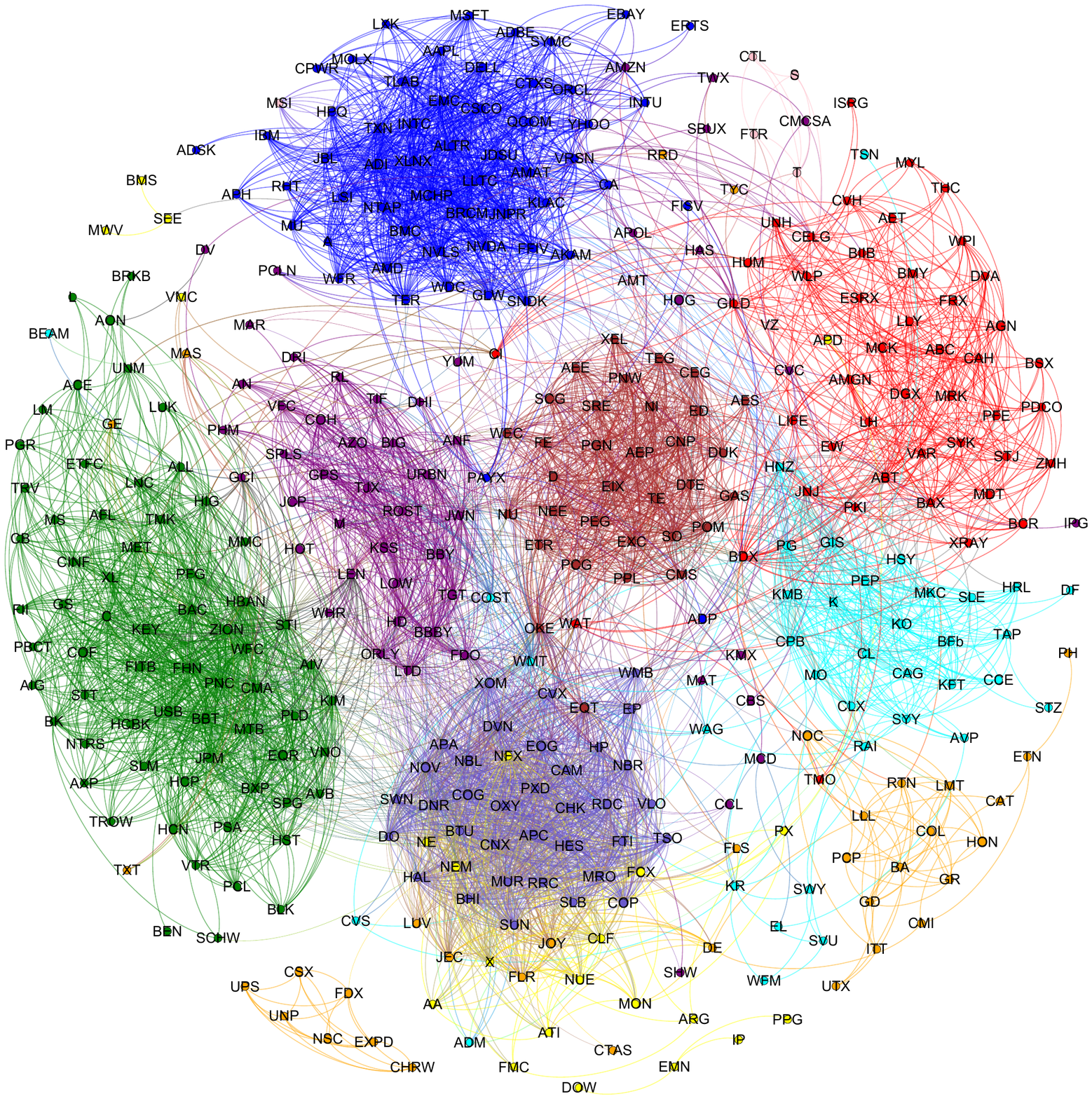}
\vskip -3cm
\caption{Asset graph for the S\&P 500 (log-returns of daily closing prices from 2001Q4 to 2011Q3). The network is generated from the correlation matrix of the constituent stocks, after taking the Fisher transform and setting a threshold at 2 standard deviations. The color of each node represents the industry sector to which that stock belongs (see Table \ref{tbl:GICScolors}). The force-based layout clearly indicates the existence of strong connections between stocks of the same industry sector, however this approach (like any other threshold-based approach) cannot identify communities of stocks that are internally more correlated than with the rest of the market, and mutually anti-correlated.}
\label{fig:assetgraph}
\end{figure*}
%%%%%%%%%

\subsubsection{Asset Graph from Fisher-transformed correlations}
As we discussed in sec.\ref{sec:AG}, imposing a threshold on the entries of a correlation matrix $\mathbf{C}$ allows us to obtain an \emph{Asset Graph} where links connect the more strongly correlated pairs of stocks \cite{ag1,maxspanasset,Econophysics}.
In fig. \ref{fig:assetgraph} we show the effect of this procedure on our S\&P 500 data.
Rather than showing the results for multiple choices of the threshold, we used a rough criterion to select a unique threshold that would in principle correspond to a standard level of statistical significance. This criterion is as follows.

Using general results in statistics \cite{1915}, one can easily show that, under the null hypothesis, two time series $X_i$ and $X_j$ of length $T$ representing $T$ realizations of two independent and normally distributed random variables, the quantity
\begin{equation}
z_{ij}\equiv \textrm{artanh } C_{ij}
=\frac{1}{2}\ln\frac{1+C_{ij}}{1-C_{ij}}
\label{eq:fisher}
\end{equation}
(where $C_{ij}$ is the sample correlation coefficient) is distributed as a normal variable with zero mean (representing the population correlation coefficient in the case of independent variables) and standard error
\begin{equation}
\sigma=(T-3)^{-1/2}.
\end{equation}
In other words, under the above null hypothesis we expect a concentration of values of $z_{ij}$ around zero, with standard error $\sigma$. 

In order to detect significant deviations from the null hypothesis, one may select a threshold $\tau$ such that only the values outside $\tau$ standard errors, i.e. $|z_{ij}|>\tau \sigma$, are considered as statistically significant.
This means that one can select a threshold $z_\tau\equiv\tau\sigma$ for $z_{ij}$.
In terms of the correlations $C_{ij}$, the corresponding critical value is
\begin{equation}
C_\tau
\equiv\tanh z_\tau
=\frac{\exp{\Big(\frac{2\tau}{\sqrt{T-3}}\Big)}-1}{\exp{\Big(\frac{2\tau}{\sqrt{T-3}}\Big)}+1}.
\end{equation}
A suitable choice of the value of $\tau$ can be used to threshold the correlation matrix into an Asset Graph at the corresponding significance level: specifically, one can draw a link only if $|C_{ij}|>C_\tau$.
The advantage of introducing the above criterion is that, at least in principle, it associates a precise statistical significance level to any value of the threshold (there are however various problems with this approach, as we briefly comment later).
This makes it possible to select a unique threshold value corresponding to a standard accepted level of significance.

We used the above approach as a rough criterion to select an indicative threshold, choosing $\tau=2$ so that only the correlations lying two standard deviations away from the null hypothesis are in principle retained. 
The resulting Asset Graph for the stocks of the S\&P 500, plotted in fig.\ref{fig:assetgraph}, was visualized using a clustered rendering ~\cite{ICWSM09154} of all the stocks that do not end up completely isolated after the filtration, according to the Fruchterman-Reingold~\cite{SPE:SPE4380211102} force-based algorithm. 
As expected, we immediately see a significant correspondence between groups of densely connected nodes and industry sectors.
%Repulsive forces stem from the nodes and attractive forces are generally the result of edges between the nodes.
%Rendering the graph using such an algorithm coaxes the system to settle, geometrically into its lowest (or close to) energy state, allowing us to visually determine the groups of nodes that maintain the highest collective correlation. 
%In terms of community detection, the resulting groups of closely packed nodes might partially resemble the clusters that contribute highly to the modularity function. 
However, there is no linear relationship between the attractive and repulsive forces defined by the graph drawing algorithm and the contribution of the corresponding correlations to the modularity. As such, the visualization of the graph cannot be directly used to partition the network into communities.

Moreover, it should be noted that the approach we have used to define a threshold has two main theoretical disadvantages: first, it assumes normally distributed log-returns (while it is well known that real log-return distributions are fat-tailed \cite{Kurtosis, fattails}); secondly, it does not introduce multiple hypothesis test corrections.
A more rigorous way to statistically validate links in a correlation-based network would be that of using numerical bootstrapping methods such as the one considered in ref.\cite{validated}.

In any case, since no search over the space of possible partitions is performed, the Asset Graph method cannot identify communities of stocks that are more strongly correlated internally than with the rest of the market.
As we anticipated in sec.\ref{sec:AG}, this leaves the problem we started with unsolved.
We also recall from sec. \ref{sec:anti} that our methodology detects residually anti-correlated communities.
This property, which we will illustrate in sec. \ref{sec:anticorrelation} for the data considered here, cannot be achieved by any threshold-based method, or any of the other available methods we described in sec.\ref{sec:existing}.

%%%%%%
\begin{figure}[t]
\includegraphics[width = 0.3\textwidth]{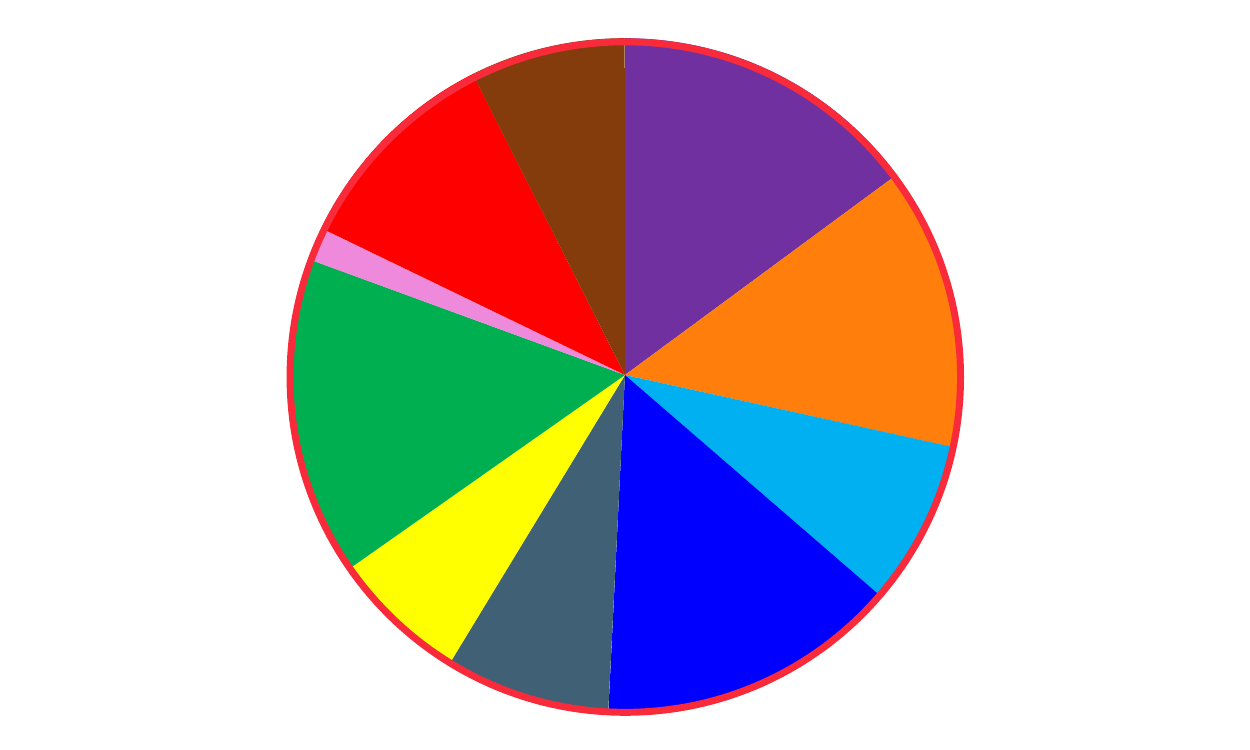}
\caption{The trivial, single community containing all stocks of the S\&P 500 (log-returns of daily closing prices from 2001Q4 to 2011Q3), obtained by either na\"ively treating the correlation matrix as a weighted network and using the ordinary network-based modularity, or alternatively using the correlation-based modularity $Q_1(\vec{\sigma})$ (i.e. without filtering the correlation matrix). In both cases, the Louvain algorithm (see Appendix) has been used. The colors represent different GICS sectors (color legend in Table \ref{tbl:GICScolors}) and span an area proportional to the number of stocks in each sector.}
\label{fig:oneCommunity}
\end{figure}
%%%%%%%%%

\subsubsection{Na\"ive application of community detection}
As another baseline reference, in fig.\ref{fig:oneCommunity} we show the result of applying to the same S\&P 500 data, the community detection described in sec. \ref{sec:inconsistency}, i.e. by treating the correlation matrix $\mathbf{C}$ as a weighted network and running an ordinary (network-based) community detection algorithm \cite{saramaki_correlations,mason_correlations}.
We see that the resulting, trivial, community is a single one spanning the entire set of stocks. 
In such a case, the pie chart depicting the community merely illustrates the distribution of industries within the  S\&P 500.
The same result is obtained if one uses the correlation-based modularity $Q_1(\vec{\sigma})$ defined in eq.\eqref{QCor} in terms of the null model  $\langle\mathbf{C}\rangle=\mathbf{1}$ (i.e. assuming that all time series are completely independent and of infinite length). 

In the first case, this result is due to the inconsistent structure of the modularity and to the resulting bias of the algorithms used to maximize it, as we discussed in sec.\ref{sec:inconsistency}.
In the second case, it is due to the inadequacy of the null model defined in eq.(\ref{eq:goodnull}) for financial correlations (see sec.\ref{sec:ourmod}): the community detection algorithm finds only a single community because of the systemic correlation of the market mode affecting all stocks simultaneously.
 
%%%%%%
\begin{figure}[t]
\includegraphics[width = 0.49\textwidth]{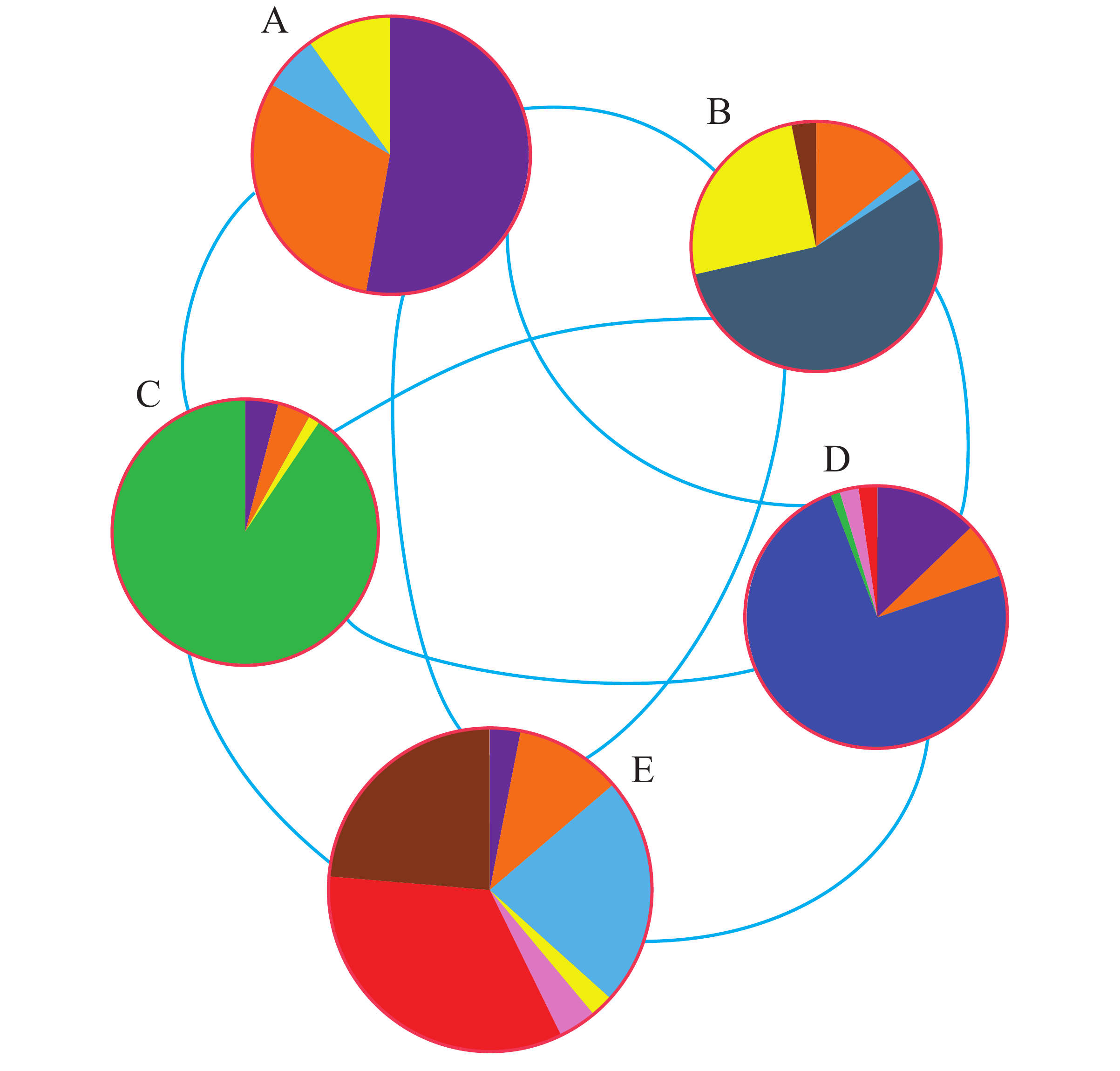}
\caption{Communities of the S\&P 500 (log-returns of daily closing prices from 2001Q4 to 2011Q3) generated using our correlation-based modularity $Q_3(\vec{\sigma})$ with the  Louvain algorithm (see Appendix). Individual communities are labeled $A$ through $E$ and the pie chart represents the relative composition of each community based on the industry sectors of the constituent stocks (color legend in Table \ref{tbl:GICScolors}). The blue inter-community link weights are negative, indicating that the communities are all residually anti-correlated. The red circles around each community indicate that the total intra-community correlations are all positive.}
\label{fig:SPCommunities}
\end{figure}
%%%%%%%%%

\subsection{Community detection using our method}
We now come to the application of our own methodology described in sec.\ref{sec:methods}. In fig.\ref{fig:SPCommunities} we show the result of the application, to the same daily S\&P 500 data, of the appropriately redefined community detection methods introduced in sec.\ref{sec:redef}. Specifically, making use of the modularity $Q_3(\vec{\sigma})$ defined in eq.\eqref{QFinance}.
Since such null models discount both random and market-wide correlations, the community detection algorithms are now able to successfully find correlations that exist in between the microscopic and macroscopic levels. 
For the S\&P 500, the result is a set of five mesoscopic communities whose relative size (the number of nodes in each community) is expressed by the size of the pie chart in the graph. The relative breakdown of the stocks in each community, classified according to their top level GICS sector (see table \ref{tbl:GICScolors}), is represented by the fraction of the pie chart for that community.

\begin{figure}[t]
\includegraphics[width = .45\textwidth]{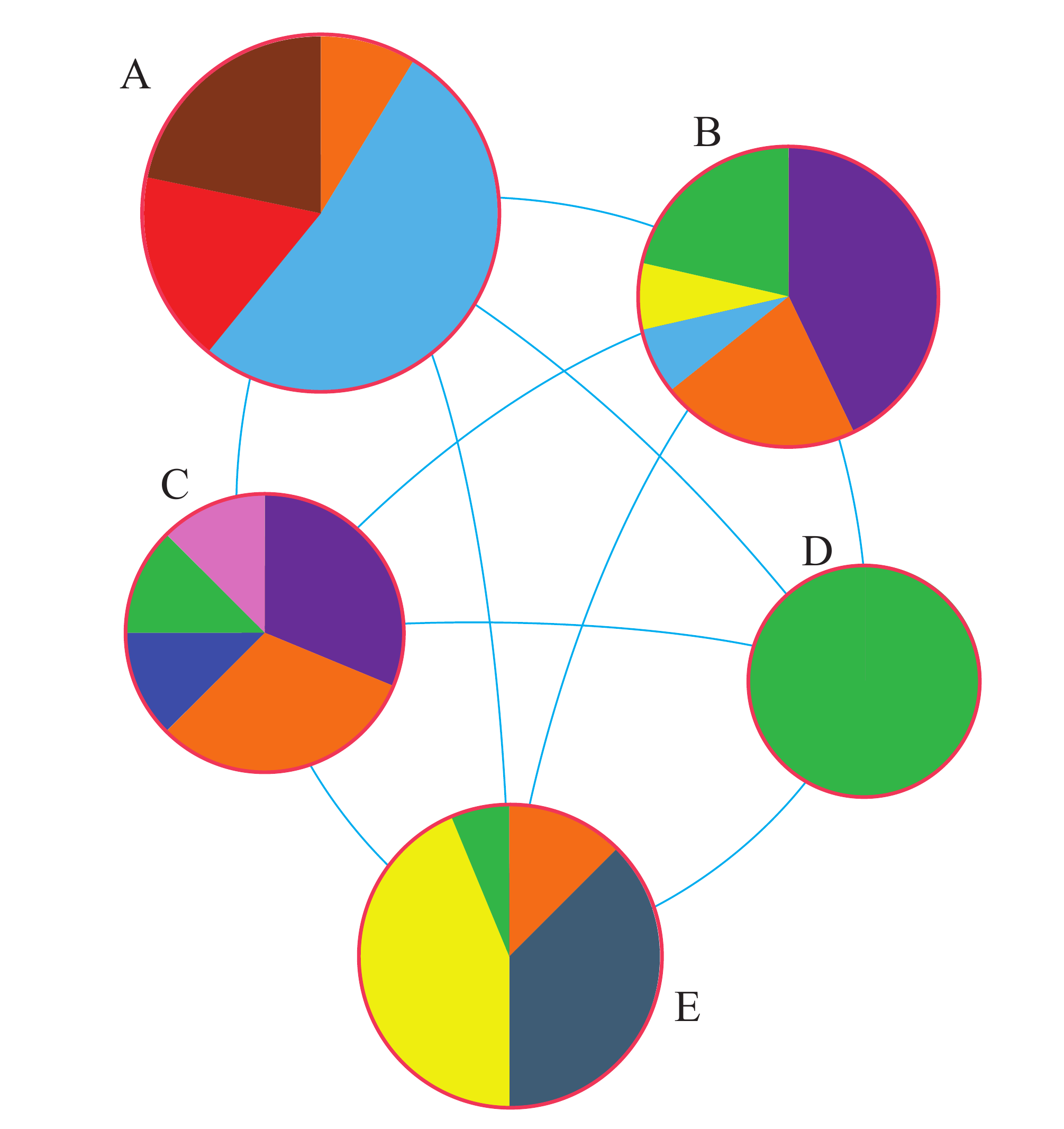}
\caption{Communities of the FTSE 100 (log-returns of daily closing prices from 2001Q4 to 2011Q3) generated using our correlation-based modularity $Q_3(\vec{\sigma})$  with the Louvain algorithm (see Appendix). Individual communities are labeled $A$ through $E$ and the pie chart represents the relative composition of each community based on the industry sectors of the constituent stocks (color legend in Table \ref{tbl:GICScolors}). The blue inter-community link weights are  negative, indicating that the communities are all residually anti-correlated. The red circles around each community indicate that the total intra-community correlations are all positive.}
\label{fig:FCommunities}
\end{figure}

\begin{figure}[t]
\includegraphics[width=.49\textwidth]{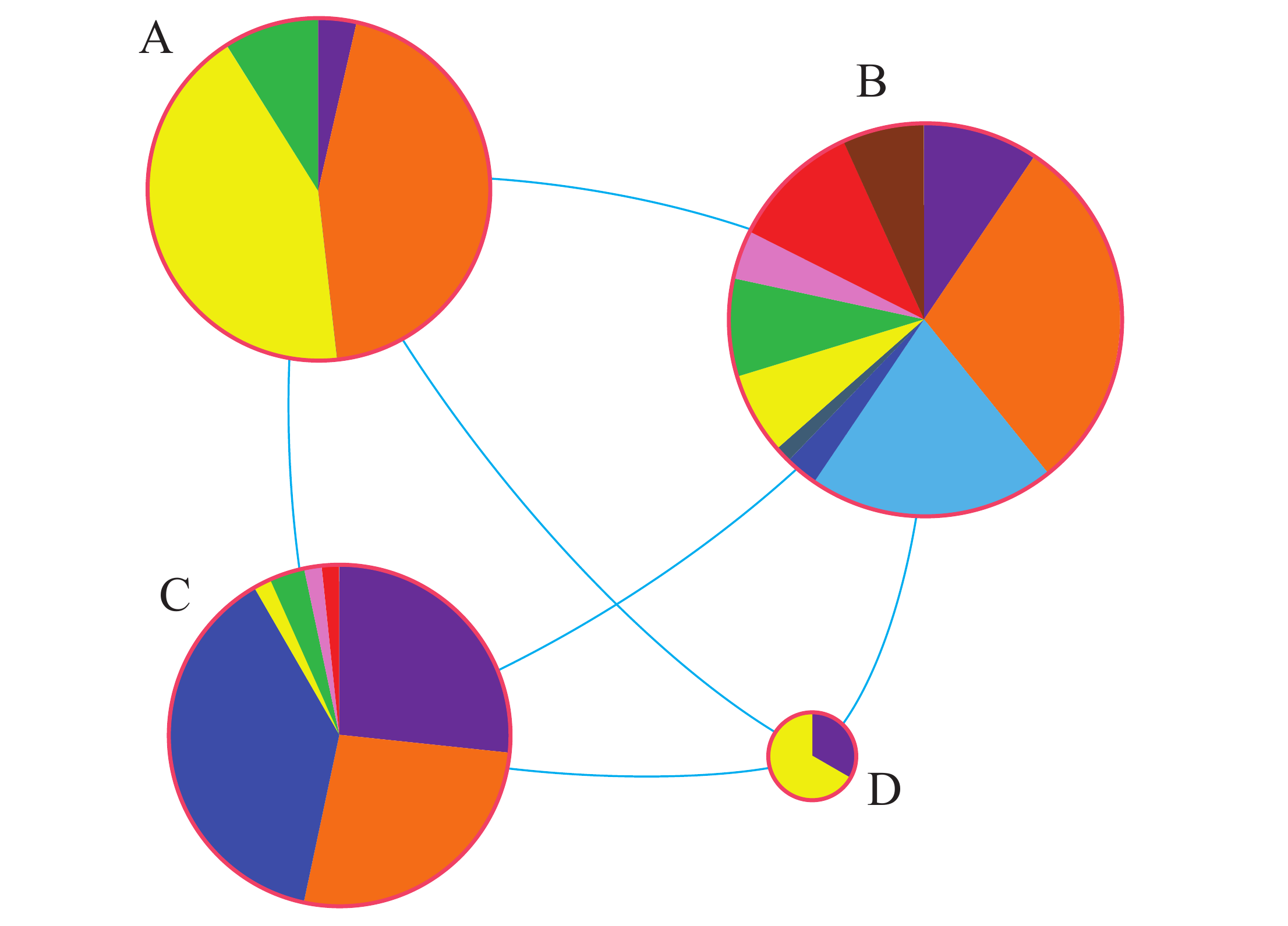}
\caption{Communities of the Nikkei 225 (log-returns of daily closing prices from 2001Q4 to 2011Q3) generated using our correlation-based modularity $Q_3(\vec{\sigma})$ with the Louvain algorithm (see Appendix). Individual communities are labeled $A$ through $D$ and the pie chart represents the relative composition of each community based on the industry sectors of the constituent stocks (color legend in Table \ref{tbl:GICScolors}). The blue inter-community link weights are  negative, indicating that the communities are all residually anti-correlated. The red circles around each community indicate that the total intra-community correlations are all positive.}
\label{fig:NCommunities}
\end{figure}

In addition to the communities presented for the S\&P 500, in figs. \ref{fig:FCommunities} and \ref{fig:NCommunities} we also provide the communities for the FTSE 100 and the Nikkei 225 respectively, again detected using the null model from eq.\eqref{QFinance}. 
As before, the na\"ive community detection would place all stocks into a single community (not shown).
For all these data sets, the values of the maximized modularity $Q_3(\vec{\sigma}^*)$ achieved by the optimal partitions will be shown later in sec.\ref{sec:comparative}.

%%%%%%%
%%%%%%%%

While at first glance it may seem as though there is no particular pattern to the community structures in the three markets (as each community contains a plethora of stocks from different industry sectors), a closer look at the industries to which the stocks belong though does in fact yield some interesting observations. 
First and foremost, some of the industry sectors tend to dominate communities, where in some cases 100\% of the stocks for a particular industry sector are in the same community, meaning that on average over the past ten years they have all remained correlated. 
Examples of this include Energy (\color{CadetBlue} $\blacksquare$\color{black}, community $B$), Financials (\color{Green}$\blacksquare$\color{black}, community $C$) and Information Technology (\color{Blue}$\blacksquare$\color{black}, community $D$) in the S\&P, Utilities (\color{Brown}$\blacksquare$\color{black}, community $A$), Health Care (\color{Red}$\blacksquare$\color{black}, community $A$),
Information Technology (\color{Blue}$\blacksquare$\color{black}, community $C$),
Telecom. Services (\color{Magenta}$\blacksquare$\color{black}, community $C$) and Energy (\color{CadetBlue} $\blacksquare$\color{black}, community $E$) in the FTSE, and finally Utilities (\color{Brown}$\blacksquare$\color{black}, community $B$),
Energy (\color{CadetBlue} $\blacksquare$\color{black}, community $B$) and Consumer Staples (\color{Aquamarine}$\blacksquare$\color{black}, community $B$) in the Nikkei.

There are also instances where top-level sectors are split among different communities according to their subclassification (Industry Group and Industry).
This is very interesting because it shows that subgroups of stocks within one sector are often more correlated with a different sector than their own sector. 

%%%%%%%
%%%%%%%%

Other interesting cross-sector correlations can be found too, with Health Care for example. In the FTSE communities, Health Care stocks (\color{Red}$\blacksquare$\color{black}) are exclusively in community $A$, whereas in the S\&P they are predominately in community $E$, with some in community $D$. 
Interestingly enough, in the latter case the one \textit{Health Care Technology} industry sector stock from the Health Care sector is in community $D$, which also happens to be the community containing all of the Information Technology (IT) stocks (\color{Blue}$\blacksquare$\color{black}), whereas all of the Pharmaceutical stocks are in community $E$, which contains the bulk of the Consumer Staples (\color{Aquamarine}$\blacksquare$\color{black}) stocks. 
The reader might at this point notice that the FTSE community $A$ containing all Health Care  (\color{Red}$\blacksquare$\color{black}) stocks also contains the bulk of the Consumer Staples (\color{Aquamarine}$\blacksquare$\color{black}) stocks. It is probably not surprising then to discover that those Health Care stocks are comprised of entirely Pharmaceuticals. 
We find an identical relationship between Pharmaceuticals and Consumer Staples in community $B$ of the Nikkei 225 as well. Furthermore, the one other Health Care stock, a \textit{Health Care Equipment \& Supplies} stock trades in the same community as the IT stocks. This might not be particularly interesting except for the fact that in the Nikkei, the vast majority of IT sector stocks are sub-classified as Electronic Equipment.  

One might continue finding interesting trends such as these, however our purpose is not to glean specific qualitative information regarding financial markets, but rather to illustrate how the underlying quantitative information can be ascertained from the raw data, through the appropriate choice of null models in conjunction with the process of community detection.
The most important result of this process is the successful identification of mesoscopic communities of correlated stocks that are irreducible to a standard sectorial taxonomy and also anti-correlated with each other, as we now discuss.

\subsection{Residually anti-correlated communities and portfolio optimization\label{sec:anticorrelation}}
The age old proverb, \emph{``Don't put all your eggs in one basket''}, could never be more insightful than when deciding how to invest one's money. Entire departments of almost every investment bank, insurance firm and hedge fund are dedicated to picking the right baskets for their customers' nest eggs. 
This process is often referred to as \emph{portfolio optimization} (or \emph{asset allocation}) and involves optimizing the way in which a sum of money is divided up between a variety of financial instruments such that one maximizes the return for a given risk, or alternatively minimizes the risk for a given return. 
According to Modern Portfolio Theory (MPT)~\cite{MPT, MPTII, MPTIA}, which is widely used in the financial world to calculate asset allocations, one of the most effective ways to accomplish this is through diversification, that is to select groups of assets which are as uncorrelated as possible, or even anti-correlated. 

Clearly, we can identify numerous parallels between MPT and our community detection method. As we anticipated in our proof of eq.(\ref{eq:split}), a key property of the correlation-based modularity is that its maximization will identify mutually anti-correlated groups of time series (where anti-correlations are intended as residual, if some filtering has been applied).
Indeed, in figs. \ref{fig:SPCommunities}, \ref{fig:FCommunities} and \ref{fig:NCommunities}, all the links connecting different communities have negative weights, i.e. all communities are mutually anti-correlated.

The (residual) anti-correlations among communities allow us to identify combinations of stocks, which on top of the overall market mode and purely random fluctuations, move in opposition to each other.
Recalling from eqs. \eqref{eq:l1}, \eqref{eq:l2} and \eqref{eq:borrowed} that
\begin{equation}
\tilde{C}_{AB}^{(l)}=\textrm{Cov}[\tilde{X}_A,\tilde{X}_B]-\langle \textrm{Cov}[\tilde{X}_A,\tilde{X}_B]\rangle_l
\end{equation}
where $\tilde{X}_A\equiv \sum_{i\in A}X_i$, we obtain a practical recipe to construct a set $\{\tilde{X}_A\}$ of community-specific indexes (each built as the sum of the time series of the stocks within a community) such that, as follows from eq.\eqref{eq:split}, 
\begin{equation}
\textrm{Cov}[\tilde{X}_A,\tilde{X}_B]<\langle \textrm{Cov}[\tilde{X}_A,\tilde{X}_B]\rangle_l\qquad\textrm{if }A\ne B.
\end{equation}
In other words, the two indexes are residually less correlated with each other than expected under the null model, i.e. their mutual filtered correlations are negative.
This is a desirable trait from the point of view of risk management and portfolio optimization.

%%%%%%%%%%
\begin{figure}[b]
\centerline{\includegraphics[width =.45 \textwidth]{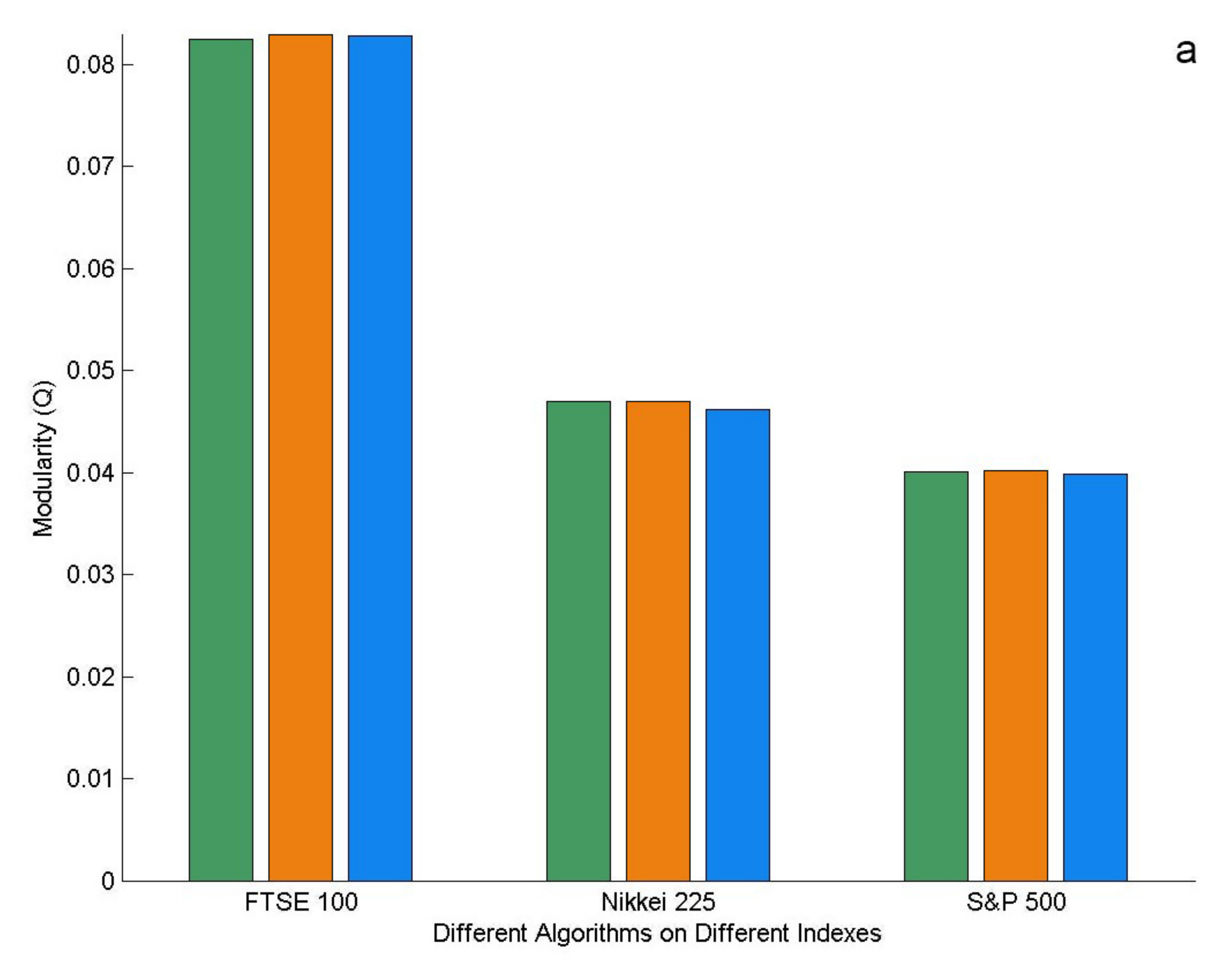}}
\caption{Maximized modularity values $Q_3(\vec{\sigma}^*)$ for each of the three markets (log-returns of daily closing prices from 2001Q4 to 2011Q3). 
Green is the Potts algorithm, orange is the Louvain algorithm and blue is the spectral algorithm.}
\label{fig:IndexMetricsA}
\end{figure}
%%%%%%%%%

\subsection{Comparative analysis of the three algorithms\label{sec:comparative}}
We now show a result that we anticipated at the beginning of this section, i.e. the fact that the three algorithms we introduced in sec.\ref{sec:redef} identify a very similar community structure on the data we considered.
This makes the results shown so far quite robust under changes of the protocol used to derive them.

%%%%%%%%%
\begin{table*}[t]
\center
    \begin{tabular}{|l|c|c|c|}
 \hline
     \textbf{S\&P 500}& \textbf{Potts} & \textbf{Louvain} & \textbf{Spectral} \\
    \hline
     \textbf{Potts} & 0 & 0.019 & 0.09 \\
\hline
     \textbf{Louvain} & 0.019 & 0 & 0.08 \\
\hline
     \textbf{Spectral} & 0.09 & 0.08 & 0\\
   \hline
   \end{tabular}%
\hskip .3cm
   \begin{tabular}{|l|c|c|c|}
 \hline
     \textbf{Nikkei 225}& \textbf{Potts} & \textbf{Louvain} & \textbf{Spectral} \\
    \hline
     \textbf{Potts} & 0 & 0.007 & 0.04 \\
\hline
     \textbf{Louvain} & 0.007 & 0 & 0.04 \\
\hline
     \textbf{Spectral} & 0.04 & 0.04 & 0\\
   \hline
    \end{tabular}%
\hskip .3cm
   \begin{tabular}{|l|c|c|c|}
 \hline
     \textbf{FTSE 100}& \textbf{Potts} & \textbf{Louvain} & \textbf{Spectral} \\
    \hline
     \textbf{Potts} & 0 & 0.11 & 0.11 \\
\hline
     \textbf{Louvain} & 0.11 & 0 & 0.05 \\
\hline
     \textbf{Spectral} & 0.11 & 0.05 & 0\\
   \hline
    \end{tabular}%
\caption{Comparison of the relative Variation of Information between the optimal partitions found by all algorithms, for the S\&P 500, the Nikkei 225 and the FTSE 100. The data are log-returns of daily closing prices from 2001Q4 to 2011Q3.}
\label{VI}
\end{table*}
%%%%%%%%

\begin{figure}[b]
\centerline{\includegraphics[width=.45\textwidth]{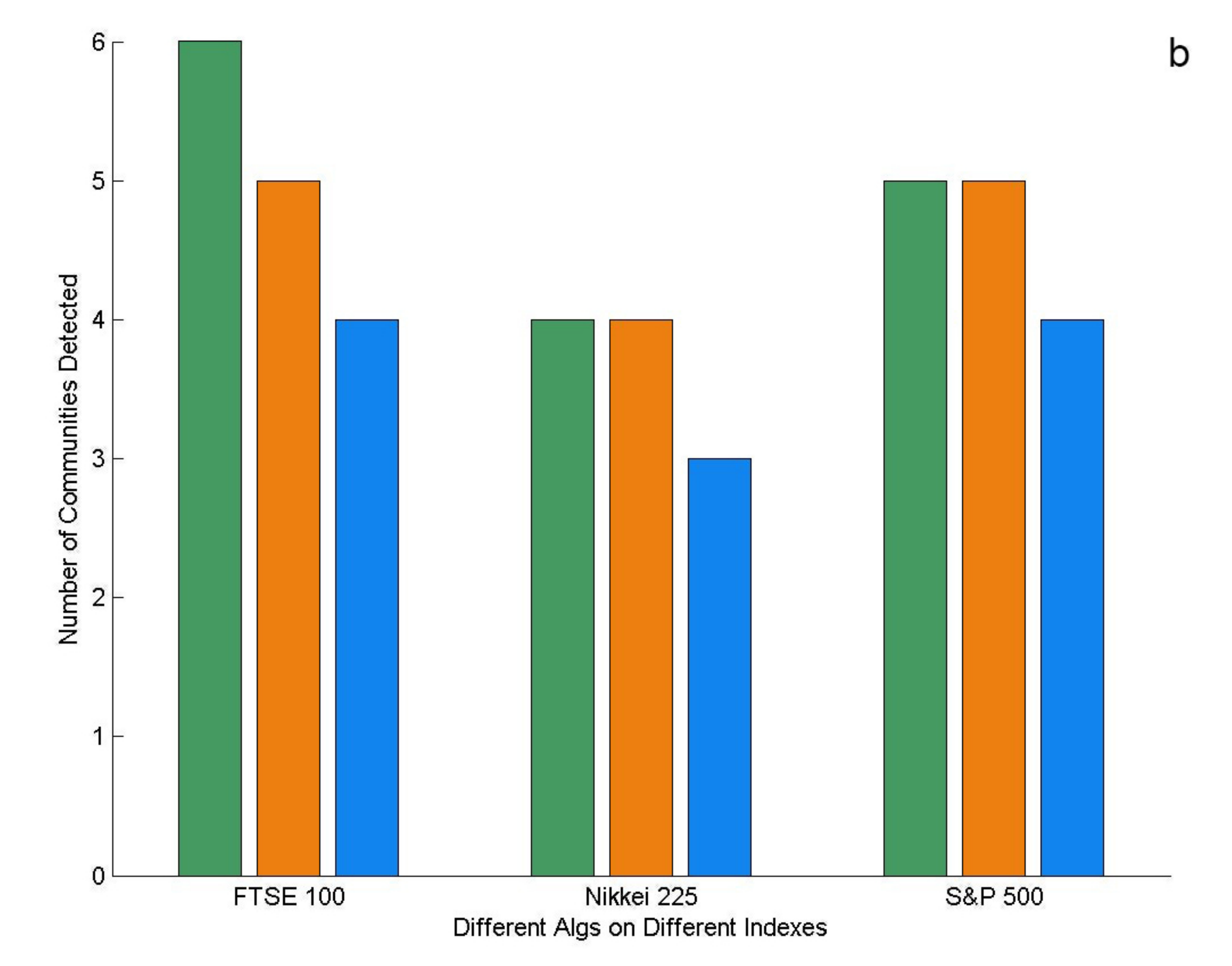}}
\caption{Number of communities detected in each of the three markets (log-returns of daily closing prices from 2001Q4 to 2011Q3). 
Green is the Potts algorithm, orange is the Louvain algorithm and blue is the spectral algorithm.}
\label{fig:IndexMetricsB}
\end{figure}

In figs. \ref{fig:IndexMetricsA} and \ref{fig:IndexMetricsB} we show the value of the maximized modularity $Q_3(\vec{\sigma}^*)$ and number of detected communities as the result of running all our three algorithms on the filtered correlation matrices for the S\&P 500, the Nikkei 225 and the FTSE 100. 
We recall from the discussion following eq.\eqref{eq:norm} and from the benchmarks studied in sec. \ref{sec:bench} that, unlike the corresponding problem in network analysis, our choice of $C_{norm}$ implies very small values of the maximized modularity, even in the  presence of well-defined communities, when the market mode is strong.
So the small values of $Q_3(\vec{\sigma}^*)$ shown in fig. \ref{fig:IndexMetricsA} do not imply a poor or weak community structure. 
It can be seen from fig. \ref{fig:IndexMetricsA} that all three algorithms perform very closely in terms of the maximized modularity value they achieve. 
Similarly, if we compare the number of communities found by the three methods (see fig.\ref{fig:IndexMetricsB}) we find that the number of communities is quite stable as well.

In table \ref{VI} we quantify more rigorously the differences in the composition of the communities detected by the three algorithms, by showing the $VI$ (see sec.\ref{sec:bench}) among all pairs of algorithms, for all the three indexes.
The values are quite low, indicating that the partitions found by  different algorithms are very similar.
%%%%%%%%%%
%%%%%%%%%

%%%%%%%%%
\begin{figure*}[t]
\includegraphics[width = .54\textwidth]{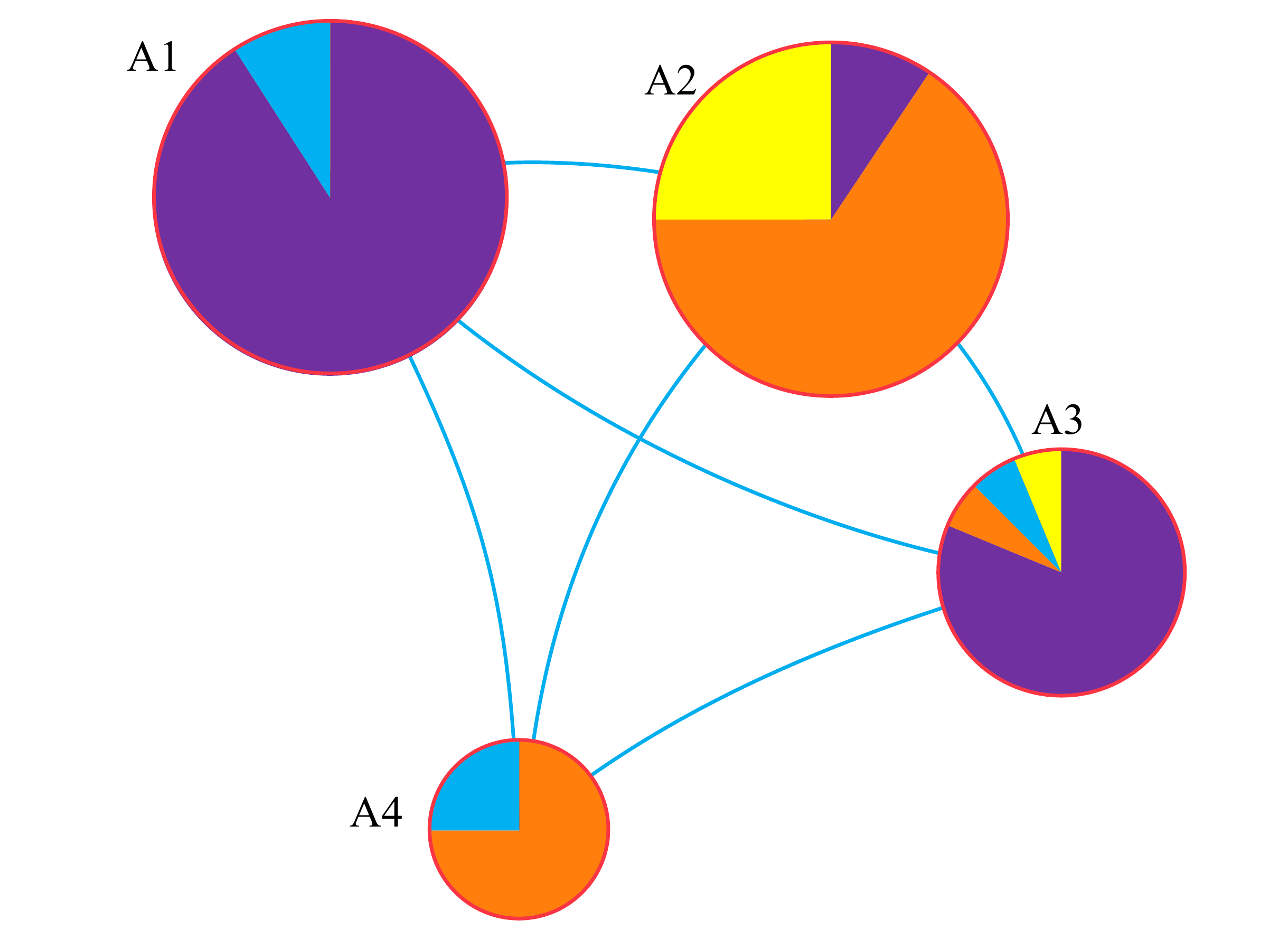}
\includegraphics[width = .45\textwidth]{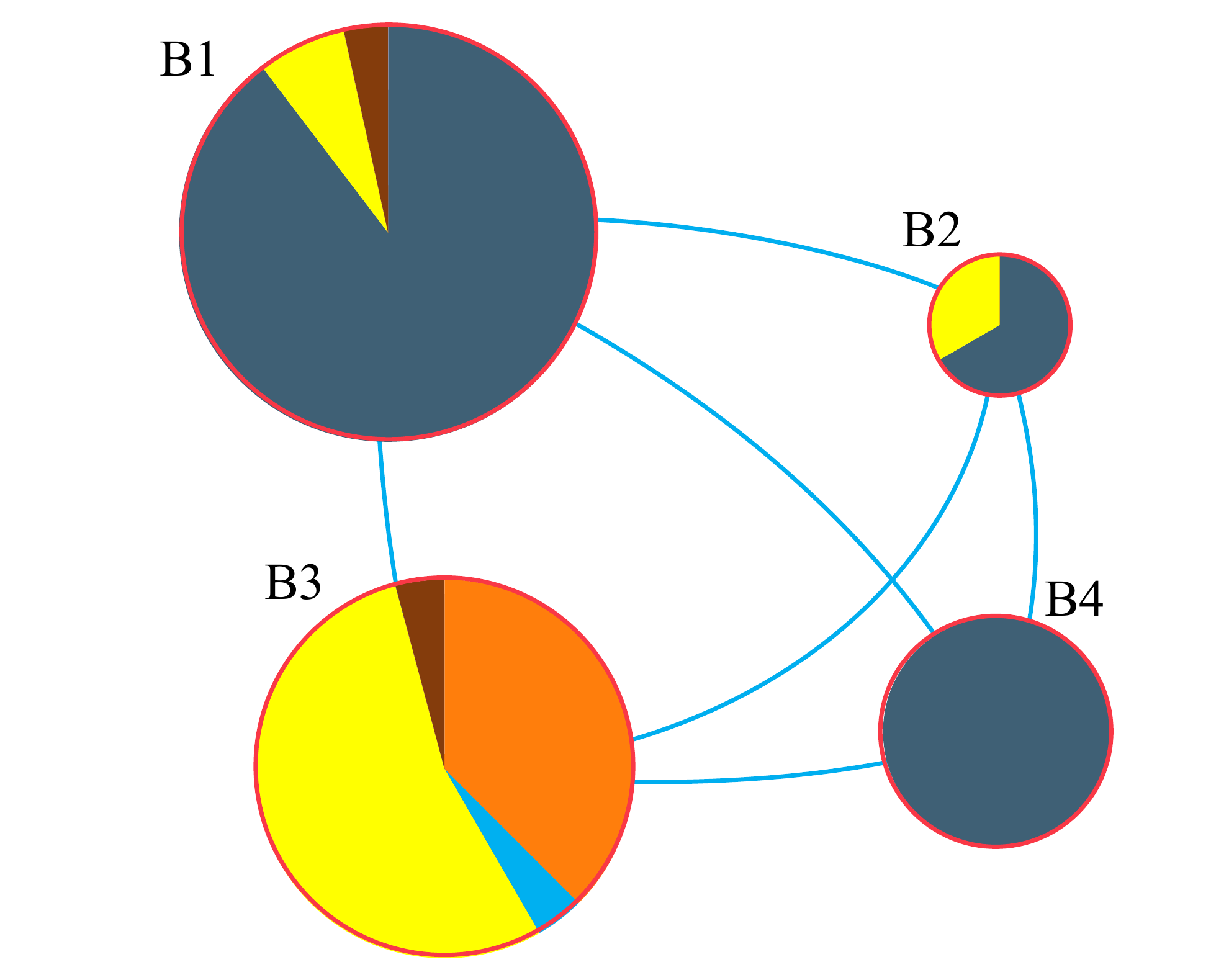}
\includegraphics[width = .42\textwidth]{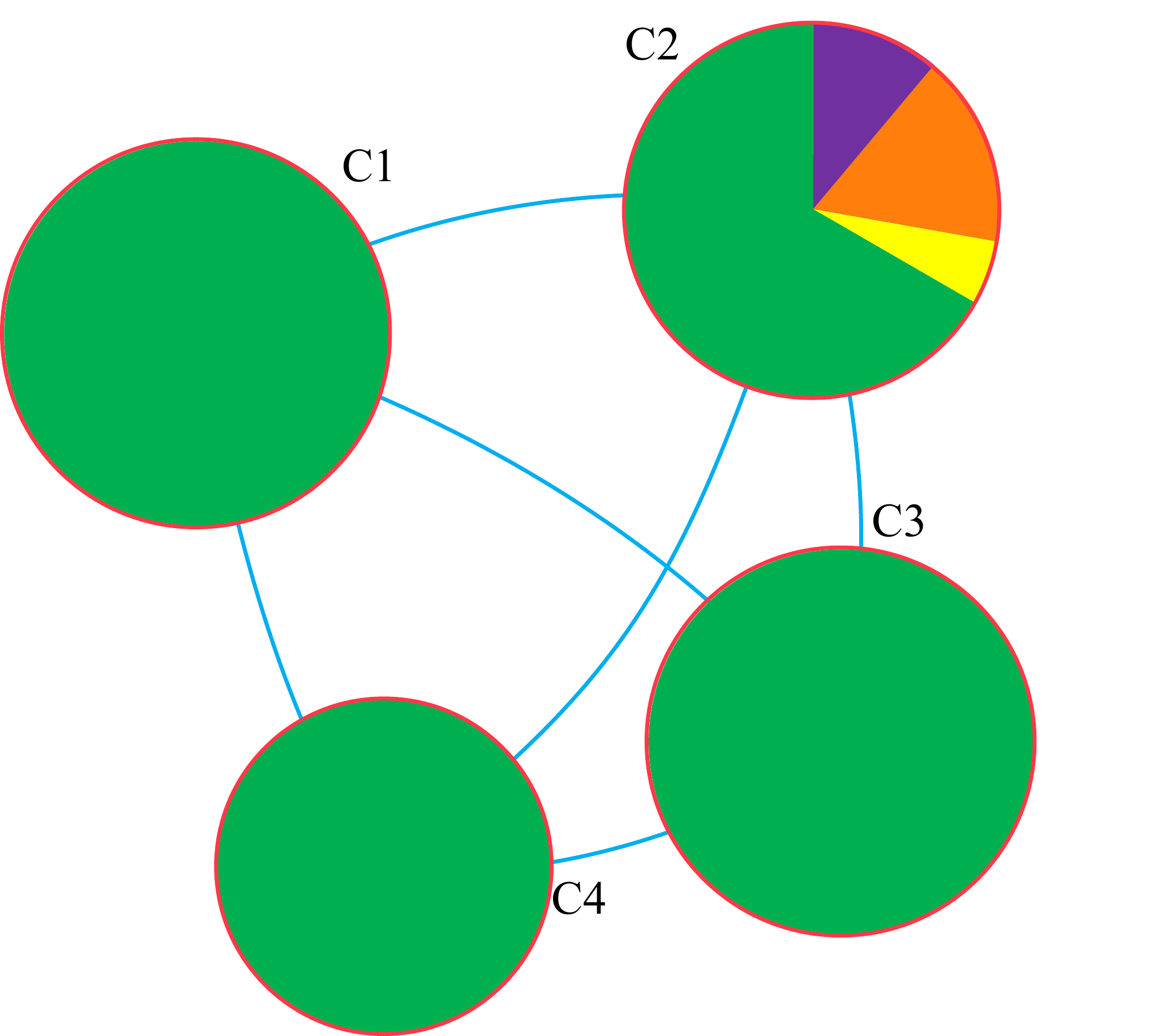}
\includegraphics[width = .45\textwidth]{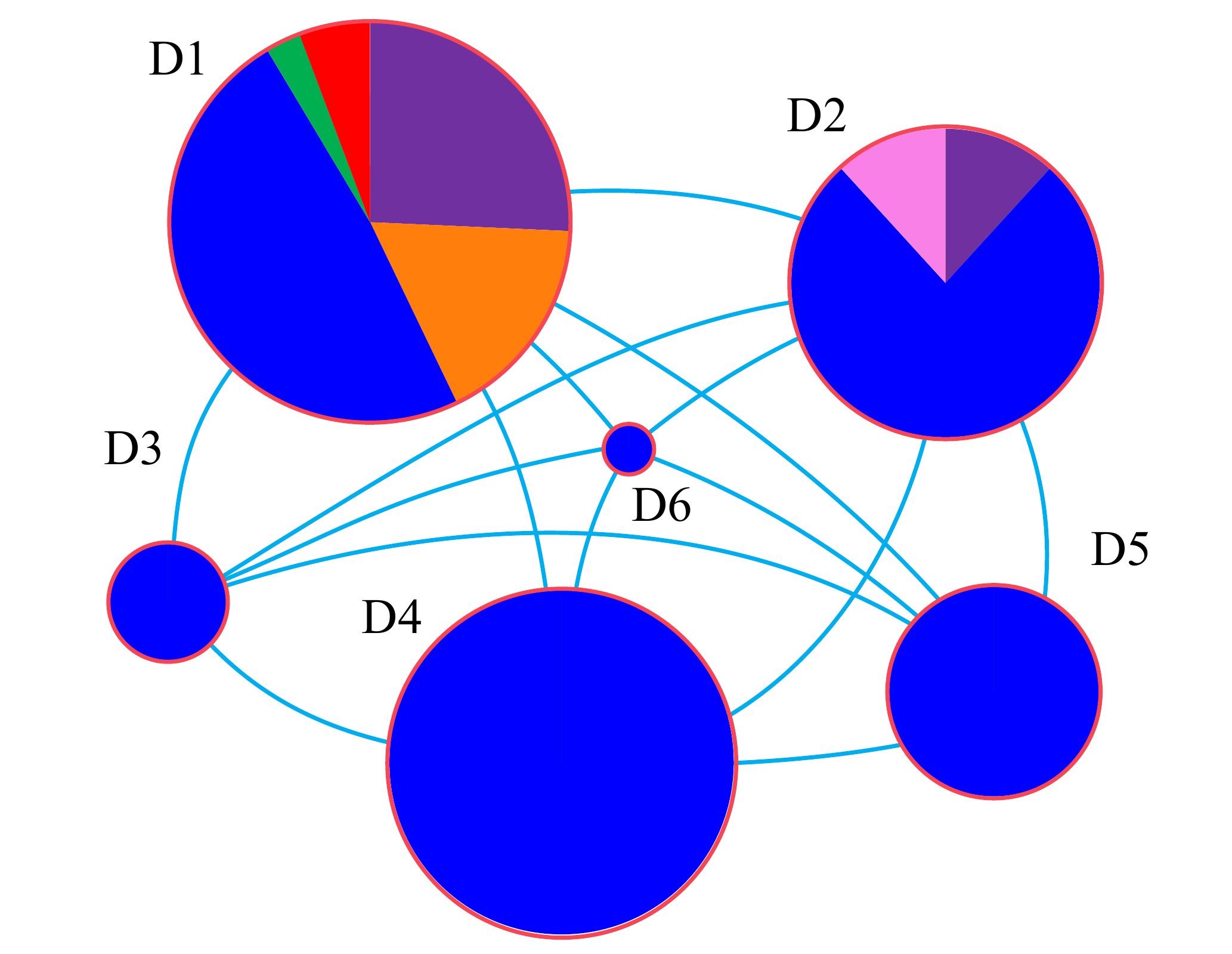}
\includegraphics[width = .5\textwidth]{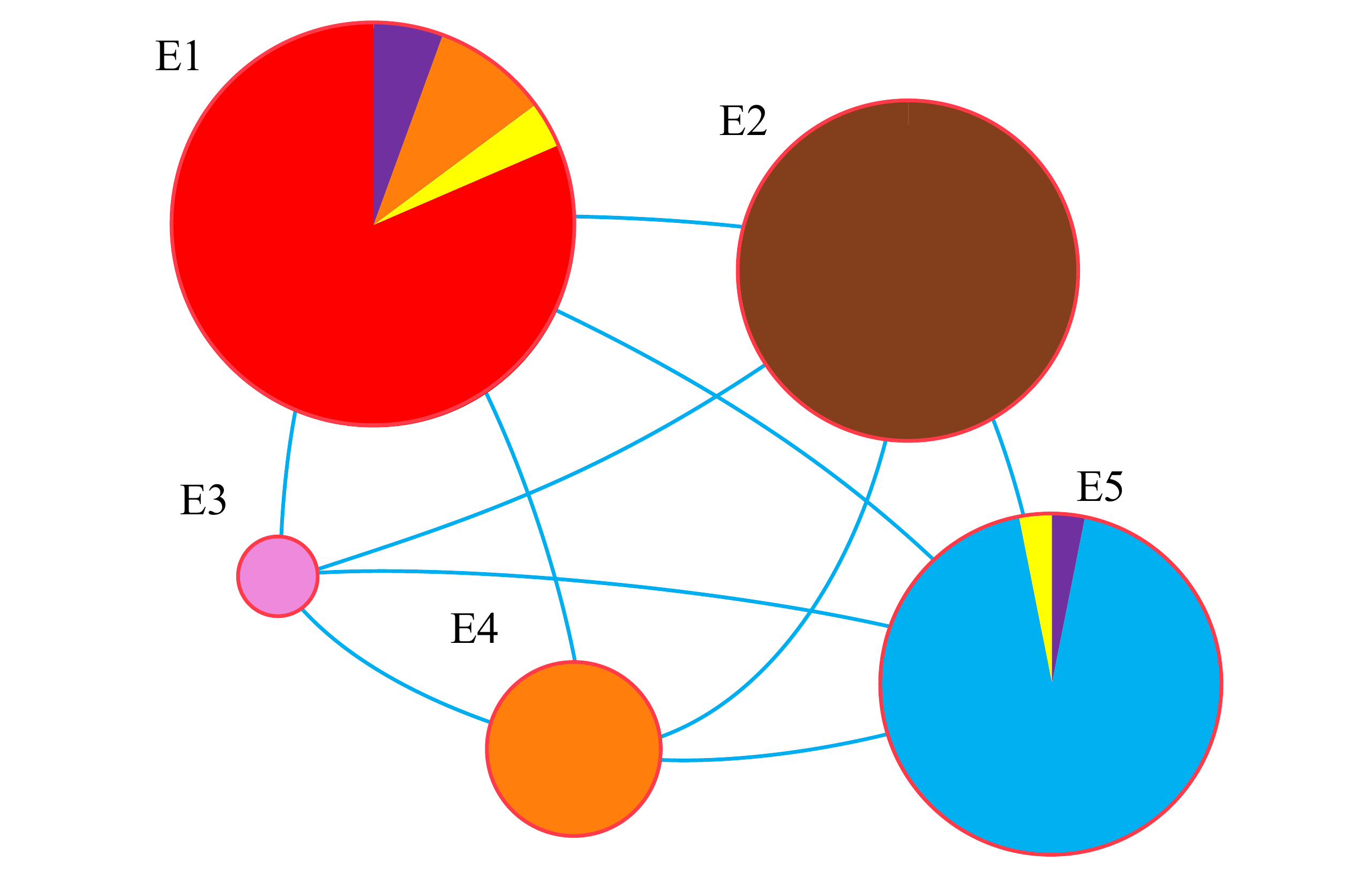}
\caption{Our multiresolution community detection method resolves the sub-community structure of the five communities of the S\&P 500 (see fig. \ref{fig:SPCommunities}).
Community $A$ mainly comprises Consumer Discretionary and Industrial stocks, $B$ all the Energy stocks, $C$ all the Finance stocks, $D$ all the IT stocks, while $E$ is highly heterogeneous but very well resolved into five separate sub-communities mainly comprising Utilities, Industrial, Health Care, Telecommunication Services, and Consumer Staples stocks respectively. 
Besides this relatively predictable partition, we note that Industrials stocks, and to a lesser extent also Materials and Consumer Discretionary stocks, are quite dispersed across different communities. (From cross-correlations of log-returns of daily closing prices from 2001Q4 to 2011Q3; the Louvain algorithm has been used).}
\label{fig:SubCommunitiesA}
\end{figure*}
%%%%%%%%%

%%%%%%%%%
\begin{figure*}[t]
\includegraphics[width = .99\textwidth]{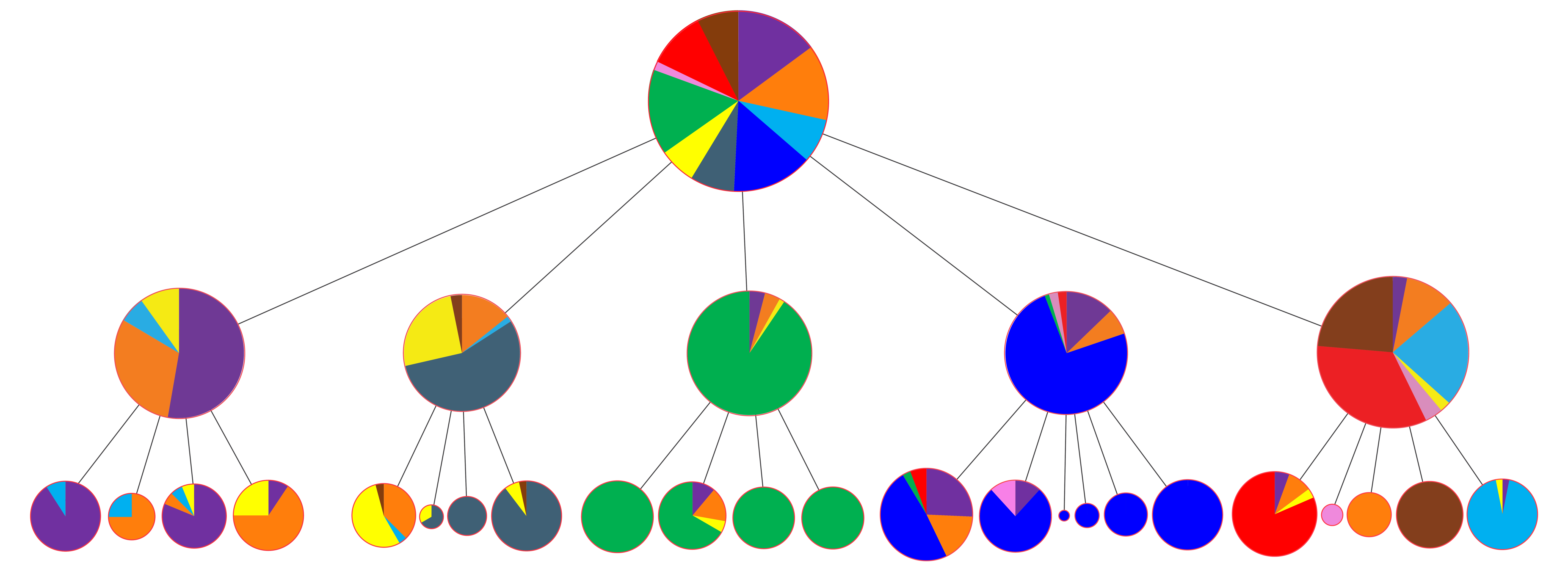}
\caption{Multiresolution community detection reveals the hierarchical structure of the communities of the S\&P 500 (from cross-correlations of log-returns of daily closing prices from 2001Q4 to 2011Q3). The dendrogram gives an alternative, combined representation of figs. \ref{fig:oneCommunity}, \ref{fig:SPCommunities} and \ref{fig:SubCommunitiesA}.
(The modified Louvain algorithm has been used).}
\label{fig:ComHierarchy}
\end{figure*}
%%%%%%%%%

\subsection{Hierarchical community structure of the market\label{sec:hiera}}
We now come to the application of the multiresolution community detection approach we introduced in sec.\ref{sec:multi}.
In the case of financial markets, the community-specific correlation responsible for the modular structure shown so far can be regarded, from the perspective of all stocks within one community, as a `micro market mode'. 
Just as the market mode discussed previously is responsible for the collective tide of an entire market, a similar force can be extrapolated at the community level. 
As discussed in sec.\ref{sec:multi}, accounting for this `community mode' in the leading eigenvalue and corresponding eigenvector of the correlation submatrix restricted to an individual community allows us to incorporate its effects, together with those of the overall market mode into the null model and again, detect any underlying structure, which surfaces upon the removal of its influence. 

Figure \ref{fig:SubCommunitiesA} shows the result of a single layer of recursion into the five communities of the S\&P 500 (depicted previously in fig. \ref{fig:SPCommunities}). Again, we note that the sub-communities are all residually anti-correlated with each other (within each parent community) but maintain an internal positive correlation. 
Although not obvious from the graph, the sub-communities tend to fall along GICS industry sector lines, with some interesting exceptions, as before. 

To call out a few examples, in community $D$ (fig.\ref{fig:SubCommunitiesA}), which contains all of the IT stocks, we see the sub-communities separating along Industry Group and Industry lines \footnote{The GICS hierarchy follows the form of ``Sector''\textbackslash ``Industry Group''\textbackslash ``Industry'' and so the examples in this section will follow the same form, e.g. ``Information Technology''\textbackslash ``Software \& Services''\textbackslash ``Software''. However, since our discussion will be predominantly focused on similarities at the Industry level, for brevity we will often omit the full GICS label. In such cases we will delineate this using italics, e.g \textit{Software} in the example above.}. Sub-community $D5$ is comprised of only \textit{Software} stocks, $D4$ contains all of the \textit{Semiconductor \& Semiconductor Equipment} stocks, and $D2$ contains all of the \textit{Internet Software \& Services} stocks. Interestingly enough though, $D2$ also contains Amazon Inc. and Priceline Inc. from the Consumer Discretionary Sector, which one could argue are quite aligned with the Internet.
Continuing the analysis further, we see that in community $C$ the Finance community sub-community $C2$ contains all of the \textit{Commercial Bank} stocks, while $C3$ contains all but one of the \textit{Insurance} companies. 
$C4$ is exclusively \textit{Real Estate Investment Trusts (REITs)} and accounts for all of them.
Similar partitions can be seen in the other sub communities and further recursion into these communities produce still further separation, close to but not exactly in line with the GICS classification. 

Figure \ref{fig:ComHierarchy} depicts the hierarchical nature of the S\&P 500 to three layers deep. The process can be continued until no single community can be partitioned further into any combination of two or more sets which are anti correlated with each other.
For instance, community $E$ (fig.\ref{fig:SubCommunitiesA}) which contains a variety of stocks from various GICS sectors separates out such that the bulk of the stocks in the different sectors find themselves in their own sub community. If we further probe into community $E1$, which contains all of the Health Care stocks we see that the sub-communities (not shown) fall very closely along Industry lines, with five communities each comprised predominantly of \textit{Pharmaceutical}, \textit{Biotechnology} and \textit{Life Science Tools}, \textit{Health Care Providers \& Services}, \textit{Health Care Equipment \& Supplies} and everything else, respectively. Albeit interesting, these results invite inspection of the stocks that end up in the ``everything else'' community. These stocks were deemed correlated with the other Health Care stocks, when they were all placed in community $E1$, and include McGraw Hill Inc., H\&R Block and Waste Management Inc. 
None of these stocks immediately stand out as being fundamentally related to Health Care. 

Similar outliers exist in the other communities as well. 
It may well be the case that there is good reason for their association, for example a shared parent company, sizable investment, common board members or some other significant relationship, or it may be purely coincidental. Gaining a better understand of this takes us to our next lines of experimentation. 

\section{Multifrequency community detection \label{sec:resol}}
Having examined the mesoscopic structure of a set of financial markets, one might be curious as to whether that structure is specific to the chosen frequency of the original time series. That is, one would like to check whether the same communities would be retrieved if the returns which comprised the original time series were calculated every minute, every half hour or every two days. To answer this `multifrequency' community detection problem, in this section we evaluate the robustness of partitions at a variety of temporal resolutions. 

\subsection{Multiple-frequency data}
In order to maintain consistency with the results previously described in this paper, we use the same time frame but, instead of working with daily log-returns, we created new data using minute log-returns for the S\&P 500 stocks. This has the initial effect of greatly increasing the amount of data being using: from 2500 data points per stock for the daily returns to approximately 900,000 for the minute returns. 
In order to accommodate some missing data from the minute returns, we had to reduce the set of 445 to 413 stocks, noting that the removed stocks were relatively evenly distributed across the top level sectors of the GICS, so as not to deplete any one particular sector. 
With the minute return time data of these 413 stocks we created nine new sets of time series, corresponding to a variety of different resolutions $\Delta_t$ spanning the same ten-year period:
\begin {equation}
\Delta_t \in \{ \text{1, 5, 10, 15 \& 30 mins, 1 hour,  0.5, 1 \& 2 days} \}. \nonumber
\end {equation}
For example, the 5-min data was created by taking the price of every stock every 5th minute throughout the day. From these nine sets of time series we then proceeded in the same fashion as was previously described for daily return data, creating correlation matrices and leveraging RMT filtering to produce the respective null models.

\subsection{Robustness over multiple frequencies}
\begin{figure}[t]
\centerline{\includegraphics[width = .45\textwidth]{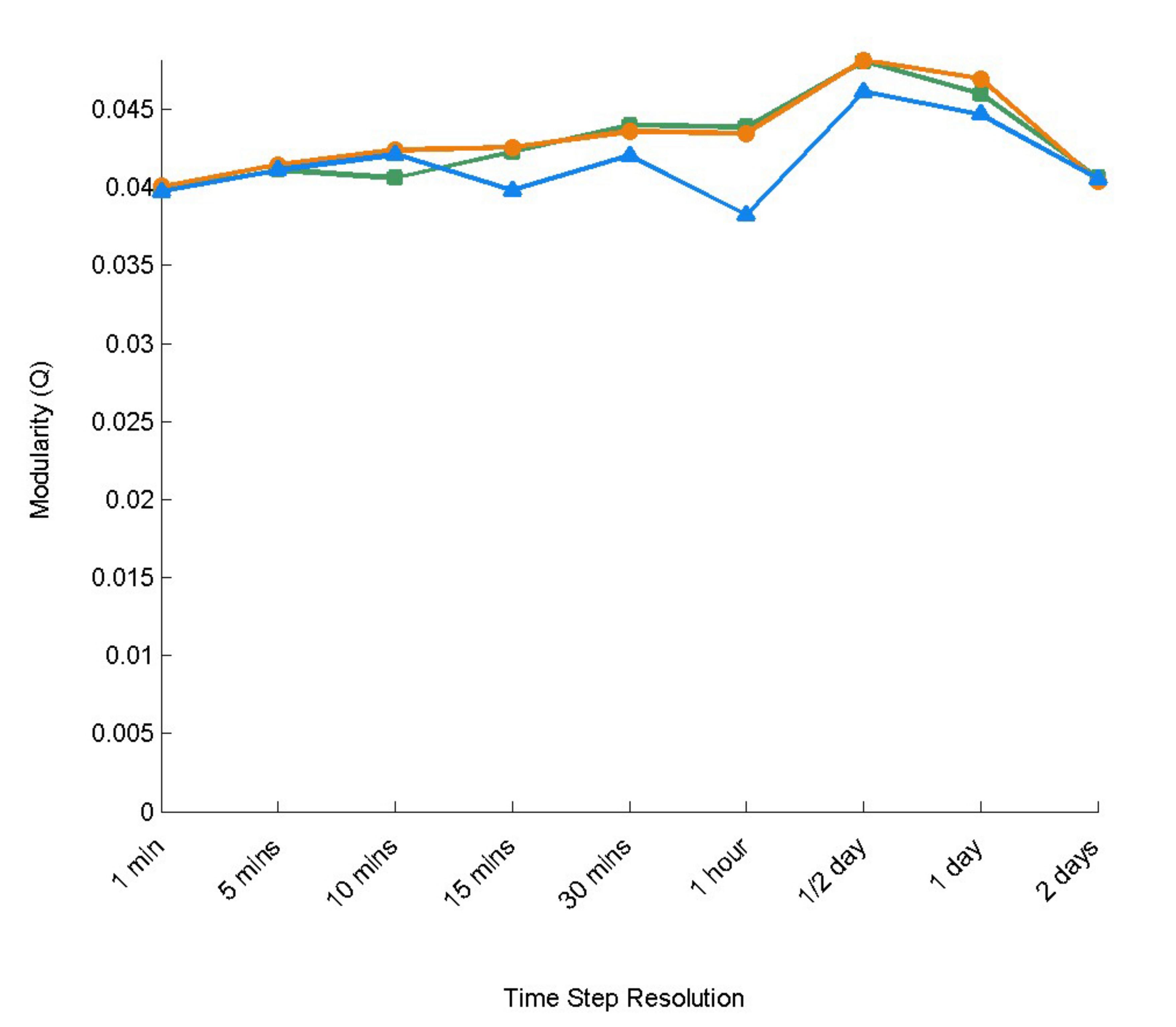}}
\caption{Multifrequency analysis of the modularity $Q_3(\vec{\sigma}^*)$ for the different methods, as the resolution goes from one-minute intervals to two-day intervals. A relatively consistent value of the modularity can been seen across all time step resolutions. (Potts algorithm in green squares, Louvain algorithm in orange circles and Spectral algorithm in blue triangles).}
\label{fig:ResQplot}
\end{figure}

\begin{figure}[t]
\centerline{\includegraphics[width = .45\textwidth]{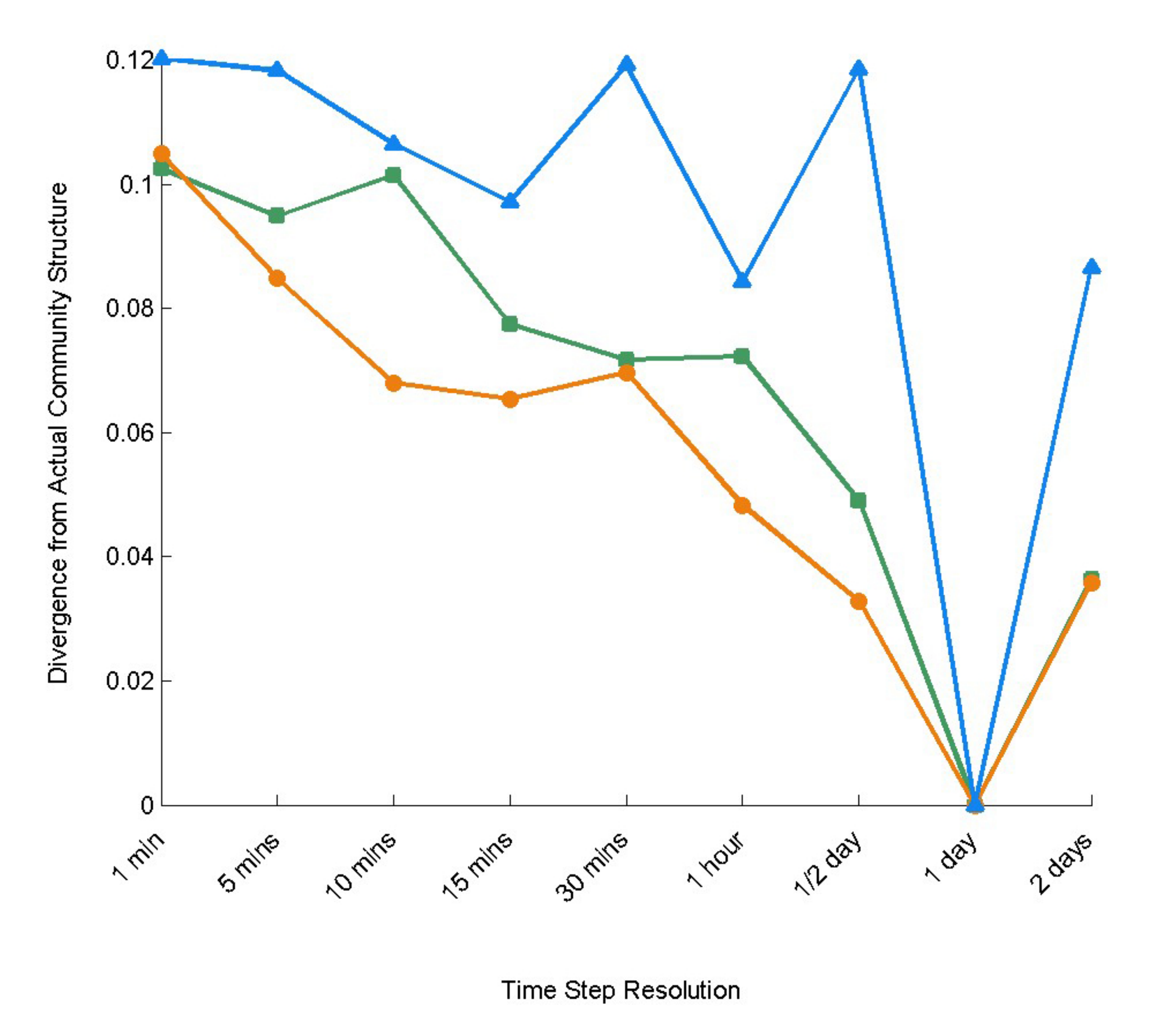}}
\caption{Multifrequency analysis of the Variation of Information between each of the nine data sets of different time resolutions and the partition for the data set of daily time steps. It can be seen that all data sets yield partitions quite similar to each other, but there is still slight degradation as the resolution gets finer. (Potts algorithm in green squares, Louvain algorithm in orange circles and Spectral algorithm in blue triangles).}
\label{fig:ResVIplot}
\end{figure}

To measure the effects of resolution we applied all three of the community detection algorithms discussed above to all nine data sets, yielding various values for the modularity $Q_3(\vec{\sigma}^*)$ of the partition (see fig. \ref{fig:ResQplot}). 
Since there can be multiple peaks within a modularity landscape ~\cite{landscape}, all yielding the same value of $Q_3(\vec{\sigma}^*)$ but exhibiting different community structures, we use $VI$ (see Sec. \ref{sec:bench}) as a measure of the difference between the partitions. 
Since $VI$ is a comparative measure, we (arbitrarily) use the community structure previously ascertained from the daily returns as the point of reference. 
Thus, as can be seen in fig. \ref{fig:ResVIplot}, the $VI$ for the 1-day returns is 0 by construction, indicating perfect similarity, whereas the community structure for every other resolution shows some level of deviation.

Overall, it can be seen from the combination of figs. \ref{fig:ResQplot} and \ref{fig:ResVIplot} that there exists a considerable amount of consistency between the communities detected at differing resolutions, with the $Q_3(\vec{\sigma}^*)$ values remaining almost constant and $VI$ deviating slightly with each resolution interval but indicating in most cases no more than a 10\% difference between the communities of a particular resolution and those of the 1-day resolution. 
This means that the correlations between large groups of stocks are not strongly dependent on the resolution of the chosen time step. 
One might expect to see fluctuations in the variance of stocks at smaller time resolutions, where the more volatile periods of trading (such as market open and market close) are captured. 
However since we are dealing with correlation matrices, this variance is normalized away. Moreover, we recall that our definition of $C_{norm}$ in eq.\eqref{eq:norm} controls for the varying volatitily (variance of the total log-return over all stocks), allowing us to focus solely on the relationships between the stocks themselves. 
%%%%%%%%%%%
%%%%%%%%%%%
%%%%%%%%%%%%
%%%%%%%%%%%%%%

\subsection{Detection of `hard' and `soft' stocks: overlapping community structure}
\begin{figure}[t]
\centerline{\includegraphics[width=0.45\textwidth]{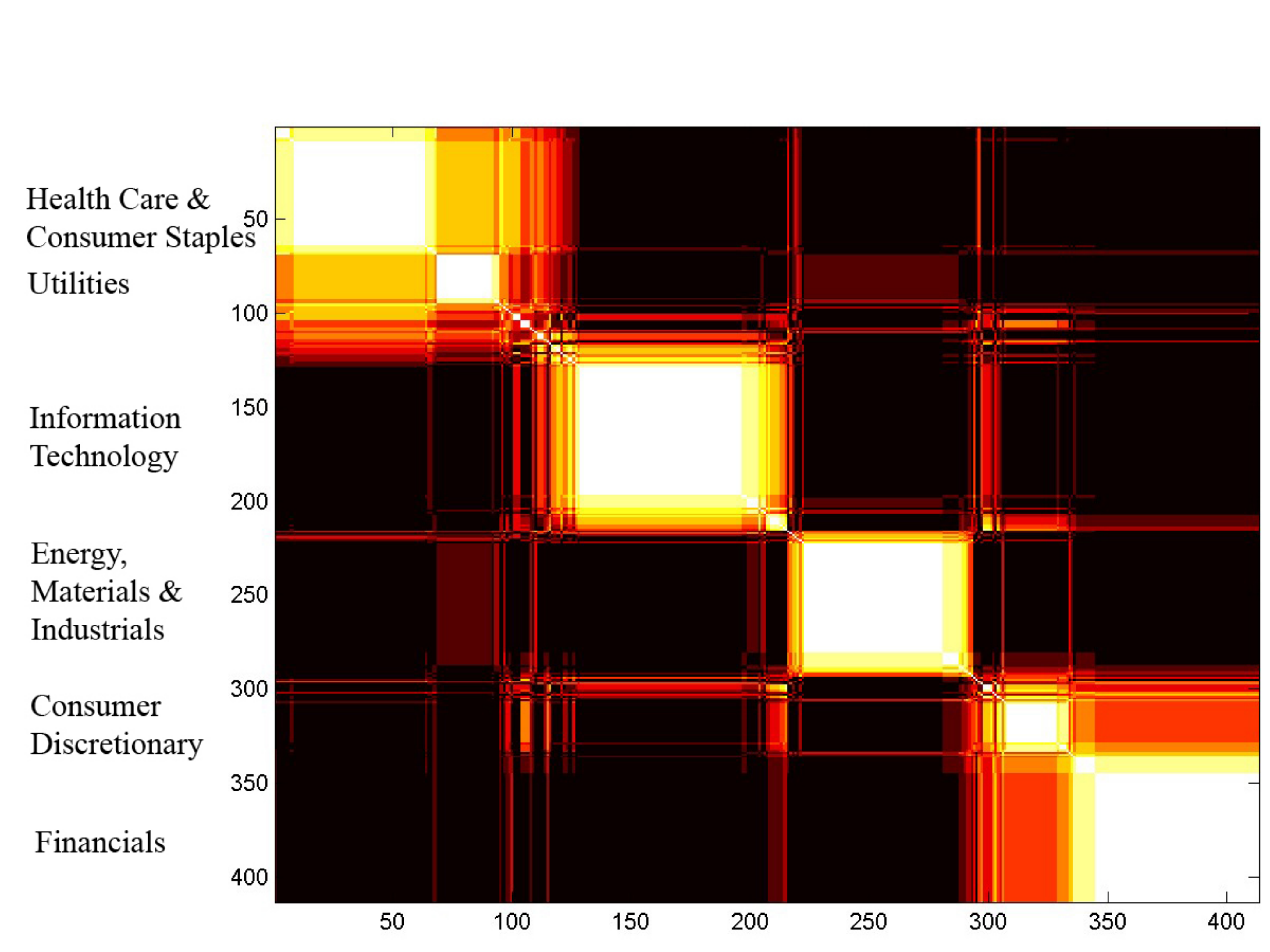}}
\caption{Multifrequency heat map showing the normalized co-occurrence of different pairs of stocks within the same community, for the same time period but over various temporal resolutions of the original time series (white: unit frequency, black: zero frequency). 
The stocks have been ordered using simulated annealing to position stocks with high degree of cross correlations next to each other. To further inform the graph, the GICS sectors have been specified, emphasizing which groupings of stocks tend to associate with a particular sector. Overall, the blocks of large black and white areas indicate a high degree of coherence of the communities at different resolutions.
However, there are two groups of `soft' stocks, one alternating across Utilities, Health Care and Consumer Staples, and one alternating across Consumer Discretionary and Financials. (Produced using the Louvain algorithm).}
\label{fig:ResolutionHM}
\end{figure}

Although the values of $Q_3(\vec{\sigma}^*)$ and $VI$ do provide reasonable insight into the robustness of community structure at the different resolutions, we take the analysis one step further and examine the communities from the perspective of the individual stocks. 
That is, we can further examine the community affiliation of individual stocks at the various resolutions to ascertain the frequency of times any two stocks find themselves in the same community as each other. 
We show the results of this analysis in fig. \ref{fig:ResolutionHM}, which is a heat map of the different stocks, such that the color of every pair indicates the frequency of co-occurrence in the same community, across all resolutions. 
For example, if two stocks were always in the same community (unit frequency) then their entry in the heat map is white, while if they were never in the same community regardless of the time step chosen (zero frequency) then their entry is black.
Stocks which share a community for some time steps are shades of red (lower frequency) or yellow (higher frequency). 

As we can see, the results are in line with the graph of $VI$ in fig. \ref{fig:ResVIplot}. That is, the communities tend to consist of a large core of `hard' stocks that are unwavering over the different resolutions, plus
a small amount of `soft' stocks that fluctuate between communities, presumably giving rise to the 10\% fluctuation in community structure observed with the $VI$ analysis. A significant finding is the existence of a group of soft stocks that alternate across the Utilities, Health Care and Consumer Staples communities, and of another group of soft stocks alternating across the Consumer Discretionary and Financials communities. 

%%%%%%%%%%%%%%%%%
%%%%%%%%%%%%%%%%

It should be noted that our identification of the `hard' stocks that are most of the time part of the core of a community and the `soft' ones that are instead alternating across communities is a way to take the potentially overlapping nature of communities into account, even if using a non-overlapping method like modularity maximization.  
This possibility has no counterpart in the standard network-based community detection problem, and is offered by the intrinsic dependence of correlation matrices on the frequency of the original time series.
In what follows, we will use the dynamical evolution of correlations to explore another dimension of variability leading to an alternative way to resolve overlapping communities of multiple time series.

\section{Time Dynamics\label{sec:dyn}}
When optimizing a portfolio, there is a constant need to choose an adequate period of history from which to try to predict future behavior of the assets in the portfolio. Choosing too short a period will inaccurately bias one's results, because extreme events are weighted too heavily. Similarly, choosing too long a history can imply stability where none exists. 
In general, analyzing the stability of communities over time provides us with reassurance that our models are in fact producing statistically significant results as well as providing insightful information about the data itself. 
For example, it is well known in finance that markets become much more globally correlated during periods of economic decline. 
Stated in the terminology we have been using throughout this paper, they fall more under the influence of the market mode and relinquish the structure provided by the group mode. 
That being the case, we would expect to see communities lose coherence during periods in the dataset that we know to have been economically troublesome, for example the tech bubble bursting in 2000 - 2001 or the sub-prime lending crisis, 2007 - 2008. 

Since we have shown that the 15-minute data set yield very similar communities to the daily data set for the S\&P, we can feel reasonably assured that we can use 15-minute data instead of the daily data, which will allow us to examine the S\&P data set using a sliding time window of 2 years, corresponding to a ratio $T/N=6$, and even look at more fine grained windows, e.g. 6 months.

\subsection{Two-year window}
\begin{figure}[t]
\centerline{\includegraphics[width=0.45\textwidth]{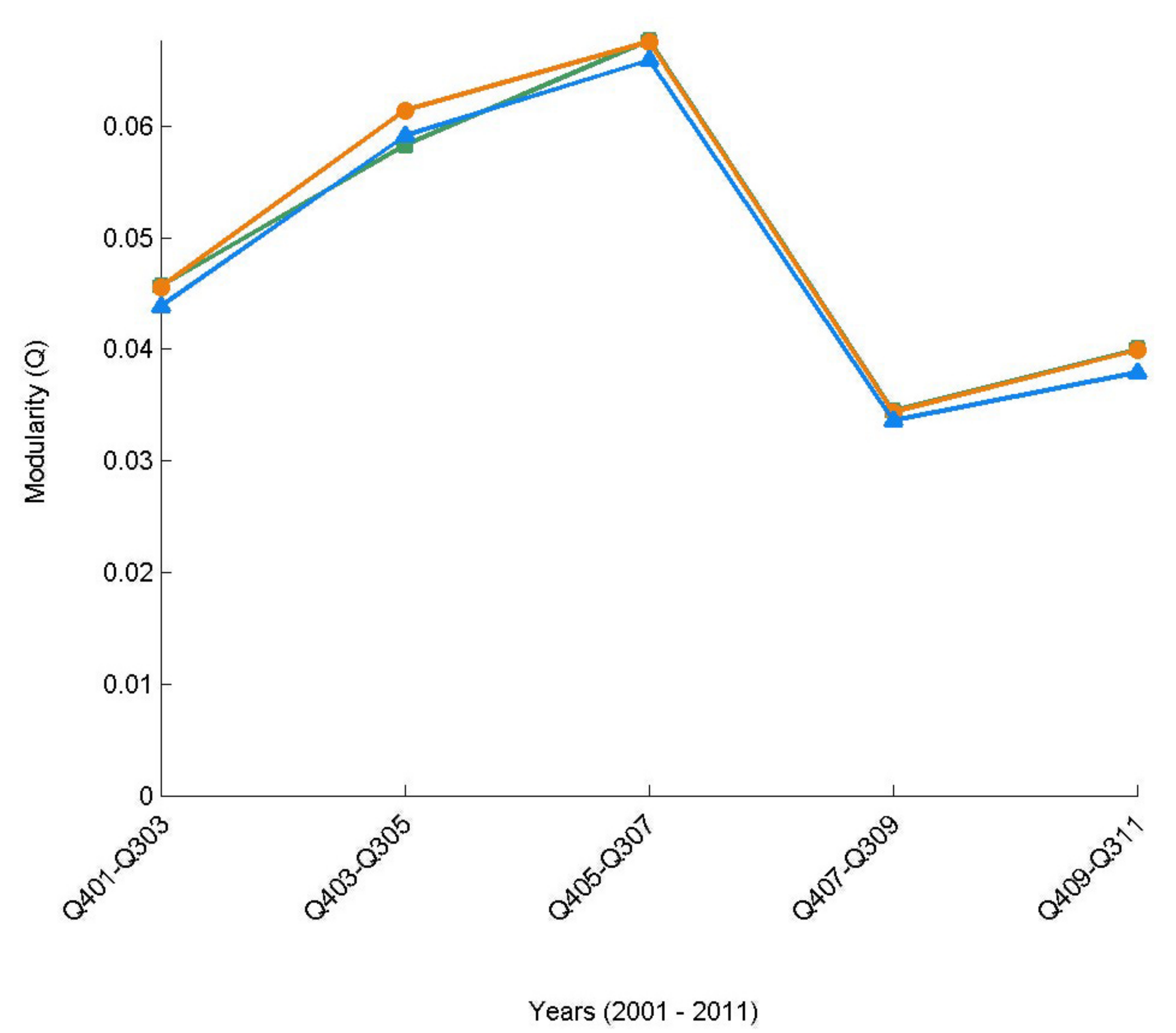}}
\caption{Temporal trend in the values of $Q_3(\vec{\sigma}^*)$ for the different methods over a 2-year sliding window spanning the time frame from Oct 2001 to Oct 2011. (Potts method in green squares, Louvain method in orange circles and Spectral method in blue triangles).}
\label{fig:QTimeSPX2Year}
\end{figure}

\begin{figure}[t]
\centerline{\includegraphics[width = 0.45\textwidth]{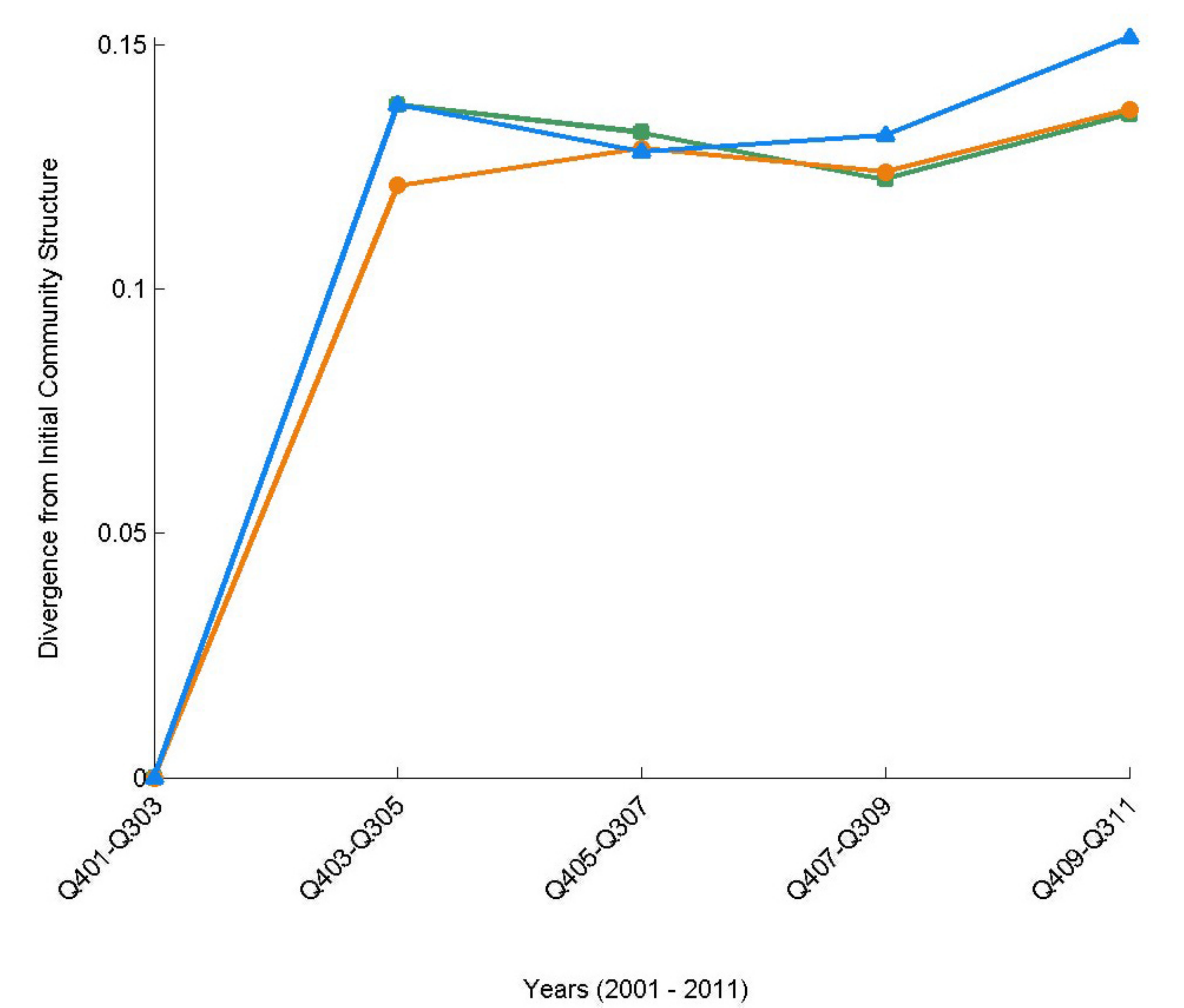}}
\caption{Temporal trend of $VI$ showing the similarity in community structure for the different algorithms over a 2-year sliding window spanning the time frame from Oct 2001 to Oct 2011. (Potts algorithm in green squares, Louvain algorithm in orange circles and Spectral algorithm in blue triangles).}
\label{fig:VITimeSPX2Year}
\end{figure}

We now seek to examine the community structure over sequential periods of two years to unearth any anomalies which might exist. 
We again apply the different methods of community detection using the two-year window time series sets of the S\&P 500 and subsequent null models created using RMT filtering. 
As we did for our analysis of resolution, here we evaluate the modularity function $Q_3(\vec{\sigma}^*)$ for each period (see fig. \ref{fig:QTimeSPX2Year}) along with the $VI$ (see fig. \ref{fig:VITimeSPX2Year}), where for $VI$ we are comparing each window with the initial two-year window.

As before, we see that all three algorithms perform in a reasonably similar manner.
However, unlike our analysis of robustness over different resolutions (which showed little change in community structure or in the modularity), here we see that $Q_3(\vec{\sigma}^*)$ fluctuates over the different windows. 
We recall again that, as we mentioned in our discussion following eq.\eqref{eq:norm}, our choice of $C_{norm}$ is already discounting (the evolution of) the volatility of the market.
Still, we see that $Q_3(\vec{\sigma}^*)$ rises slowly from the period ending in 2003 to the period ending in 2007, implying an increase in the strength of communities, and then falls by more than 50\% by the end of the period ending in 2009. This drop implies a de-coherence of the communities throughout that period, quite possibly attributed to the financial crash of 2007 - 2008. 
This seems in line with the observation that during periods of financial crisis, markets tend to become more globally correlated, overwhelming the effect of group-level correlations.
However, it is interesting that the values of $VI$ have remained quite small and stationary (fig. \ref{fig:VITimeSPX2Year}).
This indicates that, despite the fluctuating value of the modularity (i.e. of the relative intra-community correlations), the composition of the communities has remained very stable over time.

%%%%%%%%%%%%%%%%%%%
%%%%%%%%%%%%%%%

\subsection{Six-month window}
\begin{figure}[t]
\centerline{\includegraphics[width = 0.45\textwidth]{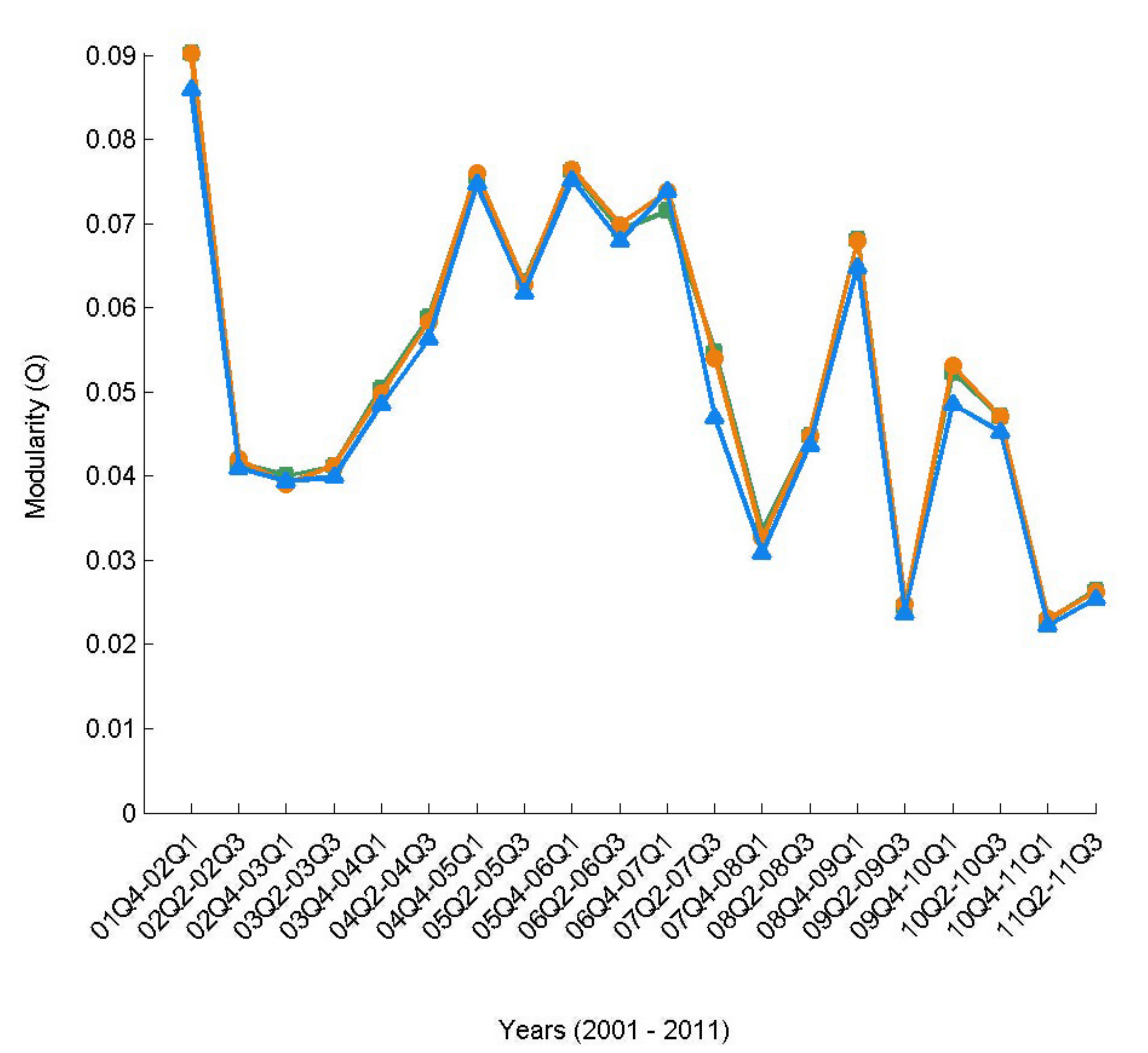}}
\caption{Temporal trend in the values of $Q_3(\vec{\sigma}^*)$ for the different algorithms over a 6-month sliding window spanning the time frame from Oct 2001 to Oct 2011. (Potts algorithm in green squares, Louvain algorithm in orange circles and Spectral algorithm in blue triangles).}
\label{fig:QTimeSPX6Month}
\end{figure}

\begin{figure}
\centerline{\includegraphics[width = 0.45\textwidth]{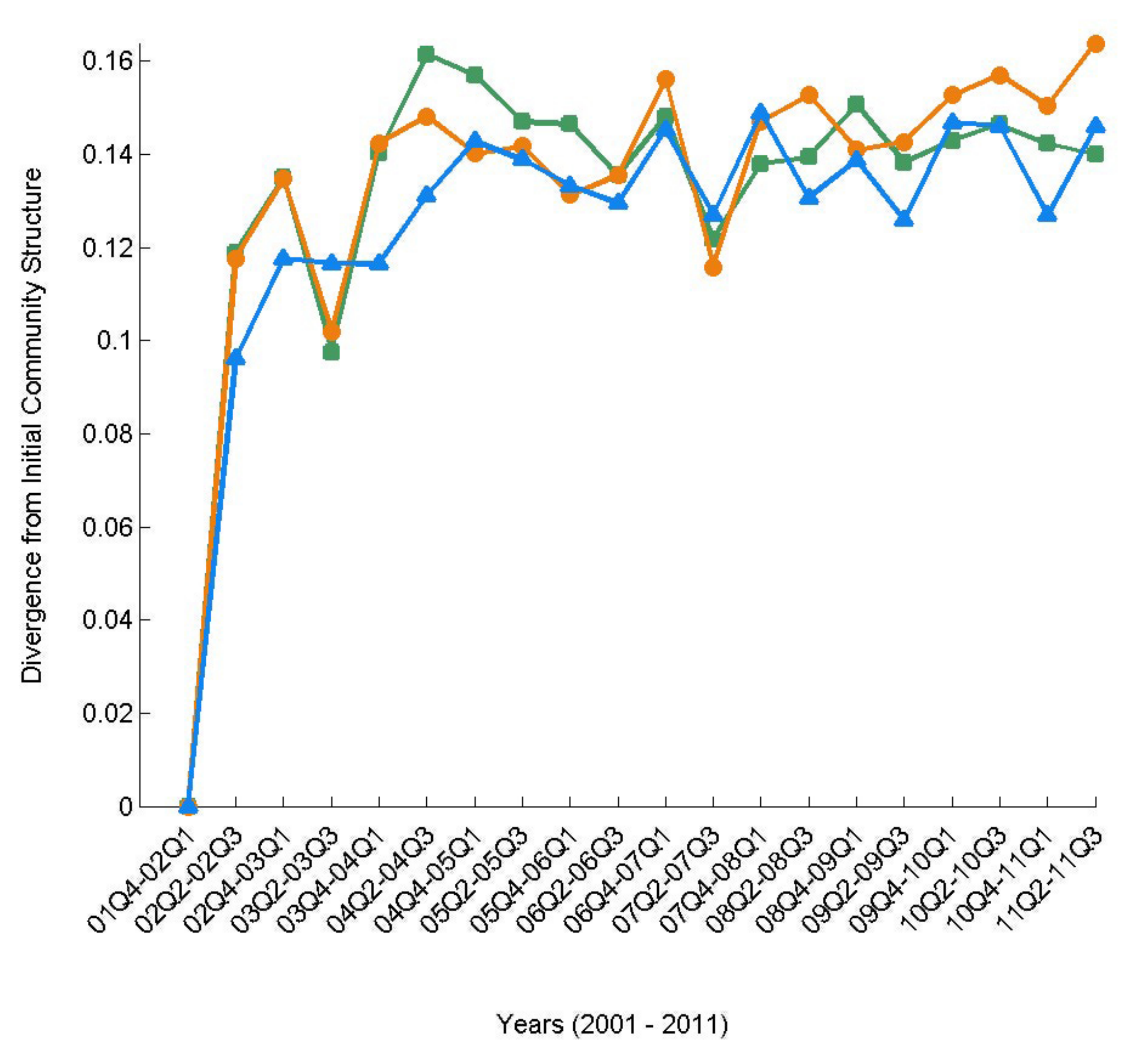}}
\caption{Temporal trend of $VI$ showing the similarity in community structure for the different algorithms over a 6-month sliding window spanning the time frame from Oct 2001 to Oct 2011. (Potts algorithm in green squares, Louvain algorithm in orange circles and Spectral algorithm in blue triangles).}
\label{fig:VITimeSPX6Month}
\end{figure}

We can continue to probe this system at a finer grained resolution of time periods, to see if the observations made with the two year window hold up. 
Again, we plot both $Q_3(\vec{\sigma}^*)$ and $VI$ for the same ten-year period of the S\&P and present the results in figs. \ref{fig:QTimeSPX6Month} and \ref{fig:VITimeSPX6Month} respectively. We can immediately see that the homogeneity of community structure that we see when probing the data using two-year time windows still exists for the most part, but there exists some fluctuations in modularity and community composition over the various six-month periods. The graph of $Q_3(\vec{\sigma}^*)$ in fig. \ref{fig:QTimeSPX6Month} reinforces the observation from fig.  \ref{fig:QTimeSPX2Year} of a significant drop in modularity around the time of the most recent financial crisis, and more accurately pinpoints it to the last half of 2007. 
The $VI$ plot in figure \ref{fig:VITimeSPX6Month} indicates again that although the strength of community structure, as measured by $Q_3(\vec{\sigma}^*)$, may have been decreasing, the overall composition of the communities remained relatively constant. 
%%%%%%%%%%%%%%%%
%%%%%%%%%%%%%%%%

\begin{figure}
\centerline{\includegraphics[width=0.49\textwidth]{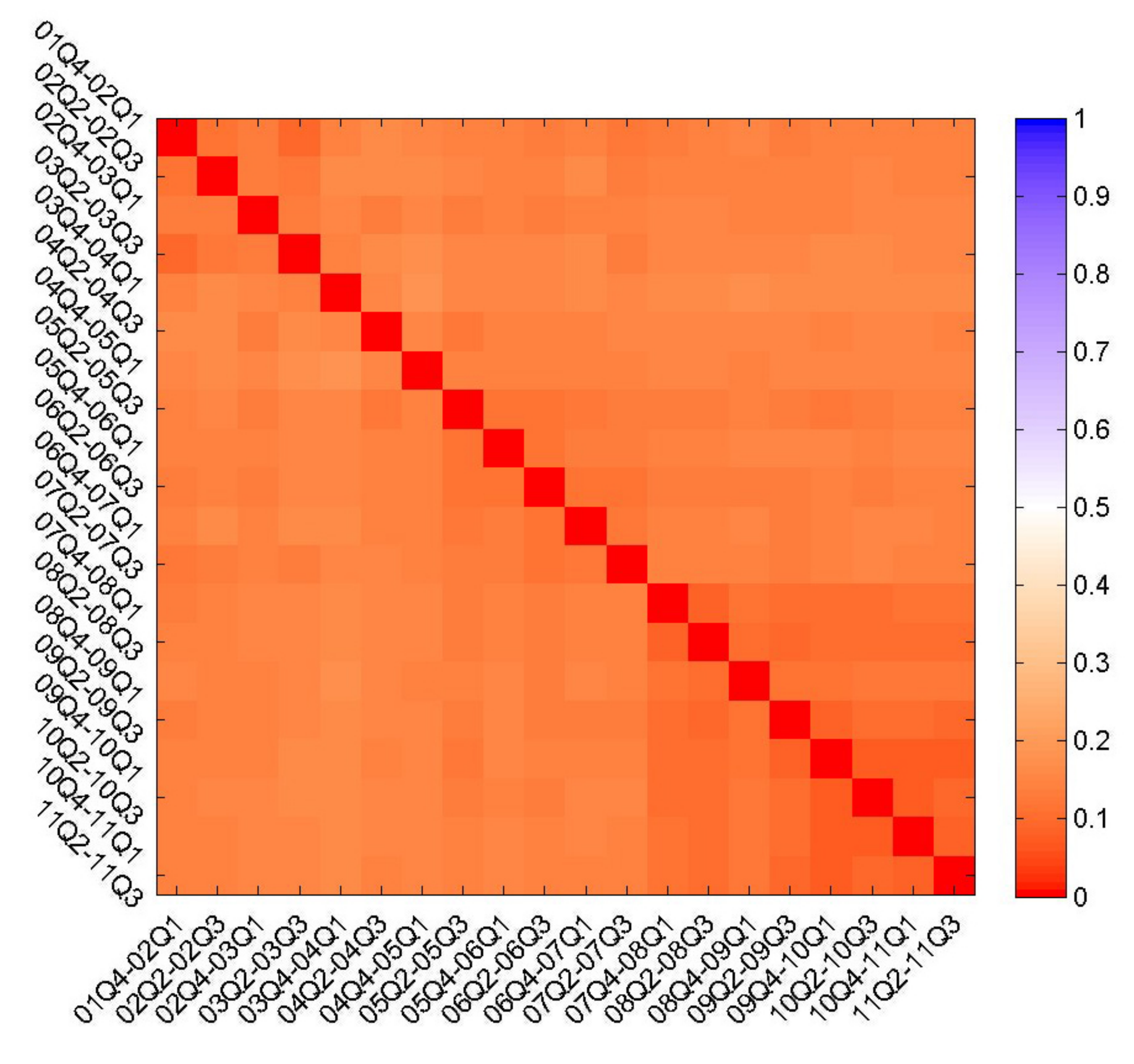}}
\caption{Heat map showing the value of $VI$ between every pair of 6-month time windows, as well as the $VI$ between each window and the total 10-year period (leftmost column and top row). Most notably, there is a slight increase in the similarity of the communities of the last 5 periods 2009 - 2011. (Produced using the Louvain algorithm).}
\label{fig:VISPX6MonthHM}
\end{figure}

To further examine the coherence and fluctuations in communities across all of the six-month windows, in fig. \ref{fig:VISPX6MonthHM} we provide a heat map showing the mutual $VI$ between every two pairs of 6-month windows. Each square in the matrix is a colored representation of the value of $VI$ between the $i^{th}$ and $j^{th}$ 6-month period. Of particular interest, we can see in the lower right corner of the image (which displays the $VI$ between the most recent time windows) that the communities are slightly more similar than communities generated from the other windows. This indicates that there was less movement of stocks between communities during the most recent couple of years of the past decade. Additionally, these periods are closer to the community structure observed when we measured the entire ten-year period. 

As far as explaining this behavior in financial and economical terms, we are again left to hypothesize. Perhaps the observed effect is due to a solidification of communities of stocks caused by the financial collapse, or perhaps it is merely the result of increased accessibility to the markets. 
With the advent of smartphones, tablets, ease of streaming and subscribing to news feeds and social networks in conjunction with faster trading systems, quantitative and high frequency trading, it is conceivable that our increased access to information and the ability to act on it in near real-time has caused a solidifying behavior of the stocks within communities. 

\subsection{Temporal coherence of communities:\\
`hard' and `soft' stocks again}
We display here one final take on the results obtained from our sliding time window, but this time with a stock-centric view, similar to that which we performed for the multifrequency analysis in sec.\ref{sec:resol}. 
In the previous sections we have alluded to how communities change with time. 
One question that should be addressed in conjunction with the previous discussion of time scales is then how the composition of a community changes over time. We have already seen from a variety of $VI$ plots that across each of the six-month periods, the sets of communities look slightly different from each other, but what changes are actually taking place? 
Are there groups of stocks that form tight knit, unwavering cores of communities? Or do they morph fluidly from one to the other, maintaining no coherence over the entire span of ten years?

%%%%%%%%%%%%%%%%
%%%%%%%%%%%%%

\begin{figure}[t]
\centerline{\includegraphics[width=.5\textwidth]{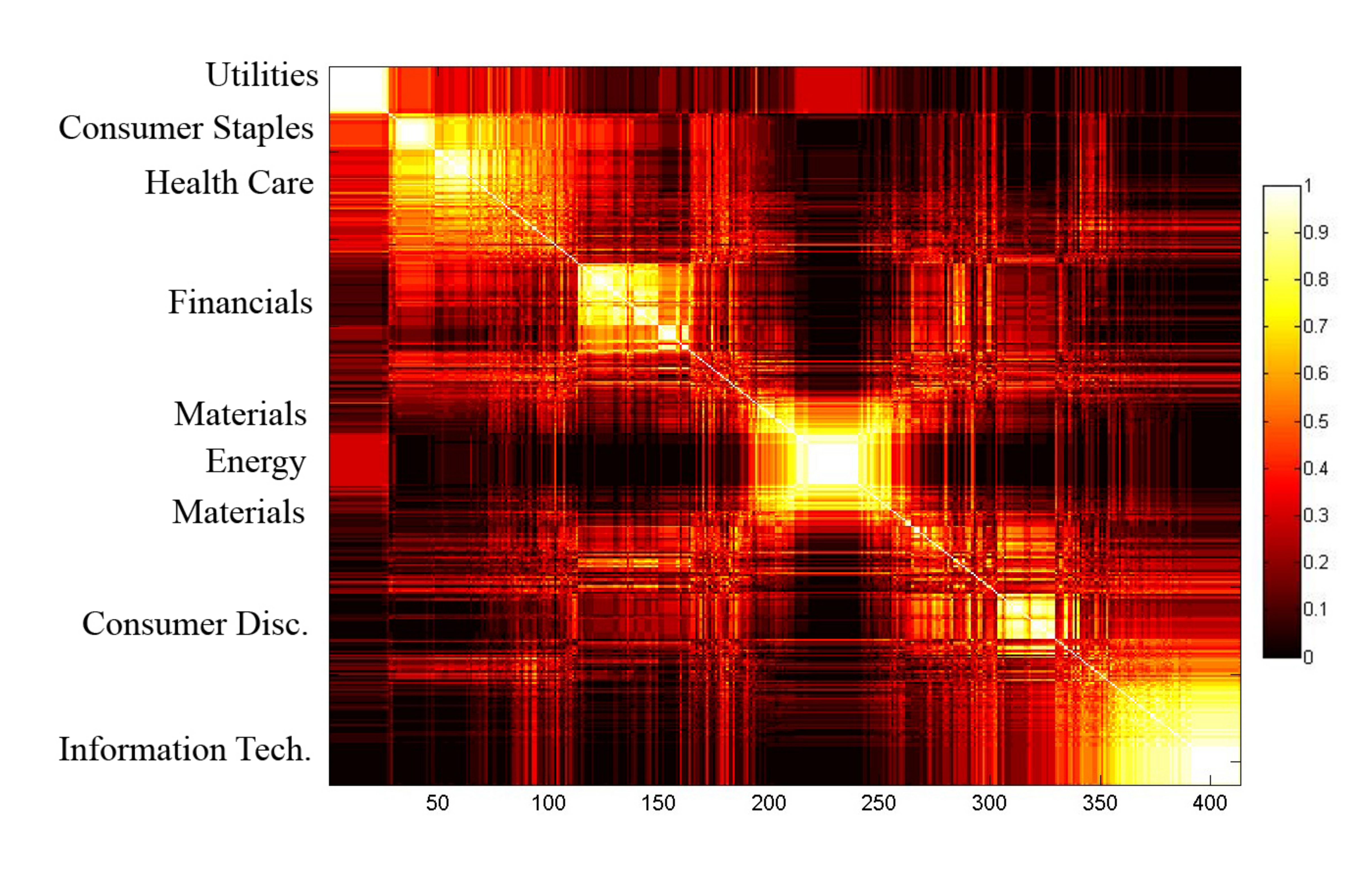}}
\caption{Coherence of communities over time.
The heat map shows the frequency of co-occurrence of different pairs of stocks within the same community over time (white: unit frequency, black: zero frequency). 
The stocks have been ordered using simulated annealing to position stocks with high degree of cross correlations next to each other. To further inform the graph, the GICS sectors have been specified, emphasizing which groupings of stocks tend to associate with a particular sector. (Produced using the Louvain algorithm).
}
\label{fig:ComHM6Months}
\end{figure}

To address this, we examined the sets of stocks that comprised the communities of each six-month time frame of the S\&P over the course of ten years and created a co-occurrence matrix like the one previously shown in fig.\ref{fig:ResolutionHM}, where we calculated the frequency of periods during which any pair of stocks resided in the same community. 
The resulting heat map is presented in figure \ref{fig:ComHM6Months}.
Again, pairs of stocks which were in the same community all the time are white, and those which were never in the same community are black. 

The list of stocks is too long to place on the figure as axis labels, but from observing the raw results we can make some very interesting observations, which we have tried to summarize by labeling again the graph with GICS industry sectors. 
We can see that over the course of ten years, the communities do exhibit strong cores which are unwavering in their construction and constantly anti-correlated with each other. For example there exists a set of core energy, IT and financial stocks which always reside in their own community,  but never share a community with each other. Groups of Energy, Materials and Utilities stocks almost always share the same community, but there have been instances when they did not. Finance is broken into a couple of segments of stocks, such as Banks, Real Estate Investment Trusts (REITs), etc. where the smaller groups always trade with each other but not necessarily aggregated together in a larger community. Similarly, Health Care stocks are fractured in subsets of highly correlated groups of Pharmaceuticals, Services and Biotech, whose allegiance to the larger industry sectors, such as IT and Consumer Staples is more fluid. These trends display interesting overlap with the hierarchical community structure of the S\&P discussed earlier in sec.\ref{sec:hiera}. We also see individual stocks from one top-level industry sector spending most of their time in communities comprised predominantly of a different top-level industry sector, for example Amazon (Consumer Discretionary) spends 90\% of the time in the IT group, as does Motorola (Telecommunications).
%%%%%%%%%%%%%
%%%%%%%%%%%%%%

%%%%%%%%%%%%%%%%%%
%%%%%%%%%%%%

%%%%%%%%%%%%%%%%
%%%%%%%%%%%%%%%%%

\section{Conclusions\label{sec:conclusions}}
In this paper we have addressed the challenging problem of the detection of communities of strongly correlated time series, whose importance resides in the possibility of identifying a mesoscopic level of organization in the dynamics of complex systems.
While the available techniques to analyze correlation matrices failed to detect such modules, we have shown how the concepts of null models, modularity and community detection developed in network theory can be appropriately modified in order to successfully cluster matrices of multiple time series. 
Our redefinitions of the standard methods solve a number of problems encountered when correlation matrices are na\"ively regarded as weighted networks and when ordinary community detection methods are used improperly.

Through the use of various financial markets as examples, we have demonstrated how community detection can be used as a tool to extract specific structural information from time series data.
By surfacing group correlations and trends of the stocks in the S\&P 500, the FTSE 100 and the Nikkei 225, we were able to isolate well-defined communities of stocks such that each community exhibited an internal positive correlation between its constituent stocks, where those same stocks exhibited an aggregate residual anti-correlation with the stocks of each of the other communities. 
While some of these communities showed an association with the more qualitative classification expressed in the GICS industry sector taxonomy, our approach was able to  uncover a host of interesting correlations between stocks of different sectors and industry groups, as well as unsuspected residual anti-correlations between stocks of the same sector. 
As such, our methods and results show that the observed patterns are irreducible to a standard taxonomy, and therefore highlight nontrivial patterns. Moreover, they could prove particularly useful in a number of different fields of finance, such as portfolio optimization and risk management.  

It is worth pointing out that our modifications to the Potts, Louvain and Spectral Optimization algorithms for community detection, although beneficial in and of themselves, act as a proof of concept opening the door to the adaptation of other techniques existing in the field of community detection, allowing e.g. for overlapping, multiresolution, or hierarchical communities \cite{Fortunato_2010,zohar1,zohar2}. 
Similarly, alternative null models controlling for additional or more sophisticated features of the data can also be developed and incorporated in our approach. The key point is that these models, unlike the na\"ive approach, which has been used so far, should always be consistent with correlation matrices. We hope that our approach will stimulate further research in this direction.
Moreover, although we have focused on financial time series as our primary example of real-world data, our general methodology can of course be applied or adapted to any type of time series data, hopefully yielding equally promising results.

We conclude by noting that, abstractly, the ordinary (network-based) community detection techniques and the (correlation-based) clustering that we have introduced can be thought of as lying at two opposite extremes of a more general problem, in the following sense. 
The network-based clustering is in the vast majority of cases aimed at identifying groups of statically linked objects (as captured by a single temporal snapshot of the network) while disregarding their possibly correlated evolution.
By contrast, the correlation-based clustering that we have introduced assumes that the community-defining features are precisely those determining synchronized trends of dynamical activity among nodes, and that the presence (if any) of static dependencies among the latter can be disregarded.
One could of course imagine a more general framework where both static linkages and temporal correlations contribute to the definition of communities, possibly overcoming the `functional versus structural' dichotomy such as the one existing in brain network analysis that we mentioned in the Introduction. 
The present work thus represents one step towards the introduction of a fundamentally more general interpolating formalism.

\begin{acknowledgments}
DG acknowledges support from the Dutch Econophysics Foundation (Stichting Econophysics, Leiden, the Netherlands) with funds from beneficiaries of Duyfken Trading Knowledge BV, Amsterdam, the Netherlands. 
This work was also supported by the EU project MULTIPLEX (contract 317532) and the Netherlands Organization for Scientific Research (NWO/OCW).
\end{acknowledgments}

\begin{appendix}
\section{Redefining community detection methods\label{sec:redefining}}
In this Appendix we show that we can successfully reformulate three of the most popular network-based community detection algorithms in order to properly detect communities of correlated time series using the modified modularity function defined in eq.(\ref{eq:Qunified}).
We stress again that even if the techniques, which we are going to describe, can be considered as three different algorithms implementing the same method of modularity maximization, they are often referred to as different `methods' in the literature. 
In what follows, we will sometimes make use of this somewhat improper terminology.
We will also necessarily use a vocabulary that applies more properly to networks than to time series: for instance, a time series will be often denoted as a `node' (or `vertex') of the `network', and the correlation between two time series will be denoted as the weight of the `link' (or `edge') between the corresponding nodes.

\subsection{Modified Potts method}
The first of the three methods we have selected is based on the so-called $q$-state Potts model \cite{potts,Reichardt_Bornholdt_2006}. It represents the system as a $q$-state spin glass, where each node maintains a spin state $\sigma_i$ (as given by some attempted partition $\vec{\sigma}$) and the weights of the edges between nodes map to coupling strengths. 
So any partition of the network is regarded as a spin configuration $\vec{\sigma}$.
In this paradigm, the modularity $Q(\vec{\sigma})$ is proportional to the negative energy $-\mathcal{H}(\vec{\sigma})$ of the system.
The goal of optimization is then to find the ground state of a spin glass, which corresponds to the maximum value for the modularity. 
The use of a multi-state super-paramagnetic model for graph clustering was first introduced by Blatt, Wiseman and Domany~\cite{PhysRevLett.76.3251} and later revised by Reichardt and Bornholdt~\cite{potts,Reichardt_Bornholdt_2006}; it is upon the latter that we base our extension to incorporate multiple time series. 

Within the $q$-state Potts spin glass model, Reichardt and Bornholdt construct a Hamiltonian by rationalizing a number of energy contributions from the edges between nodes within the same community and nodes in different communities:
\begin {eqnarray}
\mathcal{H}(\vec{\sigma}) & = & - \sum_{i,j}a_{ij}\overbrace{A_{ij}\delta(\sigma_i, \sigma_j)}^\text{internal links} + \sum_{i,j}b_{ij}\overbrace{(1-A_{ij})\delta(\sigma_i, \sigma_j)}^\text{internal non-links}\nonumber \\
& & + \sum_{i,j}c_{ij}\underbrace{A_{ij}[1-\delta(\sigma_i, \sigma_j)]}_\text{external links}\nonumber\\
& & -  \sum_{i,j}d_{ij}\underbrace{(1-A_{ij})[1-\delta(\sigma_i, \sigma_j)]}_\text{external non-links},
\label{Hamiltonian}
\end {eqnarray}
where the contributions from the various types of links can be tuned through the set of coefficients, $a_{ij}$, $b_{ij}$, $c_{ij}$, $d_{ij}$. Instead of directly maximizing the modularity $Q(\vec{\sigma})$ defined in eq.\eqref{QNewman}, Reichardt and Bornholdt minimize the Hamiltonian $\mathcal{H}(\vec{\sigma})$. The latter (under certain conditions and some simplifying assumptions) can be condensed to
\begin {equation} \label{HamiltonianCompact}
\mathcal{H}(\vec{\sigma}) = - \sum_{i,j}\big[A_{ij} - \langle A_{ij}\rangle\big]\delta(\sigma_i, \sigma_j),
\end {equation}    
where $A_{ij}$ is the observed value and $\langle A_{ij}\rangle$ is the corresponding null model for that edge. 
The actual search over spin configurations is done using Simulated Annealing \cite{Kirkpatrick83optimizationby}, which is an approximate technique that in general returns a different solution each time it is used.

In the same vein, introducing a Hamiltonian corresponding to our correlation-based modularity is equally straightforward, however we need to ensure that the logic and derivation that was used to develop the original network-based Hamiltonian holds true for a network created from time series data. 
For a correlation-based network we have the situation where every node is connected to every other node, in principle eliminating the energy contributed by non-links in eq.(\ref{Hamiltonian}) above. 
However, as is ordinarily done when applying the Potts model to weighted networks, we can replace the energy contribution of non-links with the energy contribution of links whose weight is less than expected, allowing us to immediately introduce our null model. 
Maintaining the balance between internal and external edges ($a_{ij}$ = $c_{ij}$ and $b_{ij}$ = $d_{ij}$) as was done by Reichardt and Bornholdt in their original derivation of the Hamiltonian, we end up with a variant of eq. \eqref{HamiltonianCompact} directly derived from a complete weighted network where $A_{ij}$ and $\langle A_{ij}\rangle$ are replaced by the observed correlation $C_{ij}$ and one of our three null models $\langle C_{ij}\rangle_l$ defined in sec. \ref{sec:ourmod}, giving 
\begin {equation} \label{HamiltonianCovariance}
\mathcal{H}_l(\vec{\sigma}) = - \sum_{i,j}C^{(l)}_{ij}\delta(\sigma_i, \sigma_j).
\end {equation}    
Apart from the absence of $C_{norm}$, the r.h.s. of the above expression is the opposite of the r.h.s. of eq.\eqref{eq:Qunified}. Therefore our optimization method using the Potts model will attempt to find the lowest value of the Hamiltonian, which will correspond to the highest modularity.
For the rest, our algorithm is identical to the procedure described by Reichardt and Bornholdt \cite{potts,Reichardt_Bornholdt_2006}.

Therefore the Potts model is a simple algorithm to adapt to correlation matrices, the reason being that although the modularity of the system is explained using a spin-glass model, the actual optimization process is performed using Simulated Annealing, which keeps working even if we use our redefinition of modularity as the cost function \cite{Kirkpatrick83optimizationby}.

\subsection{Modified Louvain method\label{sec:louvain}}
We now consider a second approach to the problem of modularity optimization. 
Possibly one of the most successful approaches, the Louvain method~\cite{1742-5468-2008-10-P10008} (named after the University from which it emerged) is a simple, greedy, agglomerative algorithm whose strength lies in the fact that it is computationally fast. Unlike the spin-glass model, the Louvain method does not set up a framework for its optimization problem. It simply starts from the definition of modularity specified in eq. \eqref{QNewman} and derives a new, more computationally efficient equation for testing the relative gain in modularity by moving a node from one community to another. It is this equation that allows the Louvain method to perform so well. 

The method initially considers all nodes as placed in individual communities, and then calculates (sequentially for each node $i$) the gain of modularity associated with moving node $i$ to the same community where each of its neighbours $j$ belong. 
The algorithm explores all possible such moves and implements those that give the maximum gain in modularity, and the first iteration stops when no further improvement is possible. 
Then, a new `renormalized' network is built by merging all nodes within the previously found communities into a single `hypernode', and the algorithm is iterated again until there are no more possible changes and a maximum of modularity is attained. 
To do so, the renormalized weight of the link
between two hypernodes is defined as the sum of the weight of the links between nodes
in the corresponding two communities. 
Links between nodes of the same community
lead to self-loops for the corresponding hypernode.

The key requirement of the Louvain method is that the system can be properly renormalized, i.e. that successive coarse-grainings of the system remain consistent with the meaning of the modularity at the corresponding level of aggregation. 
In an ordinary  network this is relatively straightforward to show, i.e. a hypernode obtained merging two or more nodes can be legitimately interpreted (from the point of view of the modularity function) as a coarse-grained node with a self-loop to itself and renormalized interactions to all other (hyper)nodes.
It is however not intuitively obvious whether our modularity defined in eq.(\ref{eq:Qunified}) admits an equivalently consistent definition of `renormalized time series' obtained by `merging' two or more time series.
And even if such a definition exists, one should understand how to correctly define also the renormalized interactions and self-loops.
To this end, we recall from eq.(\ref{eq:corrs}) that if $X_i$ and $X_j$ are two standardized time series then $C_{ij}=\textrm{Cov}[X_i,X_j]$. 
We can therefore exploit the fact that the covariance is a bilinear function of its arguments to calculate the following renormalized interactions between two hypernodes (communities) $A$ and $B$:
\begin{eqnarray}
\sum_{i\in A}\sum_{j\in B}C_{ij}&=&\sum_{i\in A}\sum_{j\in B}\textrm{Cov}[X_{i},X_{j}]\nonumber\\
&=&\textrm{Cov}\Big[\sum_{i\in A}X_i,\sum_{j\in B}X_{j}\Big].
\end{eqnarray}
The above formula shows that, if we define the `renormalized time series' of community $A$ as 
\begin{equation}
\tilde{X}_A\equiv\sum_{i\in A}X_i,
\end{equation}
then we can consistently define the renormalized interactions as 
\begin{equation}
\tilde{C}_{AB}\equiv \sum_{i\in A}\sum_{j\in B}C_{ij}=\textrm{Cov}\big[\tilde{X}_A,\tilde{X}_B\big]
\label{eq:l1}
\end{equation}
and the renormalized self-loops as
\begin{equation}
\tilde{C}_{AA}=\textrm{Cov}\big[\tilde{X}_A,\tilde{X}_A\big]=\textrm{Var}\big[\tilde{X}_A\big].
\end{equation}
We therefore find that, for a graph composed of financial time series, renormalized interactions have a correct interpretation in terms of covariances, rather than correlations.
They also show that the summation of a group of time series yields something that resembles an index fund of the set of stocks, so the concept of aggregating nodes maintains a strong grounding in reality.

We now have to check whether the modularity function remains consistent with the null model when defined at the level of renormalized nodes.
Note that the linearity of the definition of $\tilde{C}_{AB}$ ensures that, given any of our null models defined in sec.\ref{sec:ourmod}, we can write
\begin{equation}
\langle\tilde{C}_{AB}\rangle_l= \sum_{i\in A}\sum_{j\in B}\langle C_{ij}\rangle_l.
\label{eq:l2}
\end{equation}
This means that the filtered quantity $C_{ij}-\langle C_{ij}\rangle_l$ can be  similarly renormalized as 
\begin{equation}
\tilde{C}_{AB}^{(l)}\equiv\sum_{i\in A}\sum_{j\in B}C_{ij}^{(l)}
\label{eq:borrowed}
\end{equation}
for each of the three cases in eq.(\ref{eq:l}).
Now, imagine that in subsequent iterations of the model the hypernodes are further merged into `communities of communities'.
The resulting `metapartition' can be specified by a vector $\vec{\tilde{\sigma}}$ of dimension smaller than (or equal to, if the metapartition is trivial) any of the original vectors $\vec{{\sigma}}$. Each element $\tilde{\sigma}_A$ denotes the community to which the hypernode $A$ is placed by the metapartition. 
If $\vec{\sigma}$ denotes the underlying (node-level) partition identified by the metapartition $\vec{\tilde{\sigma}}$ (i.e. $\sigma_i=\tilde{\sigma}_A$ for all $i\in A$), we can define the renormalized modularity
\begin {eqnarray}
\tilde{Q}_l(\vec{\tilde{\sigma}})&\equiv& \frac {1}{\tilde{C}_{norm}}\sum_{A,B} \tilde{C}_{AB}^{(l)}\delta(\tilde{\sigma}_A,\tilde{\sigma}_B)\nonumber\\
&=& \frac {1}{\tilde{C}_{norm}}\sum_{A,B} \sum_{i\in A}\sum_{j\in B}C_{ij}^{(l)}\delta(\tilde{\sigma}_A,\tilde{\sigma}_B)\nonumber\\
&=& \frac {1}{\tilde{C}_{norm}}\sum_{i,j}C_{ij}^{(l)}\delta({\sigma}_i,{\sigma}_j)\nonumber\\
&=& Q_l(\vec{\sigma}),
\label{eq:Qinvariant}
\end {eqnarray}
where, in analogy with eq.\eqref{eq:norm}, we have defined
\begin{equation}
\tilde{C}_{norm}\equiv \sum_{A,B}\tilde{C}_{AB}=
\sum_{A,B}\sum_{i\in A}\sum_{j\in B}C_{ij}=
\sum_{i,j}C_{ij}=C_{norm}.\nonumber
\end{equation}
Equation \eqref{eq:Qinvariant} coincides with the original modularity defined at the level of individual nodes.
This means that the modularity is manifestly invariant under renormalization, implying that we can indeed consistently redefine a coarse-grained modularity at each iteration of the Louvain method.

The second requirement of the Louvain method is the fact that the change in the modularity obtained by adding a previously isolated node to a given pre-existing community can be easily calculated.
This ensures the computational efficiency of the algorithm.
In adapting the model to correlation-based networks, we must start from eq. \eqref{eq:Qunified} and check whether this is still the case, and if so arrive at a new corresponding expression for the modularity change.
We will do so using directly the invariant modularity defined in eq.(\ref{eq:Qinvariant}), so that we are sure that the result will hold at any aggregation level.
Given the modularity $\tilde{Q}_l(\vec{\tilde{\sigma}})$, we denote the modularity change obtained by adding the (hyper)node $I$ to the community $J$ by $\Delta \tilde{Q}_l^{(I\to J)}$ and calculate it as the difference between $\tilde{Q}_l(\vec{\tilde{\sigma}}')$ for a (meta)partition $\vec{\tilde{\sigma}}'$ where $I$ is part of the community $J$ (i.e. $\tilde{\sigma}'_I=J$) and $\tilde{Q}_l(\vec{\tilde{\sigma}}'')$ for a (meta)partition $\vec{\tilde{\sigma}}''$ where $I$ is isolated in its own community (i.e. $\tilde{\sigma}_I''\ne J$). 
Since $\tilde{\sigma}'_A=\tilde{\sigma}''_A$ for all $A\ne I$ and $\delta(\tilde{\sigma}''_I,\tilde{\sigma}''_A)=0$ for all $A\ne I$, we can write this difference as
\begin{eqnarray}
\Delta \tilde{Q}_l^{(I\to J)} & = & \tilde{Q}_l(\vec{\tilde{\sigma}}')
-\tilde{Q}_l(\vec{\tilde{\sigma}}'')\nonumber\\
&=&\frac{1}{C_{norm}}\sum_{A,B}\tilde{C}^{(l)}_{AB}
\Big[\delta(\tilde{\sigma}'_A,\tilde{\sigma}'_B)-\delta(\tilde{\sigma}''_A\tilde{\sigma}''_B)\Big]\nonumber\\
&=&\frac{1}{C_{norm}}\sum_{A}\tilde{C}^{(l)}_{IA}\Big[\delta(\tilde{\sigma}'_I,\tilde{\sigma}'_A)-\delta(\tilde{\sigma}''_I,\tilde{\sigma}''_A)\Big]\nonumber\\
& = & \frac{1}{C_{norm}} \sum_{A \in J} \tilde{C}^{(l)}_{IA}\nonumber\\
& = & \frac{\tilde{C}^{(l)}_{IJ}}{C_{norm}}.
\label{eq:K}
\end{eqnarray}
That is, the change in modularity obtained from adding a (hyper)node $I$ to a pre-existing community $J$ is simply proportional to the renormalized interaction between $I$ and $J$, i.e. the sum of the (filtered) correlations of all time series within $I$ with all those within $J$.
Note that in the above formula the notation $A\in J$ implies $A\ne I$, since $I$ does not (yet) belong to $J$.

Similarly, it is possible to calculate the change in modularity $-\Delta \tilde{Q}_l^{(I\to J')}$ obtained when a (hyper)node $I$ belonging to a community $J'$ is disconnected from the latter and placed in its own isolated community.
Combining these two contributions, we can easily calculate the change in modularity \begin{equation}
-\Delta \tilde{Q}_l^{(I\to J')}+\Delta \tilde{Q}_l^{(I\to J)}=\frac{\tilde{C}^{(l)}_{IJ}-\tilde{C}^{(l)}_{IJ'}}{C_{norm}}
\end{equation}
obtained by moving a (hyper)node $I$ from a community $J'$ to a different community $J$.
So our reformulation above satisfies also the second requirement of the Louvain method at all aggregation levels, and allows us to define a computationally efficient method to detect communities of time series.

\subsection{Modified spectral method}
We now come to the third and final method of optimizing the modularity cost function. 
Spectral Optimization is the process of using matrix eigendecomposition to recursively bisect a network into communities of nodes according to the principle of maximizing the modularity function~\cite{Newman_2006}. 
The matrix which is the subject of the eigendecomposition is the so-called \emph{modularity matrix} appearing in eq.\eqref{QNewman} and having entries $B_{ij} = A_{ij} - \frac{k_i k_j}{2m}$.
In other words, the modularity matrix $\mathbf{B}$ is the difference between the observed network, represented by the adjacency matrix $\mathbf{A}$, and the null model $\langle\mathbf{A}\rangle$. 
In the spectral method, the modularity matrix is eigendecomposed into its constituent eigenvalues and eigenvectors, the intent being to isolate the eigenvector corresponding to the largest eigenvalue and use the signs of the elements of this vector to infer an optimal partition.
Specifically, the network is split into two communities, each comprising the nodes corresponding to eigenvector components with the same sign.
The process is implemented recursively in each partition (deriving a new modularity matrix for every community), until no further increase in modularity is obtained.

We need to extend this algorithm to accommodate correlation-based networks. 
In our case, as clear from eqs.\eqref{eq:Qunified} and \eqref{eq:l}, the modularity matrix is $\mathbf{C}^{(l)}$,
i.e. the filtered matrix defined using one of our three null models. 
We fill therefore adapt the procedure outlined by Newman in the original paper, and implement the spectral optimization method by iteratively bisecting the network into two sub-communities (say $A$ and $B$).
Each such bisection can be denoted either by an appropriate partition vector $\vec{\sigma}$ or equivalently by a vector $\vec{s}$ having elements $s_i=-1$ if node $i$ belongs to (say) community $A$ and $s_i=+1$ if $i$ belongs to community $B$. 
The correspondence between these vectors is given by
\begin{equation}
\delta(\sigma_i,\sigma_j)=\frac{s_i s_j +1}{2}.
\end{equation}
Given a bisection, we can therefore rewrite our unified correlation-based modularity $Q_l(\vec{\sigma})$ defined in eq.(\ref{eq:Qunified}) as
\begin {eqnarray}
Q_l(\vec{s}) &=& \frac{1}{C_{norm}}\sum_{i,j} {C^{(l)}_{ij}} \frac{s_i s_j + 1}{2}\label {QC}\\
& =& \frac{1}{2C_{norm}}\sum_{i,j} {C^{(l)}_{ij}} s_i s_j
+\frac{1}{2C_{norm}}\sum_{i,j} {C_{ij}^{(l)}}.\nonumber
\end {eqnarray}
In Newman's original formulation the last term sums to zero, because the network-based modularity matrix $\mathbf{B}$ has the property that all of its rows sum to zero. 
However, this is not the case with our correlation-based modularity matrix $\textbf{C}^{(l)} $ defined in eq.\eqref{eq:l}.
So we retain the second term and, defining $C^{(l)}_{tot}\equiv\sum_{i,j}C^{(l)}_{ij}$, rewrite our modularity in matrix form as
\begin{equation}
Q_l (\vec{s})= \frac{\langle {s}| \textbf{C}^{(l)} | {s}\rangle }{2C_{norm}} +\frac{C^{(l)}_{tot}}{2C_{norm}}.
\end{equation}
The vector $\vec{s}$ maximizing $Q_l (\vec{s})$ is easily found as the vector matching the signs of the components of the eigenvector of $\textbf{C}^{(l)}$ corresponding to the largest eigenvalue.
Clearly, both $C_{norm}$ and $C^{(l)}_{tot}$ have no effect on the result, making the original procedure of the spectral algorithm consistent with our reformulation.

After the initial bisection, we need to calculate the potential modularity change $\Delta Q_l$ obtained by further subdividing the communities yielded in the previous step. 
Let us consider the case where one community (say $A$) among the ones obtained thus far in the algorithm is further subdivided into two new communities (say $A_1$ and $A_2$). 
If $\vec{s}$ is a vector (restricted to the vertices in $A$ only) denoting the bisection of $A$ into $A_1$ and $A_2$, then the modularity change associated with such bisection reads
\begin{eqnarray}
\Delta Q_l^{(A_1|A_2)} &=&  \frac{1}{C_{norm}}\Big[\sum_{i,j \in A_1}\!\! C^{(l)}_{ij}  + \!\!\sum_{i,j \in A_2} \!\! C^{(l)}_{ij} -\!\! \sum_{i,j \in A}\!\! C^{(l)}_{ij} \Big]\nonumber\\
&=&\frac{1}{C_{norm}}\Big [ \sum_{i,j \in A} C^{(l)}_{ij} \frac{s_i s_j +1}{2} - \sum_{i,j \in A} C^{(l)}_{ij}\Big]\nonumber\\
&=&\frac{1}{2C_{norm}}\Big [ \sum_{i,j \in A} C^{(l)}_{ij} s_i s_j -\sum_{i,j \in A} C^{(l)}_{ij}\Big]  \nonumber\\
&=&\frac{\langle s| \textbf{C}_A^{(l)} | s\rangle }{2C_{norm}} +\frac{\tilde{C}^{(l)}_{AA}}{2C_{norm}},
\label{eq:deltaQnewman}
\end{eqnarray}
where $\textbf{C}_A^{(l)}$ represents the sub-matrix of $\textbf{C}^{(l)}$ restricted to the subset of nodes within community $A$, and the notation $\tilde{C}^{(l)}_{AA}$ is borrowed from eq.\eqref{eq:borrowed}.
As with the initial bisection, $\vec{s}$ is chosen to maximize $\Delta Q_l^{(A_1|A_2)}$ by selecting its elements to match the sign of the eigenvector corresponding to the largest eigenvalue of the matrix $\textbf{C}_A^{(l)}$. 

As for the original algorithm, our modified spectral method proceeds by iterating the above procedure until no further bisection can make the modularity increase.
\end{appendix}

%\begin{small}
%\addcontentsline{toc}{section}{References}
%\bibliographystyle{apsrev4-1}
%\bibliography{biblio}
%\end{small}

\end{document}